\documentclass[opre,nonblindrev]{informs3}

\OneAndAHalfSpacedXI 

\usepackage{endnotes}
\let\footnote=\endnote
\let\enotesize=\normalsize
\def\notesname{Endnotes}%
\def\makeenmark{\hbox to1.275em{\theenmark.\enskip\hss}}
\def\enoteformat{\rightskip0pt\leftskip0pt\parindent=1.275em
  \leavevmode\llap{\makeenmark}}

\usepackage{graphicx}
\usepackage{subfigure}
\usepackage{natbib}
 \bibpunct[, ]{(}{)}{,}{a}{}{,}%
 \def\bibfont{\small}%
\usepackage{hyperref}
\hypersetup{hypertex=true,
            colorlinks=true,
            linkcolor=blue,
            anchorcolor=blue,
            citecolor=blue}
\TheoremsNumberedThrough     
\ECRepeatTheorems
\usepackage{url}
\EquationsNumberedThrough    

\usepackage{subfigure, epsfig, epstopdf}
\usepackage{graphicx,amssymb}
\usepackage{multirow, multicol}
\usepackage{amsmath,etoolbox}
\usepackage{color}
\usepackage{setspace}
\usepackage{mathrsfs}
\usepackage{dsfont}
\usepackage{bbm}
\usepackage{textcomp}

\usepackage{tikz} 
\usetikzlibrary{shapes.geometric, arrows} 
\usetikzlibrary{snakes}

\usepackage[]{algorithm}
\usepackage{algpseudocode}

\usepackage{array}
\usepackage{threeparttable}
\usepackage{graphicx}
\usepackage{verbatim}

\usepackage{multibib}
\newcites{appendix}{References}

\patchcmd{\subequations}
 {\def\theequation{\theparentequation\alph{equation}}}
 {\def\theequation{\theparentequation.\arabic{equation}}}
 {}{}

\graphicspath{{figure/}}

\begin{document}



\RUNTITLE{Online OPF planning for power systems under bushfire}

\TITLE{Online Planning of Power Flows for Power Systems Against Bushfires Using Spatial Context}

\ARTICLEAUTHORS{%
	
\AUTHOR{Jianyu Xu,$^\text{a}$ Qiuzhuang Sun,$^\text{b}$ Yang Yang,$^\text{c}$ Huadong Mo,$^\text{d}$ Daoyi Dong$^\text{e,f}$}
\AFF{$^\text{a}$ International Business School Suzhou, Xi'an Jiaotong-Liverpool University, Suzhou, China} 
\AFF{$^\text{b}$ School of Mathematics and Statistics, University of Sydney, Australia} 
\AFF{$^\text{c}$ Department of Industrial Systems Engineering and Management, National University of Singapore, Singapore} 
\AFF{$^\text{d}$ School of Systems and Computing, University of New South Wales, Canberra, Australia}
\AFF{$^\text{e}$ School of Engineering and Technology, University of New South Wales, Canberra, Australia}
\AFF{$^\text{f}$ CIICADA Lab, School of Engineering, Australian National University, ACT, Australia}
} 

\ABSTRACT{%
The 2019--20 Australia bushfire incurred numerous economic losses and significantly affected the operations management of power systems. 
A power station or transmission line can be significantly affected due to bushfires, leading to an increase in operational costs.
We study a fundamental but challenging problem of planning the optimal power flow (OPF) for power systems subject to bushfires. 
Considering the stochastic nature of bushfire spread, we develop a model to capture such dynamics based on Moore’s neighborhood model.
Under a periodic inspection scheme that reveals the in-situ bushfire status, we propose an online optimization modeling framework that sequentially plans the power flows in the electricity network.
Our framework assumes that the spread of bushfires is non-stationary over time, and the spread and containment probabilities are unknown.
To meet these challenges, we develop a contextual online learning algorithm that treats the in-situ geographical information of the bushfire as a “spatial context”.
The online learning algorithm learns the unknown probabilities sequentially based on the observed data, and then accordingly makes the OPF decision.
The sequential OPF decisions aim to minimize the regret function, which is defined as the cumulative loss against the clairvoyant strategy that knows the true model parameters.
We provide a theoretical guarantee of our algorithm by deriving a bound on the regret function, which outperforms the regret bound achieved by other benchmark algorithms.
Our model assumptions are verified by the real bushfire data from NSW, Australia, and we apply our model to two power systems to illustrate its applicability.
}%



\maketitle

%


\setcounter{equation}{0}
\setcounter{lemma}{0}
\setcounter{section}{0}
\setcounter{theorem}{0}
\renewcommand*{\theHequation}{\arabic{section}.\arabic{equation}} 
\renewcommand*{\theHlemma}{\arabic{section}.\arabic{lemma}} 
\renewcommand*{\theHsection}{\arabic{section}.\arabic{section}} 
\renewcommand*{\theHtheorem}{\arabic{section}.\arabic{theorem}} 


\section{Introduction}\label{section introduction}

The devastating bushfire in Australia from 2019 to 2020 is record-setting in terms of size and destructiveness and well known to incur severe impacts on its economy, landscape, forests, and wild animals \citep{munawar2021uav}. 
Power systems, as critical infrastructure that drive daily activities in modern society, have been significantly impacted, for instance, in both the 2018 wildfires in California, USA, and the 2019--2020 Australia bushfire season \citep{wang2022bushfire}.
According to the operational practice for bushfire prevention, a power station is forced to shut down and the transmission/distribution lines may be just-in-time disconnected at the ignition or approach of bushfires \citep{guo2018determination,Moutis2022pmu,Su2023qua}. 
Otherwise, inappropriate operations management of power systems would lead to cascading failures in other systems and cause huge economic losses \citep{Parker2019Ele,Sushil2016Dis}.
For instance, amidst the 2023
Maui fires were caused by Hawaiian Electric, which neglected to deactivate its power grid
in Lahaina, even as other fires ignited and nearly 24 hours of intense winds occurred amidst dry weather circumstances \citep{Emily2023BBC}.
Hence, the need for electricity network operational management against bushfires is becoming progressively prominent with the increasing risk of bushfires in Australia \citep{csiro20202019}. 
In particular, the random spread of bushfires makes the dynamic adjustment of the optimal power flow (OPF) for power systems quite challenging.

To optimize power flows, existing literature generally uses two types of models, i.e., the alternating current (AC) and direct current (DC) models.
The AC OPF model better reflects real-world physics, but the corresponding optimization problem is non-convex and non-linear, making it hard to be solved in real time \citep{hong2022data,aigner2023solving,Dmitry2021Mo}.
Compared with the long computational time of AC OPF models, DC OPF models can often be solved in a reasonable computational time.
Although it simplifies the physics of power flows, DC OPF models generally achieve satisfactory performance in practice and are commonly used in literature \citep{pena2020dc,rhodes2020balancing,johnson2022scalable,misra2022learning}. Moreover, the DC formulation is also widely used for fault analysis problems \citep{Xiang2021Deep}, optimizing network topology \citep{Burak2016A}, and optimizing near-real-time generator dispatch \citep{Ignacio2023Re}.
Facing bushfires, the power system aims to respond quickly to dynamic environmental conditions.
Therefore, this study adopts the DC OPF model as a building block to obtain an online dispatch solution.

Specifically, this study aims to address the OPF problem in an online manner based on the in-situ information on bushfires over time. 
We mainly answer the following three questions for the online electricity network management problem:
\begin{itemize}
\item[(i)] How to model the evolution of a bushfire?
\item[(ii)] How to model the influence of the bushfire on the operations management of power systems?
\item[(iii)] How to sequentially make online decisions of OPF for power systems to minimize the cumulative loss caused by the bushfire.
\end{itemize}

To answer the first question on the modeling of bushfires, there is a trade-off between the complexity and tractability of the bushfire spread model.
A complex statistical model can accurately capture the dynamics of bushfires over time (e.g., \citealp{wei2022physics, Vazquez2022wild}), by considering features such as meteorological, topographic, and remote sensing data, the normalized difference vegetation index, and vegetation coverage, etc.
However, the resulting model may be difficult to be integrated into an optimization framework that aids authorities in determining the OPF of an electricity network \citep{salehi2018survey}.
Using a complex statistical model becomes intractable in an online optimization setting where we need to dynamically learn the bushfire spread and accordingly adjust the OPF of the power system. 
To make the online optimization problem tractable, we consider two commonly used models for bushfire, i.e., von Neumann’s neighborhood model and Moore’s neighborhood model \citep{ito2020evaluation,wu2022simulation}.
Both models are based on the techniques of cellular automata \citep{wootton2001local}, and the effectiveness of these models has been verified by the Commonwealth Scientific and Industrial Research Organisation in Australia (CSIRO), based on the dataset ``Bureau of Meteorology Annual Climate Statement 2019'' \citep{BOM2020}. 
We adopt Moore’s neighborhood model to fit and predict the bushfire spread due to its flexibility and then plan the optimal power flow for predictable bushfires \citep{Manoj2016An,Stauffer2020Im}.
However, existing Moore’s neighborhood models do not consider the suppression and containment of the fire, possibly done by the fire management personnel. 
To improve the applicability and accuracy of Moore’s neighborhood model, this study further considers the containment rate of bushfire, i.e., the probability of the bushfire being extinguished \citep{plucinski2011effect}, and develops a more realistic model for bushfire spread. In comparison, most existing literature analyzes the impact of bushfires on power systems in a static way \citep{Dorrer2018use,rhodes2020balancing,hong2022data}.
Considering the dynamic spread of bushfires significantly complicates the OPF problem.


To answer the remaining two operation management questions, we model the influence of bushfires on the power system.
A power system typically has various hierarchical levels with many components, making it {\color{blue!80!black}challenging} to integrate a bushfire spread model into a power system management problem.
We first adopt the framework in \cite{xu2022online} that uses a distributed generation system to model the electricity network \citep{Mehdi2016Mul,Angelus2020Dis}, where the loads, conventional generators, renewable resources, feeders, and transmission lines are physically distributed and subject to bushfires in the same region.
The influence of bushfire spread on the power system can then be quantified by its impact on the power flow and the resulting increase in the operational cost \citep{xu2021bayesian}.
In particular, we embed the power system into a grid map that represents the region under the threat of bushfires.
During each time, we can observe the area on fire represented by a set of nodes on the map. 
Generators and transmission lines have to be closed down if the distance between them and the fire is too close.
As a result, the working load and maximum capacity of a generator become zero, and the capacity of a transmission line also becomes zero.
In this case, backup energy storage (power transmission measure) is activated to cover the capacity loss and keep the power flow of the system balanced.
Such operations management causes additional operational costs that quantify the negative impact of bushfire \citep{wang2018risk}. 
In Section~\ref{section practical example of bushfire spread}, we use an IEEE 57 bus system to explicitly illustrate the economic loss of a power system subject to the threat of bushfires.

The main challenges of online OPF planning result from the stochastic nature of bushfire spread.
As such, decision-makers need to adaptively adjust the power flows of the electricity network based on the in-situ bushfire information.
However, most existing literature considers the static OPF or other management problems for power systems \citep{birchfield2016statistical,xavier2021learning,zolan2021decomposing}, so they cannot be directly applied to our online problem.
\cite{Yang2022optimizing} and \cite{Kadir2023re} consider stochastic planning problems after hurricanes or wildfires, but all model parameters are assumed to be known for optimization.
However, the dynamics of bushfire spread are complex, so this study assumes that the probability of bushfire spread is unknown to decision-makers and changes over time.
Hence, we need to learn these model parameters while simultaneously predicting the future spread of bushfires.
This task is complicated due to the non-stationary nature of the bushfire evolution \citep{mccarthy2012analysis}.
If an algorithm fails to track the dynamic change of these model parameters precisely and timely, the prediction bias of the bushfire spread can be large and result in a surge in operational costs. 
To the best of our knowledge, there are no existing online optimization algorithms that can plan the power flows for an electricity network adaptively and provide a theoretical guarantee of performance.


To meet the above challenges, we make a methodological contribution by proposing an online optimization algorithm that minimizes the loss of the power system subject to bushfires.
We follow the idea of contextual online learning \citep{agrawal2016linear,simchi2022bypassing} and consider the area on fire at each decision epoch as a ``spatial context''. 
Based on the state-of-the-art adaptive change point detection method in \cite{auer2019adaptively}, we develop an algorithm that sequentially learns the transition probabilities for the bushfire spread model with the ``spatial context''.
This learning scheme leads to a precise prediction of the bushfire spread in the future, such that we can plan the power flows of the electricity network to minimize the operational cost. 
Compared with existing models that only consider offline optimization, our online model has the potential to prevent more serious losses due to cascading failures of a power system subject to the dynamic spread and containment of bushfires.

The main contributions of this study are summarized as follows.
\begin{itemize}
\item[(i)] \textit{Modeling of power systems subject to bushfire.}
Based on Moore’s neighborhood model, we develop a bushfire spread model that considers the containment rate of the fire, which is largely overlooked in prior literature.
We link this model to the planning of power flows of an electricity network.
Even though the 2019--2020 bushfire in Australia has caused numerous losses for power systems,
to the best of our knowledge, no existing study has tackled such a kind of problems.
\item[(ii)] \textit{New online optimization algorithms.} Different from most existing literature, we assume the bushfire spread and containment probabilities are unknown. 
We propose an online optimization algorithm to simultaneously learn the unknown parameters and plan the power flows.
Compared with most existing literature on the OPF of power systems, 
the proposed algorithm can adaptively adjust the operations management of an electricity network based on the in-situ state of bushfires. 
Furthermore, our algorithm is shown to retain a sublinear regret over time that is near-optimal.
\item[(iii)] \textit{Significant improvement for real-world power systems.}
We apply the proposed online optimization algorithm to real-world power systems in our numerical experiments.
The results reveal that our method significantly outperforms the existing change point detection approaches when the evolution of bushfires is non-stationary over time and can maintain excellent computational efficiency even though the complexity of the power system increases (from 11 buses to 57 buses).
Our numerical study uses a real-world bushfire dataset to validate that our model assumptions are valid.
This suggests the applicability of our model in practice.
\end{itemize}

The rest of the paper is organized as follows. Section \ref{section problem formulation} formulates the mathematical model for the OPF problem of smart power systems against bushfires. 
Section \ref{sec-online-opt-alg} proposes the main algorithm of this study and proves the performance guarantee of the algorithm. 
Section \ref{section practical example of bushfire spread} verifies our model settings and assumptions through a practical bushfire dataset.
Section \ref{section simulation study} uses numerical simulations to illustrate the advantages of our proposed algorithm by comparing it with several benchmark algorithms. Section \ref{section conclusions} presents the concluding remarks.

\section{Problem Formulation}\label{section problem formulation}

This section formulates the operational management model of the electricity network under bushfires.
Section~\ref{section electricity network operational problem} first formulates a static optimization problem that determines the OPF for the electricity network.
Section \ref{section bushfire spread model} characterizes the bushfire spread based on the classical Moore’s neighborhood model.
Section \ref{section operational management against bushfire} combines the OPF model with the bushfire spread model to formulate our dynamic online optimization problem.

\subsection{Electricity network operational problem}\label{section electricity network operational problem}

We consider a geographical region that is represented as a grid on the map. 
Each node of the grid corresponds to a geographical location in reality. 
For ease of discussion, we suppose that the region is represented by a rectangular grid network with $N$ rows and $M$ columns, so the total number of nodes is $MN$. 
We denote by $(x_{i},\,y_{i})$ the coordinates of node $i$ and
$\mathcal{G}\triangleq\{i: x_i\in \{1,\ldots,M\},~y_i\in\{1,\ldots,N\}\}$ the set of nodes in the grid. 

\begin{table}[h]
\centering
\scriptsize
\caption{\textcolor{red}{List of symbols and their meanings as presented in Section 2.1.}}
\label{tab:2.1symbols}
\begin{tabular}{p{2.2cm} p{5.5cm}} 
\hline
\textbf{Symbol} & \textbf{Meaning} \\ \hline
$N, M$ & Grid size with $N$ rows and $M$ columns \\
$(x_i, y_i)$ & Coordinates of node $i$ in the grid \\
$\mathcal{G}$ & Set of all nodes, where $x_i \in \{1, \dots, M\}$, $y_i \in \{1, \dots, N\}$ \\
$\mathcal{N}_k(i)$ & Set of $k$-neighbors of node $i$, where $\mathcal{N}_1(i)$ includes the 8 surrounding nodes (king-move neighborhood) \\
$d(i, j), d(i, \mathcal{U})$ & Distance measure, where $d(i, j)$ is the distance between nodes $i$ and $j$, and $d(i, \mathcal{U}) = \min_{j \in \mathcal{U}} d(i, j)$ \\
$\mathcal{S}, \mathcal{E}$ & Set of stakeholders (generators or consumers); Set of transmission lines \\
$L(i)$ & Working load at node $i$ \\
$U^\text{P}(i), U^\text{F}(i, j)$ & Maximum capacity of node $i$ ($U^\text{P}(i) = 0$ for consumers); Maximum flow on transmission line $(i, j)$ \\
$\texttt{PTDF}((i, j), k)$ & Power Transfer Distribution Factor of node $k$ on line $(i, j)$ \\
$\alpha(i), \beta(i, j)$ & Capacity of node $i$; Flow on line $(i, j)$ \\
$C^\text{P}(i)$ & Unit cost of capacity for node $i$ \\ \hline
\end{tabular}
\end{table}

\begin{figure}[h]
    \centering
    \includegraphics[width=0.8\textwidth]{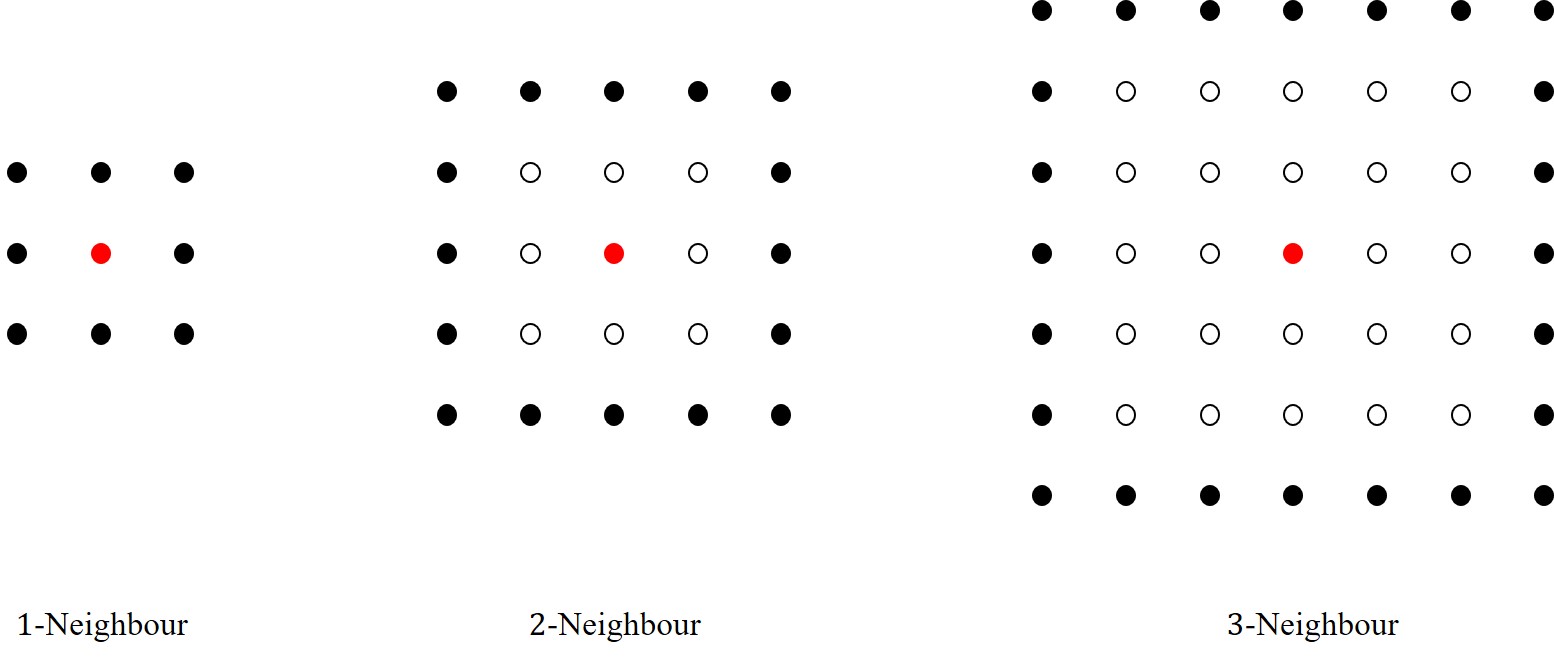}
    \caption{The $1$-neighbors to $3$-neighbors of a node.}
    \label{fig neighbor}
\end{figure}

We next define the $k$-neighbors of a node, which is denoted by $\mathcal{N}_{k}(i)$, in a recursive way as follows. 
The $1$-neighbors $\mathcal{N}_{1}(i)$ of a node $i\in\mathcal{G}$ are defined as the set of the $8$ nodes surrounding $i$ (king-move neighborhood). 
For $k\geq 2$, we say a node belongs to the $k$-neighbors of $i$ if $j\notin \mathcal{N}_{k'}(i)$ for all $k'<k$, and $j\in\mathcal{N}_{1}(j')$ for some $j'\in\mathcal{N}_{k-1}(i)$.
Here and thereafter, we call the $1$-neighbor of a node as the neighbor for simplicity.
Figure~\ref{fig neighbor} displays a graphical illustration for the $1$-neighbors to $3$-neighbors of a node. 
We define the distance between two nodes $i$ and $j$ as $d(i,\,j)=k$ if $j$ is a $k$-neighbor of $i$. 
It can be verified that $d(\cdot,\,\cdot)$ is a distance metric, i.e., it is non-negative, symmetric, and satisfies the triangle inequality. In addition, we define $d(i,\,\mathcal{U})\triangleq\min_{j\in\mathcal{U}}d(i,\,j)$ as the distance between a node $i\in\mathcal{G}$ and a set $\mathcal{U}\subset\mathcal{G}$.

Assume that there is an electricity network in the above grid network.
Let $\mathcal{S}\subset\mathcal{G}$ be a set of nodes representing all the stakeholders in the electricity network, each of which is either a power station (generator) or a consumer. 
Let $\mathcal{E}$ be a set of edges representing all the transmission lines in the network.
A generic element of $\mathcal{E}$ is $(i,\,i_{1},\,i_{2},\ldots,i_{n},j)$ that corresponds to an electricity line connecting nodes $i,\,j\in\mathcal{S}$ along the nodes $\{i_{1},\,i_{2},\ldots,i_{n}\}$, where $i_1,\ldots,i_n\notin \mathcal{S}$ and $d(i_k,i_k+1)=1$ for all $k=1,\ldots,n-1$. 
For notational convenience, we denote by $(i,\,j)$ the transmission line from node $i$ to node $j$ for $i,j\in\mathcal{S}$.
We assume that all the edges in $\mathcal{E}$ are undirected, i.e., $(i,\,j)\in\mathcal{E}$ implies $(j,\,i)\in\mathcal{E}$. 
Each node $i$ in the electricity network $\mathcal{S}$ is associated with a working load $L(i)$.
The maximum capacity of a node $i\in\mathcal{S}$ is $U^\text{P}(i)$, and we simply set $U^\text{P}(i)=0$ if node $i$ is a consumer.
Moreover, the maximum flow on transmission line $(i,j)\in\mathcal{E}$ is $U^\text{F}(i,j)$. 
The power transfer distribution factor of node $k\in\mathcal{S}$ on transmission line $(i,j) \in \mathcal{E}$ is denoted as $\texttt{PTDF}((i,j),k)$, which can be determined by the reactance and the topology of the transmission lines \citep{cheng2005ptdf}.

We consider a static OPF problem on the electricity network $\mathcal{S}$.
This static problem sets the stage for our online optimization problem in Section~\ref{section operational management against bushfire}.
Specifically, decision-makers need to decide on an operational strategy $\pi=(\alpha,\,\beta)$ for the electricity network. Here, $\alpha=(\alpha(i))_{i\in\mathcal{S}}$ denotes the capacity of all nodes $i\in\mathcal{S}$ and
$\beta=(\beta(i,j))_{(i,j)\in\mathcal{E}}$ denotes the flow on all edges $(i,\,j)\in\mathcal{E}$, where $\beta(i,\,j)<0$ means the electricity flows from node $j$ to node $i$.
Similar to \cite{Dong2016Ele}, the unit cost of capacity for station $i\in\mathcal{S}$ is $C^{\text{P}}(i)>0$.
Our objective is to minimize the operational cost of the electricity network while satisfying the working loads and the physical constraints on capacities and flows
%
\begin{subequations}\label{eq_optimization_lp}
\begin{alignat}{2}
\min_{\pi=(\alpha,\beta)} & \quad \sum_{i\in\mathcal{S}} C^{\text{P}}(i)\alpha(i)
\label{eq_optimization_lp 2-1}\\
\textup{s.t.} & \quad
\alpha(i) + \sum_{j:\,(j,\,i)\in\mathcal{E}}\beta({j,\,i}) = L(i),\qquad \forall i\in\mathcal{S}, \label{eq_optimization_lp 2-2}\\
& \quad
\alpha(i) \leq U^\text{P}(i), \qquad \forall i\in\mathcal{S}, 
\label{eq_optimization_lp 2-3}\\
& \quad
\beta(i,\,j) + \beta(j,\,i) = 0,\qquad \forall (i,\,j)\in\mathcal{E}, 
\label{eq_optimization_lp 2-4}\\
& \quad 
|\beta(i,\,j)|\leq U^\text{F}(i,\,j),\qquad \forall (i,\,j)\in\mathcal{E}, 
\label{eq_optimization_lp 2-5}\\
& \quad \beta(i,\,j) = \sum_{k \in \mathcal{S}} \texttt{PTDF}((i,j),k) [\alpha(k) - L(k)], \qquad \forall (i,\,j) \in \mathcal{E},  \label{eq_optimization_lp 2-6} \\
& \quad
\alpha(i)\geq 0, \quad \forall i\in\mathcal{S}
\label{eq_optimization_lp 2-7}.
\end{alignat}
\end{subequations}
Constraints~(\ref{eq_optimization_lp 2-2}) require the working loads to be satisfied; 
that is, we require the power generated and consumed to be balanced at each node of a distributed generation system \citep{mhanna2021exact}. 
Constraints~(\ref{eq_optimization_lp 2-3}) come from the requirement that the power generated should not exceed the maximum allowable capacity.
Constraints~\eqref{eq_optimization_lp 2-4} are by definition of power flows.
Constraints~(\ref{eq_optimization_lp 2-5}) require that the power flow cannot exceed the capacity of a transmission line. 
Constraints~(\ref{eq_optimization_lp 2-6}) align with the power flow physics and reflect the relationship between power flow on transmission lines and the capacity of nodes.

In a well-designed electricity network, it is reasonable to suppose that there is at least one strategy that satisfies Constraints (\ref{eq_optimization_lp 2-2})--(\ref{eq_optimization_lp 2-7}). 

\begin{assumption}\label{assumption balancing strategy}
Problem (\ref{eq_optimization_lp 2-1})--(\ref{eq_optimization_lp 2-7}) has at least one feasible solution and is thus solvable.
\end{assumption}
Here and thereafter, we call a strategy $\pi$ satisfying Constraints~(\ref{eq_optimization_lp 2-2})--(\ref{eq_optimization_lp 2-7}) a balancing strategy.

\subsection{Bushfire spread model}\label{section bushfire spread model}
Bushfire can occur and spread along the grid network $\mathcal G$ considered in Section~\ref{section electricity network operational problem}.
A power station or a transmission line must be shut down when catching bushfires, leading to a negative effect on the electricity network.
This section formulates a dynamic model for the bushfire spread.
Section~\ref{section operational management against bushfire} combines this model and the static decision-making model in Section~\ref{section electricity network operational problem} to formulate an online optimization problem for the electricity network management subject to bushfire.

\begin{table}[h]
\centering
\scriptsize
\caption{\textcolor{red}{List of symbols and their meanings as presented in Section 2.2.}}
\label{tab:2.2symbols}
\begin{tabular}{p{2.2cm} p{5.5cm}}  
\hline
\textbf{Symbol} & \textbf{Meaning} \\ \hline
$\mathcal{B}_t, \mathcal{B}_{h,t}$ & Area on fire at time $t$; Area on fire in partition $\mathcal{G}_h$ at time $t$ \\
$p_t^{+}(i), p_t^{-}(i)$ & Probability of node $i$ catching fire; Probability of fire dying out at node $i$ at time $t$ \\
$\mathcal{F}_t(\mathcal{U})$ & Nodes in $\mathcal{U}$ on fire at time $t$ \\
$\mathcal{N}(i), \mathcal{N}(\mathcal{U})$ & Neighbors of node $i$; Neighbors of a set of nodes $\mathcal{U}$ \\
$\mathcal{G}_h$ & Partitioned area of grid $\mathcal{G}$ \\
$p_{h,t}^{+}, p_{h,t}^{-}$ & Probability of fire spreading; Probability of fire dying out in area $\mathcal{G}_h$ during period $t$ \\
$\ell(\cdot)$ & Log-likelihood function for parameter estimation \\
$\hat{p}_{h,t}^{+}, \hat{p}_{h,t}^{-}$ & Maximum likelihood estimate of $p_{h,t}^{+}, p_{h,t}^{-}$ \\ \hline
\end{tabular}
\end{table}

\begin{figure}[h]
    \centering
    \includegraphics[width=0.8\textwidth]{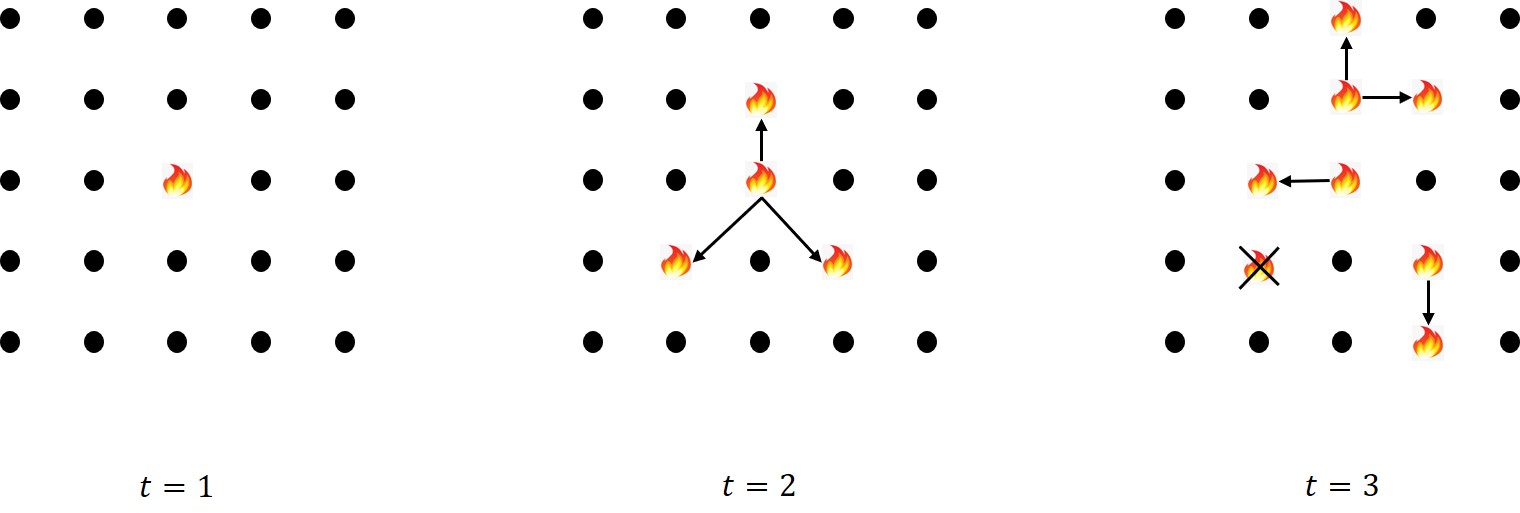}
    \caption{An example of the spread of bushfires.}
    \label{fig fire spread}
\end{figure}

We capture the bushfire spread by the well-known Moore's neighborhood model \citep{wootton2001local,trunfio2011new}, which works on a discrete time scale.
At time $t=1,2,\ldots,$ we denote by $\mathcal{B}_{t}\subseteq\mathcal{G}$ the current area that is on fire at the beginning of time $t$. 
In each period starting from $t$, we consider an arbitrary node $j\in\mathcal{B}_t$ on fire at the beginning of the period.
We assume the fire on this node can only spread to the neighbor of the node in a single period. 
This assumption is reasonable when the length of a period is relatively small, e.g., 15 minutes as in our case. 
If a neighbor of node $j$, denoted by $i\in\mathcal{N}_1(j)$, is not on fire at the beginning of time $t$ (i.e., $i\in\mathcal{G}\backslash\mathcal{B}_t$), we assume that this neighbor catches fire during a period starting from $t$ with probability {\color{blue!80!black}$p_{t}^{+}(i)$}. 
For any set of nodes $\mathcal{U}\subseteq\mathcal{G}$, let $\mathcal{F}_t(\mathcal{U})\triangleq \mathcal{U}\cap\mathcal{B}_t$ be the set of nodes in $\mathcal{U}$ that are on fire at the beginning of time $t$. 
Then, for any node $i\in\mathcal{G}\backslash\mathcal{B}_t$ that is not on fire at the beginning of time $t$, the probability of $i$ being on fire at the beginning of time $t+1$ is
$(1-[1 - p_{t}^{+}(i)]^{|\mathcal{F}_{t}(\mathcal{N}(i))|})$.
After the spread of fire, for each node $i\in\mathcal{B}_t$ that has already caught fire at the beginning of time $t$, we assume the fire dies out at the end of time $t$ with probability {\color{blue!80!black}$p_{t}^{-}(i)$}. 
A possible spreading route from a single node is illustrated in Figure \ref{fig fire spread}.

In reality, the spread of bushfires can be spatially heterogeneous due to environmental factors such as vegetation.
To consider such spatial heterogeneity, $p_t^+(i)$ and $p_t^-(i)$ should vary in $i\in\mathcal{G}$.
To make the model scalable, we divide the geographical region into several areas such that the stochastic spread of fires within each area can be treated as homogeneous.
Specifically, we partition $\mathcal{G}$ into $H$ disjoint areas $\{\mathcal{G}_h\}_{h=1}^H$ such that $\mathcal{G}=\bigcup_{h=1}^{H}\mathcal{G}_{h}$ and $\mathcal{G}_{h}\cap \mathcal{G}_{h'}=\emptyset$ for all $h\neq h'$. 
Moreover, for all $t\geq 1$ and $h=1,\ldots,H$, we suppose that
\begin{align*}
p_{t}^{+}(i)=p_{h,\,t}^{+},\:
p_{t}^{-}(i)=p_{h,\,t}^{-},\:
\forall i\in\mathcal{G}_{h}.
\end{align*}

In practice, both $p_{h,\,t}^{+}$ and $p_{h,\,t}^{-}$ are unknown so we need to use observed data for estimation. 
With a slight abuse of notation, we define the neighbor of $\mathcal{U}$ to be $\mathcal{N}(\mathcal{U})\triangleq\bigcup_{i\in\mathcal{U}}\mathcal{N}(i)\setminus\mathcal{U}$ for any subset of nodes $\mathcal{U}\subseteq\mathcal{G}$.
We let $\mathcal{B}_{h,\,t}\triangleq \mathcal{B}_{t}\cap \mathcal{G}_{h}$. Note that in area $\mathcal{G}_{h}$, $\mathcal{B}_{h,\,t}\setminus\mathcal{B}_{h,\,t-1}$ represents the nodes that newly caught fire at time $t$, 
$\mathcal{N}(\mathcal{B}_{h,\,t-1})\setminus\mathcal{B}_{h,\,t}$ represents the nodes that are neighbours of $\mathcal{B}_{h,\,t-1}$ but not on fire until $t$, $\mathcal{B}_{h,\,t}\setminus\mathcal{B}_{h,\,t+1}$ represents the nodes where fire just burnt out at time $t$, and $\mathcal{B}_{h,\,t-1}\cap\mathcal{B}_{h,\,t}$ represents the nodes that are on fire at both time $t-1$ and $t$. 
Then, the likelihood function of $(p_{h,\,t}^{+},p_{h,\,t}^{-})_{h=1}^H$ based on the observed data $\{\mathcal{B}_{t},\mathcal{B}_{t+1}\}$ is given by
\begin{align*}
&
\mathcal{L}(p_{1,\,t}^{+},\cdots,p_{H,\,t}^{+},\,p_{1,\,t}^{-},\cdots,p_{H,\,t}^{-})
\nonumber\\
= & 
\prod_{h=1}^{H}\Bigg(
\prod_{i\in\mathcal{B}_{h,\,t+1}\setminus\mathcal{B}_{h,\,t}}
\left[1-\left(1 - p_{h,\,t}^{+}\right)^{\left|\mathcal{F}_{t}(\mathcal{N}(i))\right|}\right]
\prod_{i\in\mathcal{N}(\mathcal{B}_{h,\,t})\setminus\mathcal{B}_{h,\,t+1}}
\left(1 - p_{h,\,t}^{+}\right)^{\left|\mathcal{F}_{t}(\mathcal{N}(i))\right|}\nonumber\\
& 
\times\left(p_{h,\,t}^{-}\right)^{|\mathcal{B}_{h,\,t}\setminus\mathcal{B}_{h,\,t+1}|}
\left(1-p_{h,\,t}^{-}\right)^{|\mathcal{B}_{h,\,t}\cap\mathcal{B}_{h,\,t+1}|}
\Bigg),
\end{align*}
and the log-likelihood function is
\begin{align*}
& \ell(p_{1,\,t}^{+},\cdots,p_{H,\,t}^{+},\,p_{1,\,t}^{-},\cdots,p_{H,\,t}^{-})\nonumber\\
= &
\sum_{h=1}^{H}\Bigg[
\sum_{i\in\mathcal{B}_{h,\,t+1}\setminus\mathcal{B}_{h,\,t}}
\log\left(1-\left[1 - p_{h,\,t}^{+}\right]
^{\left|\mathcal{F}_{t}(\mathcal{N}(i))\right|}\right)
+\sum_{i\in\mathcal{N}(\mathcal{B}_{h,\,t})\setminus\mathcal{B}_{h,\,t+1}}
\left|\mathcal{F}_{t}(\mathcal{N}(i))\right|\cdot
\log{\left(1 - p_{h,\,t}^{+}\right)}\nonumber\\
& +
|\mathcal{B}_{h,\,t}\setminus\mathcal{B}_{h,\,t+1}|
\sum_{i\in\mathcal{B}_{h,\,t}\setminus\mathcal{B}_{h,\,t+1}}\log{p_{h,\,t}^{-}}
+ |\mathcal{B}_{h,\,t}\cap\mathcal{B}_{h,\,t+1}|
\sum_{i\in\mathcal{B}_{h,\,t}\cap\mathcal{B}_{h,\,t+1}}\log{(1-p_{h,\,t}^{-})}
\Bigg].
\end{align*} 
The maximum likelihood estimates (MLEs) of $\{p_{h,\,t}^{+},\,p_{h,\,t}^{-}\}_{h=1}^{H}$, denoted by 
$\{\hat{p}_{h,\,t}^{+},\,\hat{p}_{h,\,t}^{-}\}_{h=1}^{H}$, can be obtained by solving 
$$(\hat{p}_{1,\,t}^{+},\cdots,\hat{p}_{H,\,t}^{+},\,
\hat{p}_{1,\,t}^{-},\cdots,\hat{p}_{H,\,t}^{-})
=\argmax_{p_{1,\,t}^{+},\cdots,p_{H,\,t}^{+},\,p_{1,\,t}^{-},\cdots,p_{H,\,t}^{-}\in[0,1]}\ell(p_{1,\,t}^{+},\cdots,p_{H,\,t}^{+},\,p_{1,\,t}^{-},\cdots,p_{H,\,t}^{-}).$$

\subsection{Operational management against bushfire}\label{section operational management against bushfire}
This section illustrates the dynamic adjustment of the electricity network in the presence of bushfires.
Our problem is challenging since we do not know the true values of $p_{h,t}^+$ and $p_{h,t}^-$ for the bushfire spread model in Section~\ref{section bushfire spread model}.
Therefore, we develop a learning framework to solve the problem in an online fashion.

We first illustrate the negative influence of bushfires on the electricity network.
We assume that a power station cannot work normally when the distance between the station and the fire is within a threshold $\bar d$ \citep{guo2018determination}.
Moreover, a transmission line cannot work when any node on the line is on fire.
Notationally, we write $i\in(\mathcal{S},\,\mathcal{E})$ for some node $i\in\mathcal{G}$ if $i\in\mathcal{S}$ or $i$ is a node on some transmission line in $\mathcal{E}$.
We then define a function $\varphi_{t}:\:(\mathcal{S},\,\mathcal{E})\mapsto \{0,\,1\}$ that indicates if node $i$ is functional ($\varphi_t(i)=1$) or fails ($\varphi_t(i)=0$) under the bushfire at time $t$ for all $t=1,\,2,\ldots,$ which is given by
\begin{align*}
    \varphi_{t}(i)=
    \begin{cases}
        1, & d(i,\mathcal{B}_{t})>\bar d,\:  i\in\mathcal{S} \text{ or } d(i,\mathcal{B}_{t})>0,\:  i\in(\mathcal{S},\,\mathcal{E})\setminus\mathcal{S}, \\
        0, & \textup{otherwise}.
    \end{cases}
\end{align*}

With the above assumptions, we consider the following dynamic scheme for decision-making.
At the beginning of time $t-1$ for any $t\geq 2$, we plan the next-period power flows of the electricity network $\pi_{t}=(\alpha_{t},\beta_{t})$ based on the observable information $\mathcal{B}_1,\ldots,\mathcal{B}_{t-1}$.
Then at time $t$, some nodes $i\in\mathcal{S}$ or edges $(i,j)\in\mathcal{E}$ may not be able to work due to bushfire.
As a result, additional resources are needed to ensure the balance of the electricity network, leading to a potential increase in the operational cost \citep{wang2018risk}.
This dynamic adjustment scheme is complex from two perspectives.
First, the bushfire evolves stochastically over time, so we need to make a prediction on $\mathcal{B}_{t}$ based on $\mathcal{B}_1,\ldots,\mathcal{B}_{t-1}$.
Second, the above prediction is further complicated due to the unknown values of $p_{h,\tau^+}$ and $p_{h,\tau^-}$ for all $h=1,\ldots,H$ and $\tau=1,\ldots,t-1$.
We remark that the assumption of planning the power flows one period ahead mainly comes from the collection process of bushfire information in practice.
For example, the status of bushfires may be inspected by a helicopter.
The helicopter reports the bushfire status periodically after inspecting all the regions that have the potential to catch fire.
As a result, the reported bushfire status may not reflect the in-situ bushfire status in all regions \citep{plucinski2011effect}.
Moreover, there could be a delay for the electricity network to implement the power flow solution obtained by the optimization model.
The length of a period is around 15 minutes in practice for the electricity network to adjust their power flows \citep{metz2018use,olivares2014centralized}, and the fire evolves stochastically during this period.
Due to the lags introduced above,
we suppose that the power flows for time $t$ are planned based on the available information up to time $t-1$, i.e., $\mathcal{B}_1,\ldots,\mathcal{B}_{t-1}$.

Based on the above setting, we need to predict if any node $i\in\mathcal{S}$ or edge $(i,j)\in\mathcal{E}$ is forced to shut down due to bushfire at time $t$ based on $\mathcal{B}_1,\ldots,\mathcal{B}_{t-1}$.
Let $L_t\triangleq(L_{t}(i))_{i\in\mathcal{S}}$, $U^\text{P}_t\triangleq(U^\text{P}_{t}(i))_{i\in\mathcal{S}}$, and $U^\text{F}_t\triangleq(U^\text{F}_{t}(i,\,j))_{(i,j)\in\mathcal{E}}$ be the working loads, maximum capacities, and maximum power flows for all the nodes and edges in the electricity network at time $t$, respectively.
When making decisions at time $t-1$, all the $L_t$, $U^\text{P}_t$, and $U_t^\text{F}$ are random variables.
Since we assume a Markovian evolution of bushfire in Section~\ref{section bushfire spread model}, their distributions can be determined given only on the observed data $\mathcal{B}_{t-1}$.
According to the indicator function $\varphi_t(i)$, we have
\begin{align}\label{eq online load}
L_{t}(i)=\begin{cases}
L(i), & \varphi_{t}(i)=1, \\
0, & \textup{otherwise},
\end{cases}
\end{align}
and
\begin{align}\label{eq online capacity}
U^\text{P}_{t}(i)=\begin{cases}
U^\text{P}(i), & \varphi_{t}(i)=1, \\
0, & \textup{otherwise},
\end{cases}
\end{align}
for all $t=1,\,2,\ldots$ and $i\in\mathcal{S}$.
Let $\varphi_{t}(i,\,j)\triangleq \prod_{i{'}\in(i,\,j)}\varphi_{t}(i{'})$ for all $(i,\,j)\in\mathcal{E}$ at time $t\geq1$, where $i{'}\in(i,\,j)$ means that $i{'}$ is a node on the transmission line $(i,\,j)\in\mathcal{E}$ but neither $i'=i$ nor $i'=j$.
Then we have 
\begin{align}\label{eq online flow}
U^\text{F}_{t}(i,\,j)=\begin{cases}
U^\text{F}(i,\,j), & \varphi_{t}(i,\,j)=1,\, \\
0, & \textup{otherwise},
\end{cases}
\end{align}
for all $t=1,\,2,\ldots$ and $(i,\,j)\in\mathcal{E}$.
If the values of $p_{h,t-1}^+$ and $p_{h,t-1}^-$ are \textit{known}, the conditional distribution of $\varphi_t(i)$ for node $i\in\mathcal{S}$ can be readily computed:
\begin{align}\label{eq state transition probability}
    & \mathrm{Pr}\left(\varphi_{t}(i)=1\,|\,\mathcal{B}_{t-1}\right)\nonumber\\
    = & 
    \prod_{h=1}^{H}\Big(\prod_{d'=1}^{\bar d}
    \prod_{j\in(\mathcal{N}_{d'}(i)\cap\mathcal{G}_{h})
    \setminus \mathcal{B}_{h,\,t-1}}
    (1-p_{h,\,t-1}^+)^{|\mathcal{F}_{t-1}(\mathcal{N}(j))\cap\mathcal{G}_{h}|}
    \cdot(p^{-}_{h,\,t-1})^{\sum_{d'=1}^{\bar d}
    |\mathcal{F}_{t-1}(\mathcal{N}_{d'}(i))\cap\mathcal{G}_{h}|}
    \Big),
\end{align} and $\mathrm{Pr}\left(\varphi_{t}(i)=1\,|\,\mathcal{B}_{t-1}\right)
=1-\mathrm{Pr}\left(\varphi_{t}(i)=0\,|\,\mathcal{B}_{t-1}\right)$. 
Then by Eqs.~\eqref{eq online load}--\eqref{eq state transition probability}, we can obtain the conditional distributions of $L_t(i)$, $U_t^\text{P}(i)$, and $U_t^\text{F}(i,j)$ given $\mathcal{B}_{t-1}$.

Let $\pi_t=((\alpha_t(i))_{i\in\mathcal{S}},(\beta_t(i,j)_{(i,j)\in\mathcal{E}})$ be the \textit{scheduled} generator outputs and power flows at time $t$; see the notation for Problem~\eqref{eq_optimization_lp}.
Based on our problem setting, $\pi_t$ is obtained at time $t-1$ given $\mathcal{B}_1,\ldots,\mathcal{B}_{t-1}$.
Due to the random spread of bushfires, some stations and transmission lines are forced to shut down, and sometimes we cannot implement $\pi_t$ at time $t$. 
In this case, the generator outputs will be automatically adjusted by automatic generation control, and the real power flows on the transmission lines will be redistributed.
During the dynamic adjustment, the power balance will have to be enforced by curtailing some load demand if the load demand cannot be satisfied, leading to the load shedding cost \citep{moreno2022microgrids}.
Directly modeling the above dynamic adjustment subject to the random spread of bushfires is too complicated to make any online learning model analytically tractable.
Hence, we follow \cite{wang2018risk} and \cite{hussain2019effort} to approximately calculate the overall load shedding cost by directly summing the power imbalance on each node based on the scheduled strategy $\pi_t$.

Due to the random shutdown of nodes and edges in the electricity network, we cannot ensure that Constraints~(\ref{eq_optimization_lp 2-2}) hold for all realizations of $L_t(i)$, $U_t^\text{P}(i)$, and $U_t^\text{F}(i,j)$.
As previously illustrated, we approximately compute the overall load shedding cost by first defining the magnitude of load shedding $\texttt{LS}^\pi(i)$ of node $i$ \citep{AEMO2020} under a strategy $\pi=(\alpha,\,\beta)$ given $U^\text{P}_t$, $U^\text{F}_t$, and $L_t$ at period $t$ as:
\begin{align*}
    \texttt{LS}^{\pi}(i;\: L_t,\,U_t^{\text{P}},\,U_t^{\text{F}})
    ~\triangleq &~ \max \{ L_t(i) - 
    \min\{\alpha(i),U_t^{\text{P}}(i)\} \\
    & ~ - \sum_{j:\,(j,\,i)\in\mathcal{E}}\min\{|\beta(j,\,i)|,U_t^{\text{F}}(j,i)\}\cdot(2\mathbb{I}\{\beta(j,i)>0\}-1) , 0 \}, \quad i\in\mathcal{S}, 
\end{align*}
where $\mathbb{I}\{A\}$ is the indicator function so that $\mathbb{I}\{A\}=1$ if statement $A$ is true and $\mathbb{I}\{A\}=0$ otherwise.
A penalty (load shedding cost), proportional to 
$\texttt{LS}^{\pi}(i;\: L_t,\,U_t^\text{P},\,U_t^\text{F})$ at node $i$, is then added to the total operational cost \citep{wang2018risk,hussain2019effort}. 
Let $C^{\text{S}}$ be the unit cost of shedding load.
Then the total operational cost under a strategy $\pi$ given $(L_t,U^\text{P}_t,U^\text{F}_t)$ is
\begin{align*}
    C(\pi;\: L_t,\,U^\text{P}_t,\,U^\text{F}_t) = 
    \sum_{i\in\mathcal{S}} C^{\text{P}}(i)\alpha(i)
    + C^{\text{S}}\sum_{i\in\mathcal{S}} \texttt{LS}^{\pi}(i;\: L_t,\,U^\text{P}_t,\,U^\text{F}_t).
\end{align*} The above display is the same as the objective function (\ref{eq_optimization_lp 2-1}) except that we further add the load shedding cost considering the potential imbalance of the electricity network due to bushfire.

We note that $C(\pi;L_t,U^\text{P}_t,U^\text{F}_t)$ is a random variable at time $t-1$ because the penalty on load shedding at time $t$ is non-deterministic based on $\mathcal{B}_{t-1}$.
The conditional distribution of $C(\pi;L_t,U^\text{P}_t,U^\text{F}_t)$ can be determined by \eqref{eq online load}--\eqref{eq state transition probability}.
We aim to minimize the expected operational cost subject to the physical constraints of the electricity network.
Based on the static Problem~(\ref{eq_optimization_lp}), we obtain $\pi_t$ by solving the following problem at the beginning of time $t-1$ if $(p_{h,t}^+,p_{h,t}^-)_{h=1}^H$ are \textit{known}:
\begin{subequations}\label{eq:power_flow}
\begin{alignat}{2}
\min_{\pi} & \quad 
\mathbf{E}\left[C(\pi;\: L_{t},\,U^\text{P}_{t},\,U^\text{F}_{t})\mid\mathcal{B}_{t-1}\right]
\label{eq-optimization-online-1}\\
\textup{s.t.}
& \quad
\alpha(i) \leq U^\text{P}(i), \qquad \forall i\in\mathcal{S}, 
\label{eq-optimization-online-2}\\
& \quad
\beta(i,\,j) + \beta(j,\,i) = 0,\qquad \forall (i,\,j)\in\mathcal{E}, 
\label{eq-optimization-online-3}\\
& \quad 
|\beta(i,\,j)|\leq U^\text{F}(i,\,j),\qquad \forall (i,\,j)\in\mathcal{E}, 
\label{eq-optimization-online-4}\\
& \quad
 \beta(i,\,j) = \sum_{k \in \mathcal{S}} \texttt{PTDF}_{t-1}((i,j),k) [\alpha(k) - L(k)], \qquad \forall (i,\,j) \in \mathcal{E}, 
\label{eq-optimization-online-5}\\ 
& \quad
\alpha(i)\geq 0, \quad \forall i\in\mathcal{S}
\label{eq-optimization-online-6}
\end{alignat}
\end{subequations}
%
where $\texttt{PTDF}_{t-1}$ is the PTDF of the power grid at time $t-1$ based on $\mathcal{B}_{t-1}$.
Problem~\eqref{eq:power_flow} is a two-stage stochastic program, 
where Constraints~\eqref{eq-optimization-online-2}--\eqref{eq-optimization-online-6} parallel Constraints~(\ref{eq_optimization_lp 2-3})--(\ref{eq_optimization_lp 2-7}).

By enumerating all possible scenarios, we can write the extensive form \citep{shapiro2021lectures} of the above stochastic program as a linear program (LP).
For notational convenience, let 
$P_{t}(i)\triangleq\mathrm{Pr}\left(\varphi_{t}(i)=1\,|\,\mathcal{B}_{t-1}\right)$ for all $i\in\mathcal{S}$ and 
$P_{t}(i,\,j)\triangleq\prod_{i'\in(i,\,j)}\mathrm{Pr}\left(\varphi_{t}(i')=1\,|\,\mathcal{B}_{t-1}\right)$ for all $(i,\,j)\in\mathcal{E}$, which separately denote the functional probability of node $i\in\mathcal{S}$ and edge $(i,j)\in\mathcal{E}$ at time $t$ given $\mathcal{B}_{t-1}$.
When $(p^{+}_{h,t-1},p^{-}_{h,t-1})_{h=1}^H$ are given, we have the following proposition.
\begin{proposition}\label{proposition lp optimal}
Given $(p^{+}_{h,t-1},p^{-}_{h,t-1})_{h=1}^H$ and conditional on $\mathcal{B}_{t-1}$, Problem~\eqref{eq:power_flow} can be reformulated as the following LP:
\begin{subequations}\label{eq:power_flow_LP}

\begin{alignat}{2}
    \min_{\alpha,\beta, \mathcal{H}} & \quad 
    \sum_{i\in\mathcal{S}} C^{\mathrm{P}}(i)\alpha(i) +
C^{\mathrm{S}}\sum_{i\in\mathcal{S}}\sum_{\mathcal{S}'\in \mathcal{P}({\mathcal{S}(i)})}P_{t}(i)
\rho_{t}(i,\,\mathcal{S}') \mathcal{H}(i,\,\mathcal{S}') \label{eq:obj_EF}
\\
\textup{s.t.}
& \quad
0\leq \alpha(i) \leq U^\mathrm{P}(i), \qquad \forall i\in\mathcal{S},\\
& \quad
\beta(i,\,j) + \beta(j,\,i) = 0,\qquad \forall (i,\,j)\in\mathcal{E},\\
& \quad 
\beta(i,\,j)\leq U^\mathrm{F}(i,\,j),\qquad \forall (i,\,j)\in\mathcal{E},
\\
 & \quad
\beta(i,\,j) = \sum_{k \in \mathcal{S}} {\emph{\texttt{PTDF}}_{t-1}}((i,j),k) [\alpha(k) - L(k)], \qquad \forall (i,\,j) \in \mathcal{E}, \\
& \quad 
\mathcal{H}(i,\,\mathcal{S}') \geq 0, \qquad \forall i\in\mathcal{S},
\:\forall  \mathcal{S}'\in \mathcal{P}({\mathcal{S}(i)}), \\
& \quad 
L(i)-\alpha(i)- \sum_{j\in\mathcal{S}'}\beta(j,\,i)\leq \mathcal{H}(i,\,\mathcal{S}'), \qquad \forall i\in\mathcal{S},\:
\forall  \mathcal{S}'\in \mathcal{P}({\mathcal{S}(i)}), 
\end{alignat}
\end{subequations}
where $\{\mathcal{H}(i,\mathcal{S}')\}_{i\in\mathcal{S},\mathcal{S}'\in\mathcal{P}(\mathcal{S}(i))}$ are auxiliary decision variables to linearize the $\{|\beta(i,j)|\}_{(i,j)\in\mathcal{E}}$, $\mathcal{S}(i)$ is the set of nodes in $\mathcal{S}$ that are connected to node $i$ by at least one transmission line for $i\in\mathcal{S}$, $\mathcal{P}({\mathcal{S}(i)})$ is the power set of $\mathcal{S}(i)$, and 
$\rho_{t}(i,\,\mathcal{S}')\triangleq \prod_{j\in\mathcal{S'}}P_{t}(j,\,i)
\prod_{j\in\mathcal{S}(i)\setminus\mathcal{S}'}\left(1-P_{t}(j,\,i)\right)$ for all $i\in\mathcal{S}$ and $\mathcal{S}'\in\mathcal{P}({\mathcal{S}(i)})$.
\end{proposition}

We make two remarks on Problem~\eqref{eq:power_flow_LP}.
First, $|\mathcal{S}(i)|$ is relatively small for most nodes in real-world electricity networks, e.g., those considered in Section~\ref{section simulation study}.
Consequently, it is not computationally difficult to obtain the power sets $\mathcal{P}(\mathcal{S}(i))$ for all $i\in\mathcal{S}$ and then solve Problem~\eqref{eq:power_flow_LP}.
Second,
although some node $i$ is shut down due to bushfires, we still consider the cost $C^\text{P}(i)\alpha(i)$ in the objective function \eqref{eq:obj_EF}.
This cost aligns with regulations found in certain actual energy markets. 
For instance, in the Australian Energy Market Operator's framework, participants present their selling offers and purchasing bids within the competitive electricity market. 
Once these offers and bids are accepted, contracts are established, and payment obligations persist irrespective of any potential shutdown involving node $i$. Such payment settlement periods generally last 5 to 30 minutes, and this operational practice facilitates cost management.
Moreover, if we intend for node $i$ to produce a significant number of power units in a period, yet node $i$ is shut down due to bushfires within that period, then this triggers a power system imbalance, resulting in associated losses.
In this case, we can regard $C^\text{P}(i)\alpha(i)$ as a penalty term for inappropriately planning node $i$ to produce a significant number of power units.
Excluding this cost term in the objective function when node $i$ is shut down will overlook such penalties, possibly leading to some misleading scheduling results.
Our regret bounds derived in Section~\ref{sec-online-opt-alg} do not rely on this assumption, and our model can be readily generalized to the setting where only working generators incur the generation cost by incorporating a second-stage term in Problem~\eqref{eq:power_flow}.

The main challenge in our problem is that $(p^{+}_{h,t-1},p^{-}_{h,t-1})_{h=1}^H$ are unknown,
while Problem~\eqref{eq:power_flow_LP} is formulated for given $(p^{+}_{h,t-1},p^{-}_{h,t-1})_{h=1}^H$, which determine $P_{t}(i)$ for $i\in\mathcal{S}$.
We hence insert our problem in an online learning framework.
As with most existing literature in online optimization \citep{huh2014online,lu2021online,xu2022online}, we evaluate our solution by the regret function.
Specifically, when the true values of $(p^{+}_{h,t-1},p^{-}_{h,t-1})_{h=1}^H$ are known, we can find the optimal solution $\pi_t^*$ by solving Problem~\eqref{eq:power_flow_LP} at time $t-1$.
In contrast, when $(p^{+}_{h,t-1},p^{-}_{h,t-1})_{h=1}^H$ are unknown, we need to use the estimates of $(p^{+}_{h,t-1},p^{-}_{h,t-1})_{h=1}^H$ to approximate $(P_{t}(i))_{i\in\mathcal{S}}$ in Problem~\eqref{eq:power_flow_LP} and then solve the problem to obtain a solution $\pi_t$.
The resulting expected operational cost under the \textit{true values} of $(p^{+}_{h,t-1},p^{-}_{h,t-1})_{h=1}^H$ is $\mathbf{E}[C(\pi_t;L_t,U_t^\text{P},U^\text{F}_t)\mid \mathcal{B}_{t-1}]$.
Given a time horizon $[0,T]$, the regret function of a sequence of strategies
$\{\pi_t\}_{t=1}^{T}$ is then defined as:
\begin{align*}
R(T)\triangleq \sum_{t=1}^{T}\mathbf{E} \left[
\left(C(\pi_{t};\: L_{t},\,U^\text{P}_{t},\,U^\text{F}_{t})
-C(\pi^{*}_{t};\: L_{t},\,U^\text{P}_{t},\,U^\text{F}_{t}\right)\mid \mathcal{B}_{t-1}\right],
\end{align*}
where the expectation is taken with respect to both the randomness of $(L_t, U^\text{P}_{t}, U^\text{F}_{t})$ and possibly that of the strategy. 
Without loss of generality, we suppose that all the nodes $i\in\mathcal{S}$ and edges $(i,j)\in\mathcal{E}$ are working at time $t=1$, such that $\pi^{*}_{1}$ can be obtained directly by solving Problem~\eqref{eq_optimization_lp}.
We aim to find an online optimization algorithm that makes $R(T)$ as small as possible.

We close this section by noting that our online optimization problem shares a similarity to the denial-of-service attack in the electricity network security area \citep{chen2019distributed,ding2019distributed,xu2021bayesian}.
However, our problem is more challenging due to its unique characteristics from two perspectives. 
First, the bushfire resembles the ``attack'' in the above literature while the attacks in our problem are unobservable to decision-makers and have spatial-temporal dependence. Hence, we need to make the prediction a period ahead.
Second, the evolution of bushfires is non-stationary over time $t$.
This makes a precise prediction of bushfire spread difficult.
Nevertheless, by imposing some additional assumptions on the evolution probabilities $(p^{+}_{h,t-1},p^{-}_{h,t-1})_{h=1}^H$, we can bound the regret function $R(T)$ as shown in the next section.

\section{Online Optimization Algorithm}\label{sec-online-opt-alg}

\subsection{{Estimation of model parameters}}
In the previous sections, we formulate the operational problem of the electricity network under the threat of bushfires. 
To minimize the regret function $R(T)$, we develop an online algorithm to obtain a sequence of strategies $\{\pi_t\}_{t=1}^T$ that retains a sub-linear regret $R(T)$.
Since $(p^{+}_{h,t-1},p^{-}_{h,t-1})_{h=1}^H$ are unknown in practice, we start by showing that we can use some estimates of $(p^{+}_{h,t-1},p^{-}_{h,t-1})_{h=1}^H$ for Problem~\eqref{eq:power_flow_LP} to obtain an approximately optimal strategy.
Let $\hat\pi_t^*$ be the strategy obtained by solving Problem~\eqref{eq:power_flow_LP} with $(p^{+}_{h,t-1},p^{-}_{h,t-1})_{h=1}^H$ replaced with their estimates (not necessarily the MLEs).
The difference between the resulting expected cost and the optimal expected cost under $\pi_t$ can be bounded by the estimation biases of $(p^{+}_{h,t-1},p^{-}_{h,t-1})_{h=1}^H$ in the following sense. 
\begin{theorem}\label{theorem approximate optimal strategy}
For $t\geq 2$ and $h=1,\ldots,H$, let $\hat{p}^{+}_{h,\,t}$ and $\hat{p}^{+}_{h,\,t}$ be some estimates of $p^{+}_{h,\,t}$ and $p^{-}_{h,\,t}$, respectively.
Let $\hat{\pi}^{*}_{h,\,t}$ be the solution of \eqref{eq:power_flow_LP} by separately replacing $p^{+}_{h,\,t-1}$ and $p^{-}_{t-1}$ with $\hat{p}^{+}_{h,\,t-1}$ and $\hat{p}^{-}_{h,\,t-1}$.
Then, we have
\begin{align*}
& \mathbf{E}\left[C(\hat{\pi}^{*}_{t};\: L_{t},\,U^\mathrm{P}_{t},\,U^\mathrm{F}_{t})\mid \mathcal{B}_{t-1}\right]
-\mathbf{E}\left[C(\pi^{*}_{t};\:L_{t},\,U^\mathrm{P}_{t},\,U^\mathrm{F}_{t})\mid 
\mathcal{B}_{t-1}\right]\nonumber\\
\leq &  K^{+}\max_{h=1,\ldots,H}\left|p^{+}_{h,\,t-1}-\hat{p}^{+}_{h,\,t-1}\right|
+ K^{-}\max_{h=1,\ldots,H}\left|p^{-}_{h,\,t-1}-\hat{p}^{-}_{h,\,t-1}\right|,
\end{align*}
where 
\begin{align*}
K^{+} & \triangleq 
2C^{\mathrm{S}}\sum_{i\in\mathcal{S}}({|\mathcal{P}({\mathcal{S}(i)})|+3})
\left[\Delta(i)+2(2\bar{d}+1)^{2}\right]\left(U^\mathrm{P}(i) + L(i) + \sum_{j:\,(j,\,i)\in\mathcal{E}}U^\mathrm{F}(j,\,i)\right),\nonumber\\
K^{-} & \triangleq 
2C^{\mathrm{S}}\sum_{i\in\mathcal{S}} |\mathcal{P}({\mathcal{S}(i)})|
\left[(2\bar{d}+1)^{2}+\Delta(i)\right]
\left(U^\mathrm{P}(i) + L(i) + \sum_{j:\,(j,\,i)\in\mathcal{E}}U^\mathrm{F}(j,\,i)\right),
\end{align*}
and $\Delta(i)\triangleq |\{i{'}:\: \exists j\in\mathcal{S}(i),\,
i{'}\in(j,\,i)\}|$ is the number of nodes on any edge connected to $i\in\mathcal{S}$.
\end{theorem}

\noindent\textit{Remark}. 
The coefficients $K^{+}$ and $K^{-}$ in Theorem \ref{theorem approximate optimal strategy} depend on model parameters and may vary under different settings. 
In particular, they rely on the scale of the network through the term $\sum_{i\in\mathcal{S}} |\mathcal{P}({\mathcal{S}(i)})|\Delta(i)$. 
In most real-world electricity networks, each stakeholder is only connected to a limited number of others compared to the total number of stakeholders. 
This implies that $|\mathcal{P}({\mathcal{S}(i)})|\Delta(i)$ would be small for most $i\in\mathcal{S}$.
Furthermore, both coefficients are derived by a worst-case analysis such that all the nodes in the electricity network are forced to shut down due to bushfires in this worst case.
The proportion of nodes that are shut down due to bushfires is normally not large in practice, and thus the difference between the two strategies can be much smaller than the upper bound in Theorem~\ref{theorem approximate optimal strategy}.
We can decrease the values of $K^+$ and $K^-$ if we further assume that the proportion of shut-down nodes is bounded from above. 
Last, the simulation studies in Section~\ref{section simulation study} illustrate that the performance of our algorithm significantly outperforms the benchmarks.

Theorem \ref{theorem approximate optimal strategy} implies that we can well approximate the optimal strategy using accurate estimates of $p_{h,\,t}^{+}$ and 
$p_{h,\,t}^{-}$. It remains to capture the time-varying dynamics of $p_{h,\,t}^{+}$ and $p_{h,\,t}^{-}$ over time $t$. This study assumes that the true values of both $p_{h,\,t}^{+}$ and $p_{h,\,t}^{-}$ are piecewise constant over time $t$, and the change points of $p_{h,\,t}^{+}$ and $p_{h,\,t}^{-}$ can be mutually different, respectively. To explain this assumption, we note that the evolution of fire is mainly dominated by environmental conditions like wind speed.
Such an environmental condition can be stable for a period of time but generally changes over time \citep{mccarthy2012analysis}.
The piecewise constant assumption captures such dynamic changes, and we further verify this assumption by using a real-world dataset in Section \ref{section practical example of bushfire spread}.

Natural choices for the estimates of $p_{h,\,t}^{+}$ and $p_{h,\,t}^{-}$ are their MLEs obtained in Section \ref{section bushfire spread model}.
By the maximum likelihood estimation theory \citep{shao2003mathematical}, $\hat{p}^{+}_{h,\,t}-p^{+}_{t}$ and $\hat{p}^{-}_{h,\,t}-p^{-}_{t}$ asymptotically follow Gaussian distributions for all $h=1,\ldots,H$. 
The variances of the two Gaussian distributions are determined by the Fisher information matrix. 
Specifically, for all $h=1,\ldots,H$, we have
\begin{align}\label{eq mle consistency 1}
\sqrt{|\mathcal{N}(\mathcal{B}_{t})\cap\mathcal{G}_{h}|}
\left(\hat{p}_{h,\,t}^{+}-p_{h,\,t}^{+}\right)
\overset{d}{\rightarrow}N\left(0,\,(I_{h,\,t}^{+})^{-1}\right) 
\textup{ and }
\sqrt{|\mathcal{N}(\mathcal{B}_{t})\cap\mathcal{G}_{h}|}
\left(\hat{p}_{h,\,t}^{-}-p_{h,\,t}^{-}\right)
\overset{d}{\rightarrow}N\left(0,\,(I_{h,\,t}^{-})^{-1}\right),
\end{align}
where
\begin{align*}
I_{h,\,t}^{+} =\mathbf{E}[-{\partial^{2}
\ell(p_{1,\,t}^{+},\cdots,p_{H,\,t}^{+},\,p_{1,\,t}^{-},\cdots,p_{H,\,t}^{-})}
/{\partial (p_{h,\,t}^{+})^{2}}];\\
I_{h,\,t}^{-} = \mathbf{E}[-{\partial^{2}
\ell(p_{1,\,t}^{+},\cdots,p_{H,\,t}^{+},\,p_{1,\,t}^{-},\cdots,p_{H,\,t}^{-})}
/{\partial (p_{h,\,t}^{-})^{2}}].
\end{align*}
By Equation~(\ref{eq mle consistency 1}), we have
\begin{align}\label{eq mle consistency 2}
\mathrm{Pr}\left(\left|\hat{p}_{h,\,t}^+-p_{h,\,t}^{+}\right|\geq x\right)
\approx
\frac{2}{\sqrt{2\pi}}\int_{x}^{\infty}
\exp{\left\{ -{2I_{h,\,t}^{+}|\mathcal{N}(\mathcal{B}_{t})\cap\mathcal{G}_{h}|}
{y^{2}} \right\}} dy
\leq 
2\exp{\left\{ -{2I_{h,\,t}^{+}|\mathcal{N}(\mathcal{B}_{t})\cap\mathcal{G}_{h}|}
{x^{2}} \right\}}; \\
\mathrm{Pr}\left(\left|\hat{p}_{h,\,t}-p_{h,\,t}^{+}\right|\geq x\right)
\approx
\frac{2}{\sqrt{2\pi}}\int_{x}^{\infty}
\exp{\left\{ -{2I_{h,\,t}^{-}|\mathcal{N}(\mathcal{B}_{t})\cap\mathcal{G}_{h}|}
{y^{2}} \right\}}dy
\leq 
2\exp{\left\{ -{2I_{h,\,t}^{-}|\mathcal{N}(\mathcal{B}_{t})\cap\mathcal{G}_{h}|}
{x^{2}} \right\}}.
\end{align}
Based on the above asymptotic property, we make the following assumption.

\begin{assumption}\label{assumption mle bias}
At each time $t$, for all $h=1,\ldots,H$, the MLEs $(\hat p^{+}_{h,t-1},\hat p^{-}_{h,t-1})_{h=1}^H$ satisfy
\begin{align*}
\hat{p}_{h,\,t}^{+} = p_{h,\,t}^{+} + \varepsilon_{h,\,t}^{+} 
\textup{ and }
\hat{p}_{h,\,t}^{-} = p_{h,\,t}^{-} + \varepsilon_{h,\,t}^{-} 
,
\end{align*}
where both $\varepsilon_{h,\,t}^{+}$ and $\varepsilon_{h,\,t}^{-}$ are sub-Gaussian random variables denoting estimation biases. 
That is, 
\begin{align*}
    \mathrm{Pr}\left(\varepsilon_{h,\,t}^{+}\geq x\right)\vee
    \mathrm{Pr}\left(\varepsilon_{h,\,t}^{+}\leq -x\right)
    \leq \exp{\left\{-\frac{x^{2}}{2\nu_{h,\,t}^{+}}\right\}};~
    \mathrm{Pr}\left(\varepsilon_{h,\,t}^{-}\geq x\right)\vee
    \mathrm{Pr}\left(\varepsilon_{h,\,t}^{-}\leq -x\right)
    \leq \exp{\left\{-\frac{x^{2}}{2\nu_{h,\,t}^{-}}\right\}},
\end{align*}
where $x\vee y\triangleq\max\{x,y\}$, and
\begin{align*}
\nu_{h,\,t}^{\pm}\triangleq 
\left[\left|\mathcal{N}(\mathcal{B}_{t})\cap\mathcal{G}_{h}\right|
\cdot \mathbf{E}\left[-\frac{\partial^{2}\ell(p_{1,\,t}^{+},\cdots,p_{H,\,t}^{+},\,p_{1,\,t}^{-}
,\cdots,p_{H,\,t}^{-})}{\partial (p_{h,\,t}^{\pm})^{2}}\right]
\Bigg|_{p_{h,\,t}^{^{\pm}}=\hat{p}_{h,\,t}^{\pm}}\right]^{-1}.
\end{align*}
\end{assumption}

Because the true values of $p^{+}_{h,\,t}$ and $p^{-}_{h,\,t}$ are unknown,
Assumption~\ref{assumption mle bias} follows the common practice that uses $\hat{p}^{+}_{h,\,t}$ and $\hat{p}^{-}_{h,\,t}$ to evaluate variances of the estimation errors. 
We assume sub-Gaussian estimation errors because this assumption is general and includes the Gaussian estimation errors as a special case.
Since both $\hat{p}^{+}_{h,\,t}-p^{+}_{h,t}$ and $\hat{p}^{-}_{h,\,t}-p^{-}_{h,t}$ are bounded for all $h=1,\ldots,H$, the sub-Gaussian assumption is more reasonable than the Gaussian assumption for the biases.

Although the MLEs $\hat{p}^{+}_{h,\,t}$ and $\hat{p}^{-}_{h,\,t}$ have good asymptotic properties, they are obtained by using the observed data in a single period, as demonstrated in Section~\ref{section bushfire spread model}. The sample size of the data from a single period can be small and thus the variance can be large. 
A common practice is to use the average of $\hat{p}^{+}_{h,\,1},\ldots,\hat{p}^{+}_{h,\,t}$ up to $t$ as the estimator to reduce the variance. 
However, these estimates overlook the piecewise constant assumption of $p^{+}_{h,\,t}$ and $p^{-}_{h,\,t}$. Directly replacing $p^{+}_{h,\,t}$ and $p^{-}_{h,\,t}$ by such estimators in Problem~\eqref{eq:power_flow_LP} may lead to a large regret $R(T)$. A straightforward remedy is to impose some parametric assumptions on the change points and then simultaneously estimate $p^{+}_{h,\,t}$ and $p^{-}_{h,\,t}$, and the change points based on the data observed hitherto. 
Then, by using the average of the MLEs between successive change points as the estimator, one may overcome the above drawback.
However, this approach requires precisely tracking the change points in an online way and may make the statistical estimation very complicated because change points are unobservable. 
Furthermore, it makes it more difficult to derive a theoretical guarantee on the regret function in an online optimization setting. 
In the next section, we introduce an adaptive change point detection method to develop an efficient online algorithm for the OPF problem above, which admits a performance guarantee on the regret function.

\subsection{Online learning algorithm}\label{section online operational algorithm}
We now provide our main algorithm that sequentially decides the operational strategy in an online manner that retains a sublinear regret function over time.
Since we assume the true $p^{+}_{h,\,t}$ and $p^{-}_{h,\,t}$ are piecewise constant, the major challenge of our problem translates into finding the unobservable change points of $\{p_{h,\,t}^{+}\}_{t=1}^{\infty}$ and $\{p_{h,\,t}^{-}\}_{t=1}^{\infty}$ simultaneously for all $h=1,\ldots,H$. Then, the algorithm uses the average of $\hat{p}^{\pm}_{h,\,t}$'s between the current time and the latest detected change point of the sequence as the estimator. Detecting change points of temporal sequences has been studied in literature; see \cite{niu2016multiple} for a comprehensive review.
However, few of the change point detection methods can be directly applied to an online optimization setting.
For example, most of the change point detection approaches suffer an $O(1)$ error when tracking the exact change point locations, such that the resulting regret may not be sublinear in the time horizon $T$ in an online learning problem.
To the best of our knowledge,  \cite{auer2019adaptively} first proposed an adaptive change point tracking approach specially designed for online learning problems. 
This approach achieves a regret function with an optimal convergence order for the classical bandit problems. 
However, the approach cannot be directly applied to our work because it was specially designed for bandit problems where the sequential decision is to select from a finite set of objectives.
Our problem is more difficult because we need to solve an optimal strategy from a linear program whose feasible region is compact.
Therefore, we tackle the above challenges and develop an online algorithm based on \cite{auer2019adaptively} to simultaneously track the change points of all $p^{+}_{h,\,t}$'s and $p^{-}_{h,\,t}$'s and decide the operational strategy of the electricity network. 

Our proposed algorithm proceeds in episodes. 
Specifically, the algorithm collects the data $\mathcal{B}_t$ sequentially and then computes the MLEs $\hat p_{h,\,t}^{+}$ and $\hat p_{h,\,t}^{-}$ for all $h=1,\ldots,H$ as in Section~\ref{section bushfire spread model}.
Meanwhile, we keep tracks of possible change points of $p^{+}_{h,\,t}$'s and $p^{-}_{h,\,t}$'s and dynamically update the change points over time.
Whenever we detect a change point of any sequence of $p^{+}_{h,\,t}$'s and $p^{-}_{h,\,t}$'s, the algorithm will restart a new episode for this sequence and only use the observations in the current episode for estimation. Notationally, we define
\begin{align*}
\Hat{p}^{+}_{h,\,t_{1},\,t_{2}}\triangleq
\frac{1}{t_{2}-t_{1}+1}\sum_{t=t_{1}}^{t_{2}}\hat{p}^{+}_{h,\,t},\quad
\Hat{p}^{-}_{h,\,t_{1},\,t_{2}}\triangleq
\frac{1}{t_{2}-t_{1}+1}\sum_{t=t_{1}}^{t_{2}}\hat{p}^{-}_{h,\,t},\quad
\forall 1\leq t_{1}\leq t_{2}\leq T.
\end{align*}
Then, at each time $t\geq 2$ and for all $h=1,\ldots,H$, let $\tau^{+}_{h}$ and $\tau^{-}_{h}$ be the starting time of the current episodes for 
$\{p_{h,\,t}^{+}\}_{t=1}^{\infty}$ and $\{p_{h,\,t}^{-}\}_{t=1}^{\infty}$, respectively.
We then separately use $\hat{p}^{+}_{h,\,\tau^{+}_{h},\,t-1}$ and $\hat{p}^{-}_{h,\,\tau^{-}_{h},\,t-1}$ as the current estimates of $p_{h,\,t}^{+}$ and $p_{h,\,t}^{-}$.
These values are used to estimate $(P_t(i))_{i\in\mathcal{S}}$ and $\rho_t(i,\mathcal{S}')$, which are then plugged into Problem~\eqref{eq:power_flow_LP} to obtain the corresponding strategy for the electricity network operations management.

Fixing $h$ throughout the discussion below, we illustrate how to update the change points of $\{p_{h,\,t}^{+}\}_{t=1}^{\infty}$ and $\{p_{h,\,t}^{-}\}_{t=1}^{\infty}$. Intuitively, a change point is likely to occur when the MLEs 
$\{\hat{p}_{h,\,t}^{+}\}_{t=1}^{\infty}$ and 
$\{\hat{p}_{h,\,t}^{-}\}_{t=1}^{\infty}$ change significantly over consecutive time periods. Then each time we observe $\mathcal{B}_{t}$, the algorithm will judge if time 
$t-1$ is a change point of $\{p_{h,\,t}^{+}\}_{t=1}^{\infty}$ or 
$\{p_{h,\,t}^{-}\}_{t=1}^{\infty}$ by checking the following two conditions for all $\tau^{+}_{h}< t_{1}\leq t_{2}\leq t$ and $\tau^{-}_{h}< s_{1}\leq s_{2}\leq t-1$:
\begin{align}\label{eq condition p+}
\left|\Hat{p}^{+}_{h,\,\tau^{+},\,t} - \Hat{p}^{+}_{h,\,t_{1},\,t_{2}}\right|
\geq \frac{4\sqrt{\sum_{t'=\tau^{+}_{h}}^{t-1}\nu^{+}_{h,\,t'}\ln{2T}}}
{t - \tau^{+}_{h}}
+ \frac{4\sqrt{\sum_{t'=t_{1}}^{t_{2}}\nu^{+}_{h,\,t'}\ln{2T}}}
{t_{2}-t_{1}+1}
\end{align}
and
\begin{align}\label{eq condition p-}
\left|\Hat{p}^{-}_{h,\,\tau^{-},\,t} - \Hat{p}^{-}_{h,\,s_{1},\,s_{2}}\right|
\geq \frac{4\sqrt{\sum_{t'=\tau^{-}_{h}}^{t-1}\nu^{-}_{h,\,t'}\ln{2HT}}}
{t - \tau^{-}_{h}}
+ \frac{4\sqrt{\sum_{t'=s_{1}}^{s_{2}}\nu^{-}_{h,\,t'}\ln{2HT}}}
{s_{2}-s_{1}+1}.
\end{align}

Here, the right-hand side of both inequalities includes bounds derived from some concentration inequalities.
If there is no change point, both inequalities hold with a small probability. 
Specifically, due to Assumption \ref{assumption mle bias}, for all $h=1,\ldots,H$ and $t\geq 1$, both the estimation biases $\hat{p}^{+}_{h,\,t}-p^{+}_{h,\,t}$ and $\hat{p}^{-}_{h,\,t}-p^{-}_{h,\,t}$ are sub-Gaussian random variables. 
Therefore, $\left|\Hat{p}^{+}_{h,\,\tau^{+},\,t} - \Hat{p}^{+}_{h,\,t_{1},\,t_{2}}\right|$ and $\left|\Hat{p}^{-}_{h,\,\tau^{-},\,t} - \Hat{p}^{-}_{h,\,s_{1},\,s_{2}}\right|$ are summations of independent sub-Gaussian random variables with mean $0$. 
Based on this observation, we can use concentration inequalities to show that each of the above inequalities holds with probability at most $1/(HT^{4})$ when there is no change point.

If the condition in (\ref{eq condition p+}) holds, then the algorithm finds a change point of $\{p_{h,\,t}^{+}\}_{t=1}^{\infty}$, restarts an episode for 
$\{p_{h,\,t}^{+}\}_{t=1}^{\infty}$, and updates $\tau^{+}_{h}=t-1$.
The same procedures are applied to $\{p_{h,\,t}^{-}\}_{t=1}^{\infty}$ and $\tau^{-}_{h}$ when \eqref{eq condition p-} holds. With the above change point detection procedure, the whole online optimization algorithm is summarized in Algorithm \ref{algorithm 1}.

\begin{algorithm}[tbh]
\caption{{Online algorithm for operational strategy against bushfire}}
\begin{algorithmic}[1]
\Require $\mathcal{B}_{1}$, $\mathcal{N}$, $\mathcal{G}$, $\mathcal{E}$, $C^{\text{S}}$, $L(i)$, $U^\text{P}(i)$, $U^\text{F}(i,\,j)$, $C^{\text{P}}(i)$, and $C^\text{F}(i,\,j)$, for all $i\in\mathcal{S}$ and $(i,\,j)\in\mathcal{E}$.
\State \textbf{Initialization}: $\tau^{+}_{h}=\tau^{-}_{h}=1$.
\State Evaluate $\pi_{1}$ from Problem \eqref{eq_optimization_lp}.
\For{$t=2,\ldots,T$}
\State Calculate $\Hat{p}^{+}_{h,\,\tau^{+},\,t}$ and $\Hat{p}^{-}_{h,\,\tau^{-},\,t}$, $\forall h=1,\ldots,H$.
\State \textbf{Detecting change points}:
\For{$h=1,\ldots,H$, $\tau^{+}\leq t_{1}\leq t_{2}\leq t-1$ and $\tau^{-}\leq s_{1}\leq s_{2}\leq t-1$}
\State Calculate $\Hat{p}^{+}_{h,\,t_{1},\,t_{2}}$ and $\Hat{p}^{-}_{h,\,t_{1},\,t_{2}}$.
\If {$\left|\Hat{p}^{+}_{h,\,\tau^{+},\,t} - \Hat{p}^{+}_{h,\,t_{1},\,t_{2}}\right|
\geq \frac{4\sqrt{\sum_{t'=\tau^{+}_{h}}^{t-1}\nu^{+}_{h,\,t'}\ln{2T}}}
{t - \tau^{+}_{h}}
+ \frac{4\sqrt{\sum_{t'=t_{1}}^{t_{2}}\nu^{+}_{h,\,t'}\ln{2T}}}
{t_{2}-t_{1}+1}$}
\State Update $\tau^{+}_{h}\leftarrow t-1$.
\State \textbf{return}.
\EndIf
\If{$\left|\Hat{p}^{-}_{h,\,\tau^{-},\,t} - \Hat{p}^{-}_{h,\,s_{1},\,s_{2}}\right|
\geq \frac{4\sqrt{\sum_{t'=\tau^{-}_{h}}^{t-1}\nu^{-}_{h,\,t'}\ln{2T}}}
{t - \tau^{-}_{h}}
+ \frac{4\sqrt{\sum_{t'=s_{1}}^{s_{2}}\nu^{-}_{h,\,t'}\ln{2T}}}
{s_{2}-s_{1}+1}$}
\State Update $\tau^{-}_{h}\leftarrow t-1$.
\State \textbf{return}.
\EndIf
\EndFor
\State \textbf{Evaluating strategy}:
\State Use $\Hat{p}^{+}_{h,\,\tau^{+},\,t}$ and $\Hat{p}^{-}_{h,\,\tau^{-},\,t}$ to calculate $\hat{P}_{t}(i)$ and $\hat{\rho}_{t}(i,\,\mathcal{S}')$ for all $i\in(\mathcal{S},\,\mathcal{E})$ and $\mathcal{S}'\subset\mathcal{S}(i)$.
\State Substitute all $\hat{P}_{t}(i)$ and $\hat{\rho}_{t}(i,\,\mathcal{S}')$ in Problem \eqref{eq:power_flow_LP} and solve $\pi_{t}$ as the optimal solution.
\EndFor
\Ensure Operational strategies $\{\pi_{t}\}_{t=1}^T$ for the electricity network.
\end{algorithmic}
\label{algorithm 1}
\end{algorithm}

Let $\{\pi_{t}\}_{t=1}^{T}$ be the sequence of strategies obtained by Algorithm \ref{algorithm 1}.
The following theorem provides the performance guarantee of Algorithm \ref{algorithm 1} by showing an upper bound on the regret function $R(T)$ that is sublinear in $T$.

\begin{theorem}\label{theorem regret}
Let $\Lambda^{+}_{h}$ and $\Lambda^{-}_{h}$ be the total numbers of change points of 
$\{p_{h,\,t}^{+}\}_{t=1}^{\infty}$ and $\{p_{h,\,t}^{-}\}_{t=1}^{\infty}$ for all $h=1,\ldots,H$, respectively. Then, the regret function $R(T)$ of the sequence of strategies $\{\pi_{t}\}_{t=1}^{T}$ generated by Algorithm \ref{algorithm 1} can be bounded by
\begin{align*}
R(T) \leq 
12\left[K^{+}\max_{h=1,\ldots,H}
\sqrt{\Lambda^{+}_{h}\max_{1\leq t'\leq T}\nu^{+}_{h,\,t'}}
+ K^{-}\max_{h=1,\ldots,H}\sqrt{\Lambda^{-}_{h}
\max_{1\leq t'\leq T}\nu^{-}_{h,\,t'}}\right]\sqrt{T\ln{2T}}
+ \frac{2\left(K^{+} + K^{-}\right)}{T}.
\end{align*}
\end{theorem}

\noindent\textit{Remark}. 
Our performance guarantees are informative because they give a growing rate of regret over time. 
In practice, bushfires may spread for a long time such as the 2019--2020 Australia bushfire season.
Our algorithm reveals a sub-linear regret over a long decision horizon, and the coefficients $K^+$ and $K^-$ do not play a critical role when the decision horizon is long.
We do not prove if the regret bound in Theorem \ref{theorem regret}, which is of the order $\sqrt{T\ln{T}}$, is optimal. 
Nevertheless, this bound is near-optimal based on the following intuitive heuristic reasoning.
Consider the case where there is no change point for $\{p_{h,t}^{+}\}_{t=1}^{\infty}$ or $\{p_{h,t}^{-}\}_{t=1}^{\infty}$. 
The operator can then use the mean values $\sum_{t'=1}^{t}\Hat{p}^{+}_{h,t'}/t$ and $\sum_{t'=1}^{t}\Hat{p}^{-}_{h,t'}/t$ to estimate $(p^{+}_{h,t-1},p^{-}_{h,t-1})_{h=1}^H$, respectively. 
The estimation biases for both $p_{h,t}^{+}$ and $p_{h,t}^{-}$ are of the order $O(1/\sqrt{t})$ at each time $1\leq t\leq T$.
By considering the relation $\sqrt{T}\leq \sum_{t=1}^{T}(1/\sqrt{t})\leq 2\sqrt{T}$, Theorem~\ref{theorem approximate optimal strategy} then implies a total regret of the order at least $O(\sqrt{T})$. 
By ignoring the insignificant term $\sqrt{\ln{T}}$, the order of the upper bound in Theorem~\ref{theorem regret} is consistent with that without change points. 
Moreover, the classical change point tracking approach suffers an $O(1)$ bias for the detection of each change point \citep{niu2016multiple}, which may possibly lead to a regret proportional to the total number of change points, i.e., $\Lambda^{+}_{h}$'s and $\Lambda^{-}_h$'s.
In contrast, our regret bound in Theorem \ref{theorem regret} relies on the number of change points through $\sqrt{\Lambda^{+}_{h}}$'s and $\sqrt{\Lambda^{-}_{h}}$'s, which outperforms the classical change point detection algorithms.

\section{Real Data on Bushfire Spread}
\label{section practical example of bushfire spread}

\begin{figure}[h]
	\centering
	\subfigure[At $t=24$ hour for selected day 1.]{
	\begin{minipage}[t]{0.4\linewidth}
		\centering
		\includegraphics[width=\textwidth]{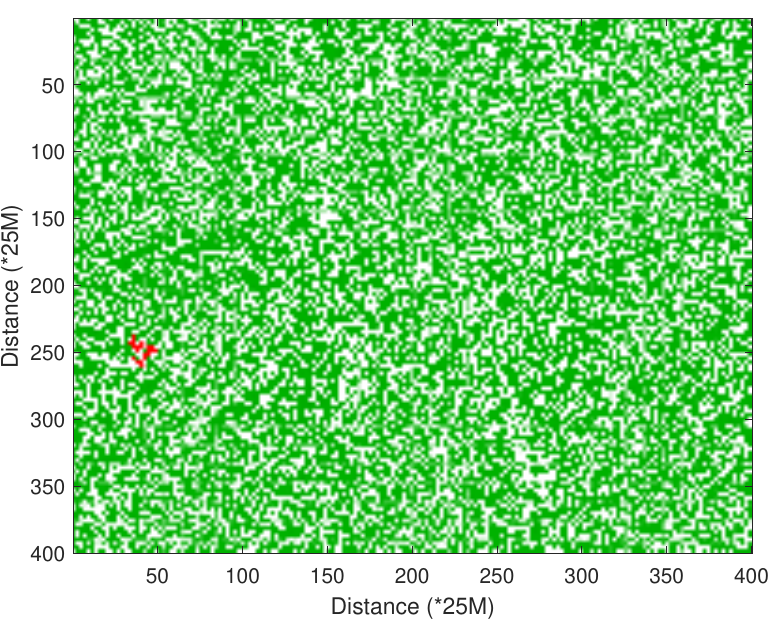}
		\label{fig scenario 1 8 hour}
	\end{minipage}%
	}%
	\subfigure[At $t=48$ hour for selected day 1.]{
		\begin{minipage}[t]{0.4\linewidth}
		\centering
		\includegraphics[width=\textwidth]{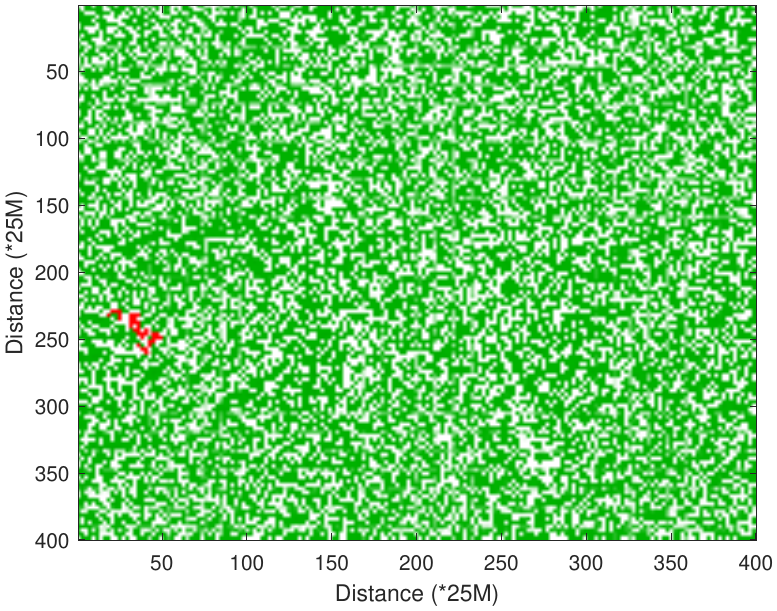}
		\label{fig scenario 1 24 hour}
		\end{minipage}%
	}%
	
	\subfigure[At $t=24$ hour for selected day 2.]{
	\begin{minipage}[t]{0.4\linewidth}
		\centering
		\includegraphics[width=\textwidth]{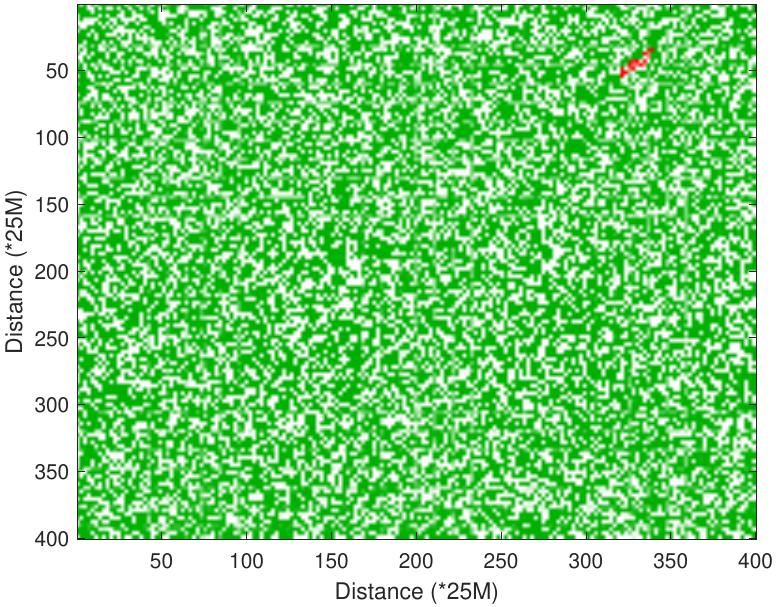}
		\label{fig scenario 2 8 hour}
	\end{minipage}%
	}%
	\subfigure[At $t=48$ hour for selected day 2.]{
	\begin{minipage}[t]{0.4\linewidth}
		\centering
		\includegraphics[width=\textwidth]{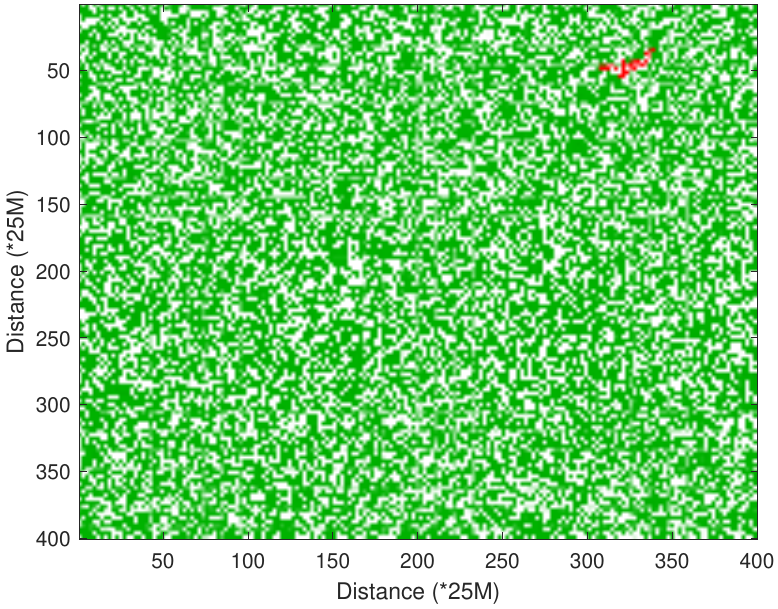}
		\label{fig scenario 2 24 hour}
	\end{minipage}%
	}%
 
 	\subfigure{
           \begin{minipage}[t]{0.8\linewidth}
		\centering
		\includegraphics[width=\textwidth]{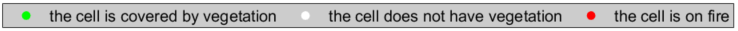}
			\end{minipage}%
	}%
	\caption{The spread of bushfires at different times in two randomly selected days.
	The time $t$ here refers to the elapsed time since the start of the 48-hour observation horizon.} 
	\label{fig scenarios evolution}
\end{figure}

This section uses real-world bushfire spread data to verify our model assumptions. 
The data are sourced from operational personnel who were heavily involved in suppression operations management of Australian bushfires. 
The dataset includes the locations of fires that occurred during 2019--2020 bushfire season within a specific region of New South Wales, Australia \citep{NSWcurrent}.
The region is mainly covered by one type of vegetation, and the corresponding grid map $\mathcal{G}$ is divided into $H=2$ types of areas (no vegetation and vegetation).
The size of the region is $10\times 10$ km$^2$, and the size of each grid cell is $25\times 25$ m$^2$, i.e., $M=N=400$.
The data were collected every 15 minutes, which is also the period length to adjust the power flows of the electricity network.
Based on the period length, we examine the dynamic change of $p_{h,t}^+$ and $p_{h,t}^-$ for all the areas within a 48-hour time horizon for two randomly selected days.
A representative table of the data is provided as Table~\ref{table sample}, where ``time'' refers to the elapsed time since the start of the 48-hour observation window.
{Figure~\ref{fig scenarios evolution} shows a graphical illustration of the fire spread on two randomly selected days at $t=24$ and $t=48$.}

%


%
\begin{table}[h]
\centering
\small
\caption{A sample of the bushfire spread data.}
\label{table sample}
\begin{tabular}{cccccc}
\hline
Time & $30$ min & $45$ min & $\cdots$ & $75$ min & $\cdots$ \\ \hline
No.\ of cells on fire
& $1$  & $4$ & $\cdots$ & $8$  & $\cdots$ \\ 
Fire coordinates
& \multicolumn{1}{l}{$(52,\,15)$} 
& \multicolumn{1}{l}{\begin{tabular}[c]{@{}l@{}}$(51,\,14)$, 
{$(51,\,16)$},\\ $(52,\,14)$, $(52,\,15)$
\end{tabular}} 
& $\cdots$ 
& \multicolumn{1}{l}{\begin{tabular}[c]{@{}l@{}}$(50,\,14)$, $(51,\,13)$, $(51,\,14)$, $(51,\,15)$,\\ 
$(52,\,13)$, $(52,\,14)$, $(52,\,15)$, $(53,\,14)$
\end{tabular}} & $\cdots$ \\ \hline
\end{tabular}
\end{table}
\begin{figure}[!h]
	\centering
	\subfigure[$\Hat{p}^{+}_{t}$ for area 1.]{
	\begin{minipage}[t]{0.5\linewidth}
		\centering
		\includegraphics[width=\textwidth]{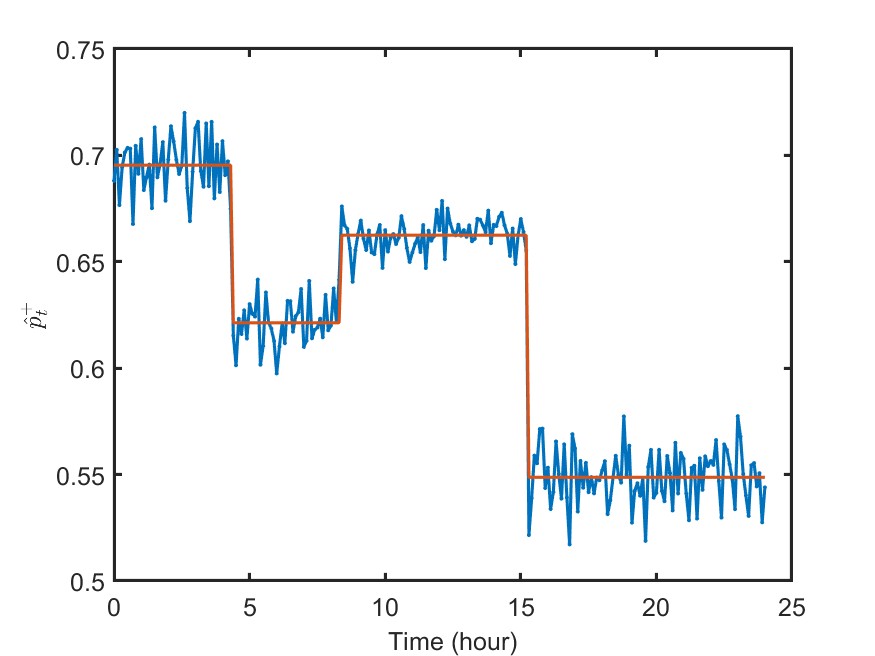}
		\label{fig scenario 1 p+}
	\end{minipage}%
	}%
	\subfigure[$\Hat{p}^{-}_{t}$  for area 1.]{
		\begin{minipage}[t]{0.5\linewidth}
		\centering
		\includegraphics[width=\textwidth]{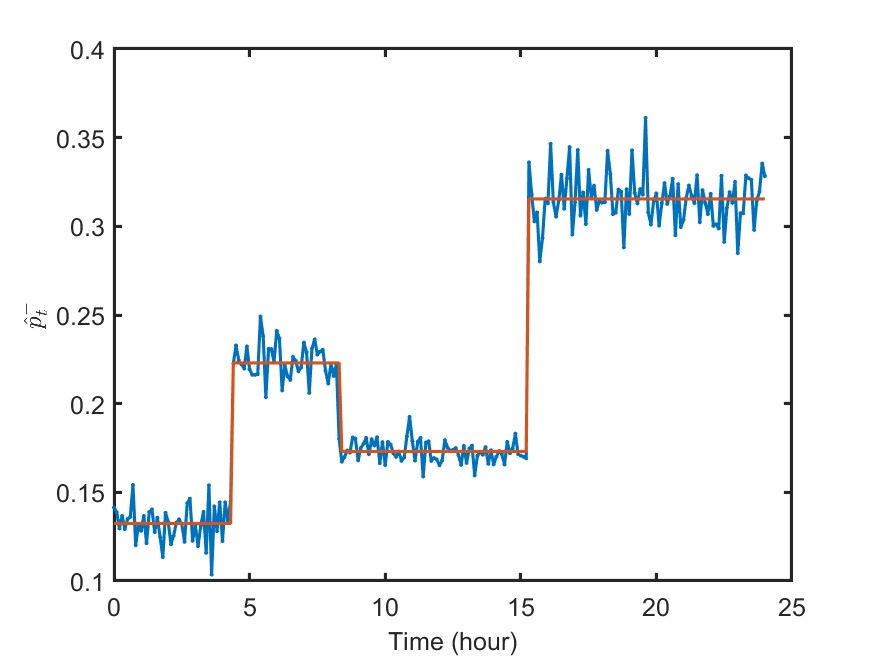}
		\label{fig scenario 1 p-}
		\end{minipage}%
	}%
	
	\subfigure[$\Hat{p}^{+}_{t}$  for area 2.]{
	\begin{minipage}[t]{0.5\linewidth}
		\centering
		\includegraphics[width=\textwidth]{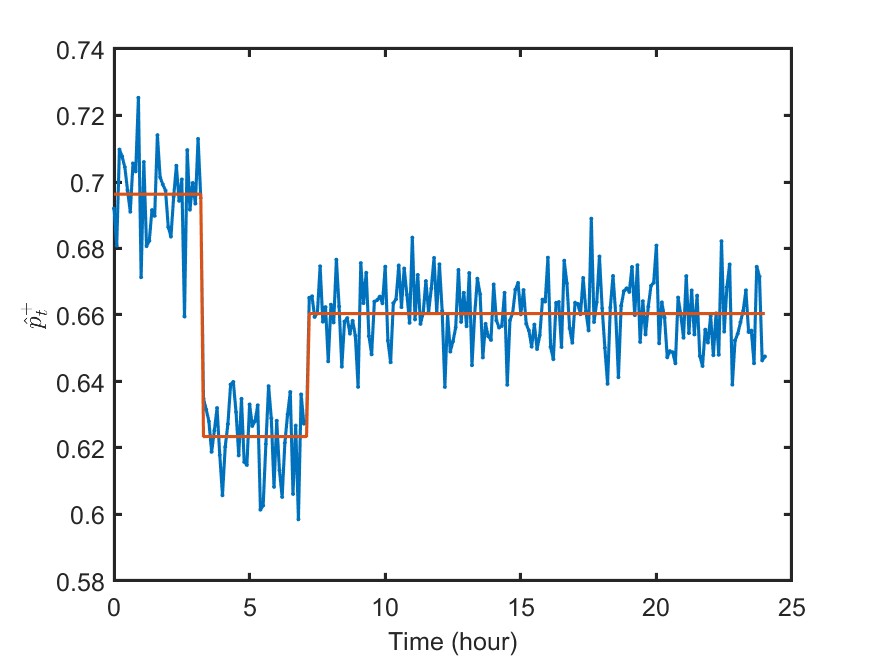}
		\label{fig scenario 2 p+}
	\end{minipage}%
	}%
	\subfigure[$\Hat{p}^{-}_{t}$  for area 2.]{
	\begin{minipage}[t]{0.5\linewidth}
		\centering
		\includegraphics[width=\textwidth]{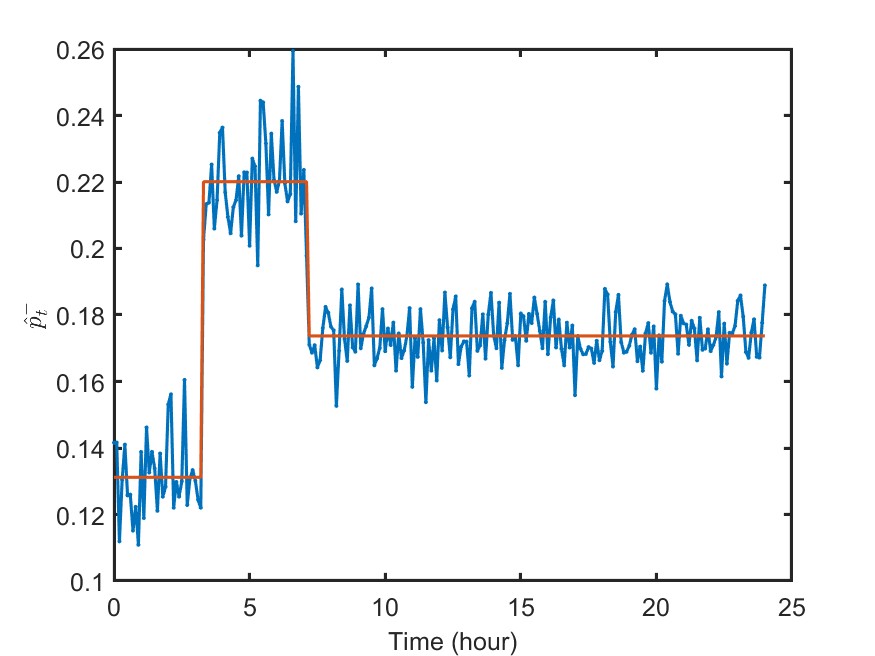}
		\label{fig scenario 2 p-}
	\end{minipage}%
	}%
	\caption{The MLEs $\Hat{p}^{+}_{t}$ and $\Hat{p}^{-}_{t}$ within $24$ hours for two geographical {regions}.
	Red curves indicate the mean of $\hat p_t^+$ or $\hat p_t^-$ during each specified time interval.} 
	\label{fig scenarios p}
\end{figure} 

\begin{figure}[h]
	\centering
	\subfigure[$\Hat{p}^{+}_{t}$ for selected region 1.]{
	\begin{minipage}[t]{0.5\linewidth}
		\centering
		\includegraphics[width=\textwidth]{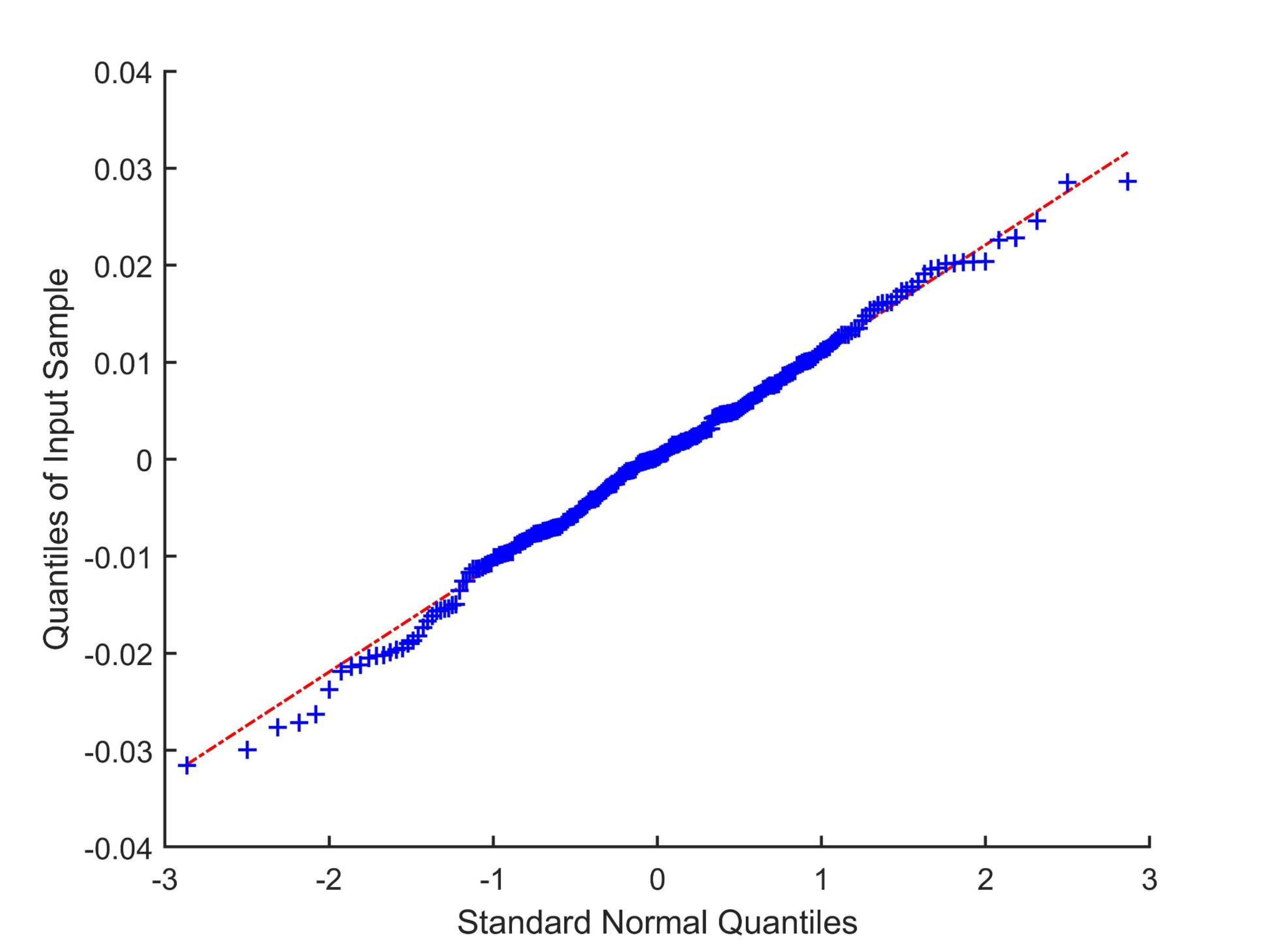}
		\label{qq-scenario-1-p+}
	\end{minipage}%
	}%
	\subfigure[$\Hat{p}^{-}_{t}$  for selected region 1.]{
		\begin{minipage}[t]{0.5\linewidth}
		\centering
		\includegraphics[width=\textwidth]{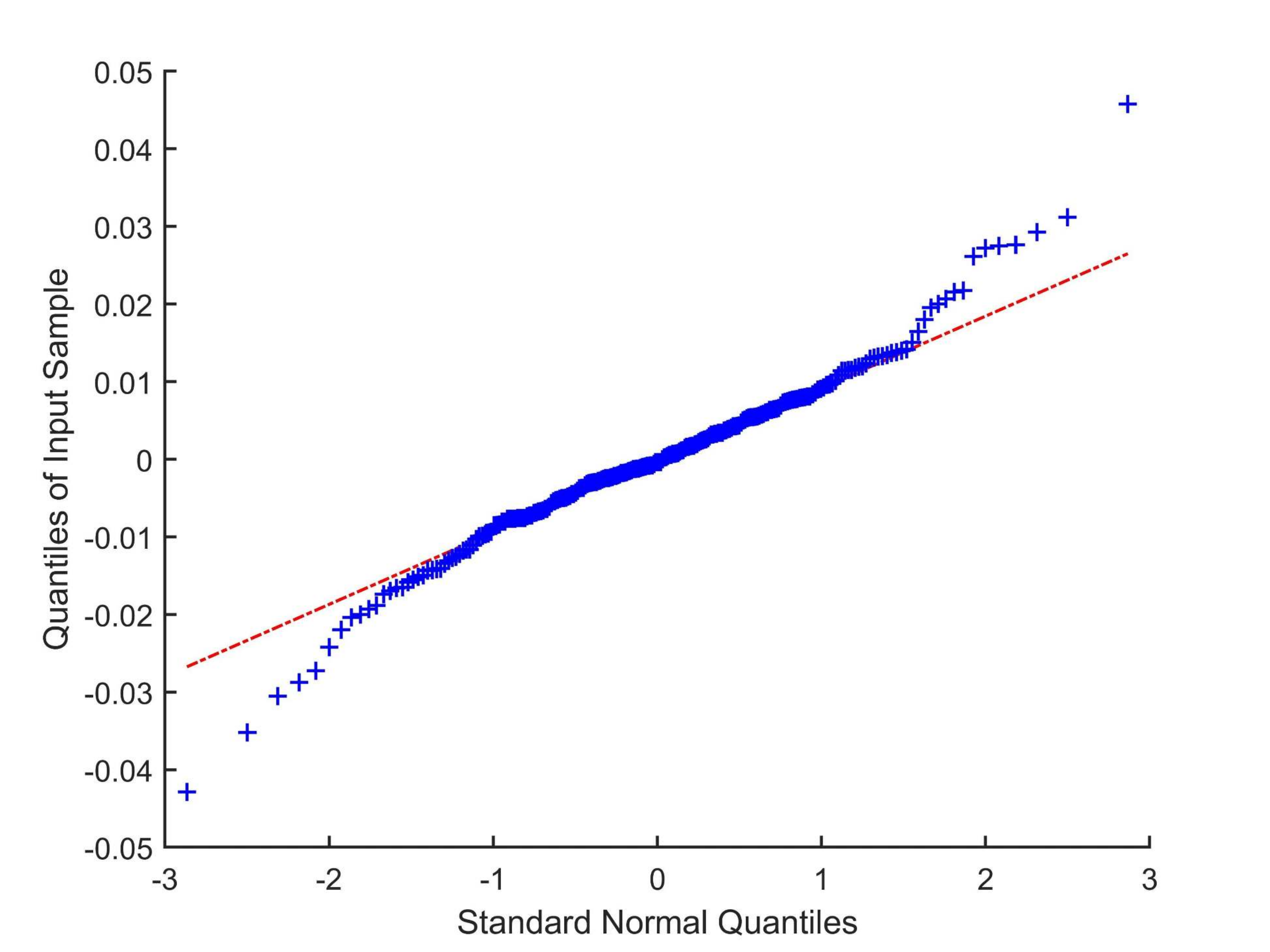}
		\label{qq-scenario-1-p-}
		\end{minipage}%
	}%
	
	\subfigure[$\Hat{p}^{+}_{t}$  for selected region 2.]{
	\begin{minipage}[t]{0.5\linewidth}
		\centering
		\includegraphics[width=\textwidth]{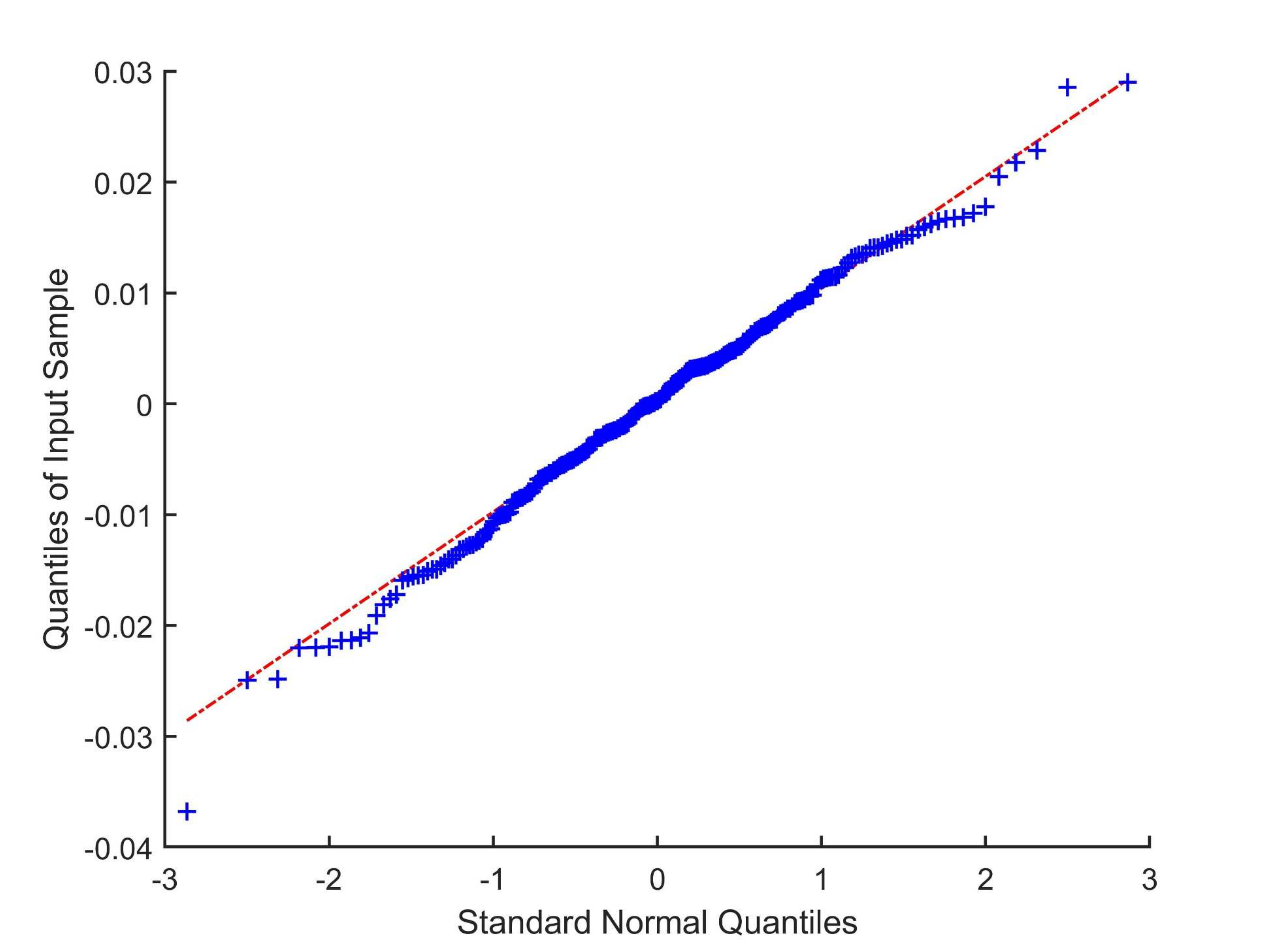}
		\label{qq-scenario-2-p+}
	\end{minipage}%
	}%
	\subfigure[$\Hat{p}^{-}_{t}$  for selected region 2.]{
	\begin{minipage}[t]{0.5\linewidth}
		\centering
		\includegraphics[width=\textwidth]{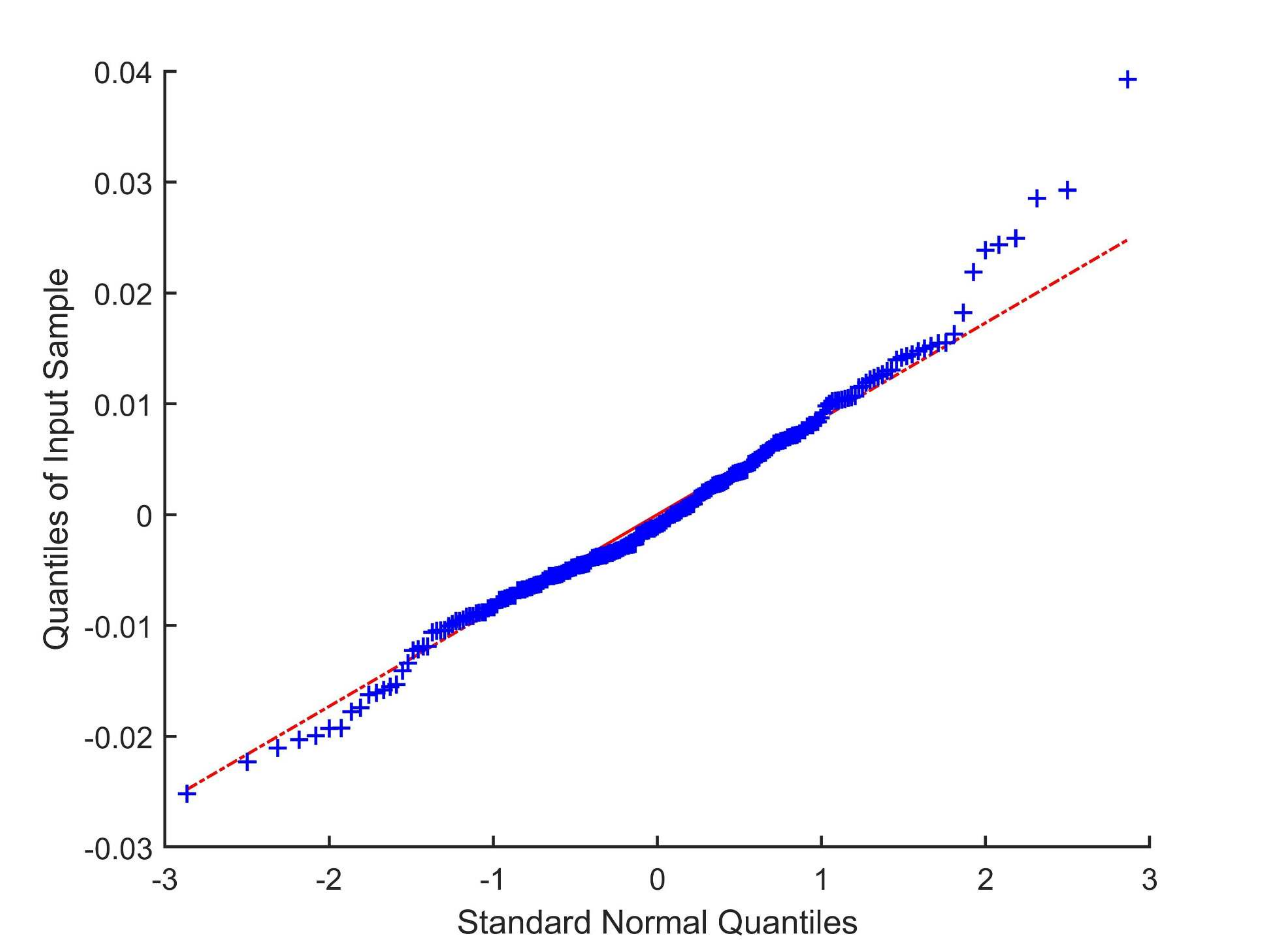}
		\label{qq-scenario-2-p-}
	\end{minipage}%
	}%
	\caption{The Q-Q plots for the estimation biases $\Hat{p}^{+}_{t}-{p}^{+}_{t}$ and $\Hat{p}^{-}_{t}-{p}^{-}_{t}$ within the $24$-hour observation window for two randomly selected regions.} 
	\label{fig-qq}
\end{figure} 

We compute the MLEs of $p_{h,t}^+$ and $p_{h,t}^-$ for all the areas based on the procedures in Section~\ref{section bushfire spread model}.
The estimation results for two types of {regions} on a randomly selected day are shown in Figure~\ref{fig scenarios p}, where {regions} 1 and 2 are mainly covered by shrubs and forests, respectively. 
We can see that both $\hat p_{h,t}^+$ and $\hat p_{h,t}^-$ of these two geographical {regions} exhibit a piecewise constant pattern.
This verifies our model assumption that the true $\{p_{h,t}^+\}_{t=1}^\infty$ and $\{p_{h,t}^-\}_{t=1}^\infty$ are piecewise constant over $t$ for all areas.
We then verify the assumption that the estimation errors are sub-Gaussian.
To this end, we make the quantile-quantile (Q-Q) plots for the estimation errors $p_{h,t}^+ - \hat p_{h,t}^+$ and $p_{h,t}^- - \hat p_{h,t}^-$, respectively, on each of the two randomly selected geographical regions.
The results are shown in Figure~\ref{fig-qq}.
Figures~\ref{qq-scenario-1-p+} and \ref{qq-scenario-2-p+} suggest that the estimation biases approximately follow the standard normal distribution.
In addition, the tail patterns in Figures~\ref{qq-scenario-1-p-} and \ref{qq-scenario-2-p-} reveal that $p_{h,t}^- - \hat p_{h,t}^-$ are more concentrated in zero compared with the standard normal distribution.
All the above results support our assumption that the estimation biases are sub-Gaussian.

\section{Simulation Study}
\label{section simulation study}

Due to confidentiality, we cannot obtain the information on the real-world power system in the region that caught bushfire in Section~\ref{section practical example of bushfire spread}.
Therefore, we first use a $11$-node radial smart power system \citep{xu2021bayesian} to illustrate the application of our online algorithm in Section~\ref{section-simulation-study-11}. 
We then apply our algorithm to a more complex system with a mesh structure, an IEEE 57 Bus System, in Section~\ref{section simulation study 57}.
\subsection{Case Study 1: IEEE 11 Bus System}\label{section-simulation-study-11}
This 11-node radial smart power system is extracted by IEEE from the test feeder in Elia Grid, Belgium. 
In the power system, each node is either a generator of one of the following four different types, namely, gas, biomass, photovoltaic (PV) and wind generator, or a consumer. 
Moreover, there is an energy storage subsystem in the network to cover the shedding load. 
To implement our online algorithm for this power system, we embed the electricity network into a grid map $\mathcal{G}$.
According to the actual size of the original power system in Elia Grid, we choose a $400\times 400$ grid as the whole region.
We let $H=1$ for simplicity and use $p_t^+$ and $p_t^-$ as shorthand notations of $p_{h,t}^+$ and $p_{h,t}^-$, respectively, in this section. 
For optimization, we let $\bar{d}=3$.
We set different maximum capacities and capacity costs for each kind of generators.
In practice, both the maximum capacity and unit capacity cost of a generator usually follows the descending order: natural gas plant; biomass plant; PV farm; wind farm.
Therefore, we set the model parameters as follows. 
For all $i\in\mathcal{S}$,
\begin{align*}
U^\text{P}(i)=\begin{cases}
4, & i \textup{ is a natural gas plant},\\
3, & i \textup{ is a biomass plant},\\
2, & i \textup{ is a PV farm or wind farm},\\
0, & \textup{otherwise},
\end{cases}\quad\quad
C^{\text{P}}(i)=\begin{cases}
10, & i \textup{ is a natural gas plant},\\
8, & i \textup{ is a biomass plant},\\
6, & i \textup{ is a PV farm or wind farm},\\
0, & \textup{otherwise},
\end{cases}
\end{align*}
and $L(i)=3$ if node $i$ corresponds to ``load'' (see Appendix~\ref{appen:PS}). 
For all $(i,\,j)\in\mathcal{E}$, we set $U^\text{F}(i,\,j)=4$ and $C^\text{F}(i,\,j)=1$. 
The unit cost of load shedding is set to be $C^{\text{S}}=20$. 
Appendix~\ref{appen:PS} provides graphical illustrations of the structure of the power system and the coordinates of different components of the smart power system in the grid $\mathcal{G}$.

We then conduct comprehensive simulation studies on the above power system.
Based on our estimation results using real data in Section~\ref{section practical example of bushfire spread}, there are around two to five change points per day; see Figure~\ref{fig scenarios p}.
Accordingly, we consider two numbers of change points in simulations, i.e., $\Lambda^+,\Lambda^-\in\{50,100\}$, which can represent the number of change points in most bushfire spread cases over 20 days, and conduct a set of simulations for each value. 
In each trial, the change points are randomly chosen from all time periods within the time horizon.
At each change point, we choose $p^{+}_{t}$ randomly from $[0.2,\,0.6]$ and $p^{-}_{t}$ randomly from $[0.1,\,0.4]$.
In each simulation, we consider two scenarios, where scenario 1 (scenario 2) randomly selects one node (two nodes) in $\mathcal{G}$ as the origin of bushfires.
Then, based on the values of $p_t^+$ and $p_t^-$ over $t$, we simulate the spread of fire within the grid map $\mathcal{G}$.
We separately use our online algorithm and three benchmarks illustrated later to optimize the power flows of the electricity network at each time.
{\color{blue!80!black}It takes several seconds to solve Problem~\eqref{eq:power_flow_LP}.}
To evaluate the regret function $R(T)$ under different algorithms (which is an expectation), 
we use the sample average where we generate 100 sequences of $\{p_t^+\}_{t=1}^T$ and $\{p_t^-\}_{t=1}^T$, and then replicate the simulation 1,000 times for each sequence and each scenario.
We set the simulation horizon to be $T=2,000$.

We consider three commonly used benchmarks, i.e., ``Algorithm with naive estimation'', ``Algorithm ignoring change points'', and ``Algorithm with traditional change point detection'', for comparison to evaluate the performance of our online algorithm.
The first benchmark is a naive algorithm that just replaces $(p_t^+,p_t^-)$ for Problem~\eqref{eq:power_flow_LP} with $(\hat{p}^{+}_{t},\hat{p}^{-}_{t})$ obtained in Section~\ref{section bushfire spread model} to plan the power flows at each time $t$.
The second benchmark simply uses the sample averages of the estimates $\hat p^{+}_{t}$ and $\hat p^{-}_{t}$ available hitherto for optimization.
That is, this benchmark does not conduct change point detection.
The third benchmark uses the classical likelihood-ratio-based method \citep{niu2016multiple} to track possible change points of $\{\hat{p}^{+}_{t}\}_{t=1}^{T}$ and $\{\hat{p}^{-}_{t}\}_{t=1}^{T}$.
After the detection of change points, the benchmark then uses the same procedure as our algorithm for optimization; 
namely, the benchmark only averages the estimates of $p_t^+$ and $p_t^-$ in the same episode (see Section~\ref{section online operational algorithm}) and plugs the estimates into Problem~\eqref{eq:power_flow_LP} for OPF planning.
To detect change points based on the likelihood ratio, the third benchmark calculates the following quantities for all $\tau^{+}< t'\leq t$ and $\tau^{-}< t'\leq t$ at time $t$:
\begin{align*}
\Gamma^{+}_{t'}=
\frac{(\hat{p}^{+}_{\tau^{+},\,t'-1}-\hat{p}^{+}_{t',\,t})^{2}}
{\left(t'-\tau^{+}\right)^{-1} + \left(t-t'+1\right)^{-1}};
\quad
\Gamma^{-}_{t'}=
\frac{(\hat{p}^{-}_{\tau^{-},\,t'-1}-\hat{p}^{-}_{t',\,t})^{2}}
{\left(t'-\tau^{-}\right)^{-1} + \left(t-t'+1\right)^{-1}},
\end{align*}
where $\tau^+$ and $\tau^-$ are the most recent detected change points for $p_t^+$ and $p_t^-$, respectively.
Let $t^{+}\triangleq \argmax_{\tau^{+}< t'\leq t}\Gamma^{+}_{t'}$ and $t^{-}\triangleq \argmax_{\tau^{-}< t'\leq t}\Gamma^{-}_{t'}$.
Then $\Gamma^{+}_{t^{+}}$ and $\Gamma^{+}_{t^{-}}$ are used as test statistics.
The time $t^+$ or $t^-$ is set as the change point for $p_t^+$ or $p_t^-$ if the corresponding testing statistics $\Gamma^{+}_{t^{+}}$ or $\Gamma^{+}_{t^{-}}$ exceed a threshold (which can be determined by bootstrap or the asymptotic distribution of the testing statistics).
Interested readers may refer to \cite{niu2016multiple} for the rationale of this likelihood-ratio-based method.

\begin{figure}[h]
	\centering
	\subfigure[One origin of fire; $\Lambda^{+}=\Lambda^{-}=50$.]{
	\begin{minipage}[t]{0.5\linewidth}
		\centering
		\includegraphics[width=\textwidth]{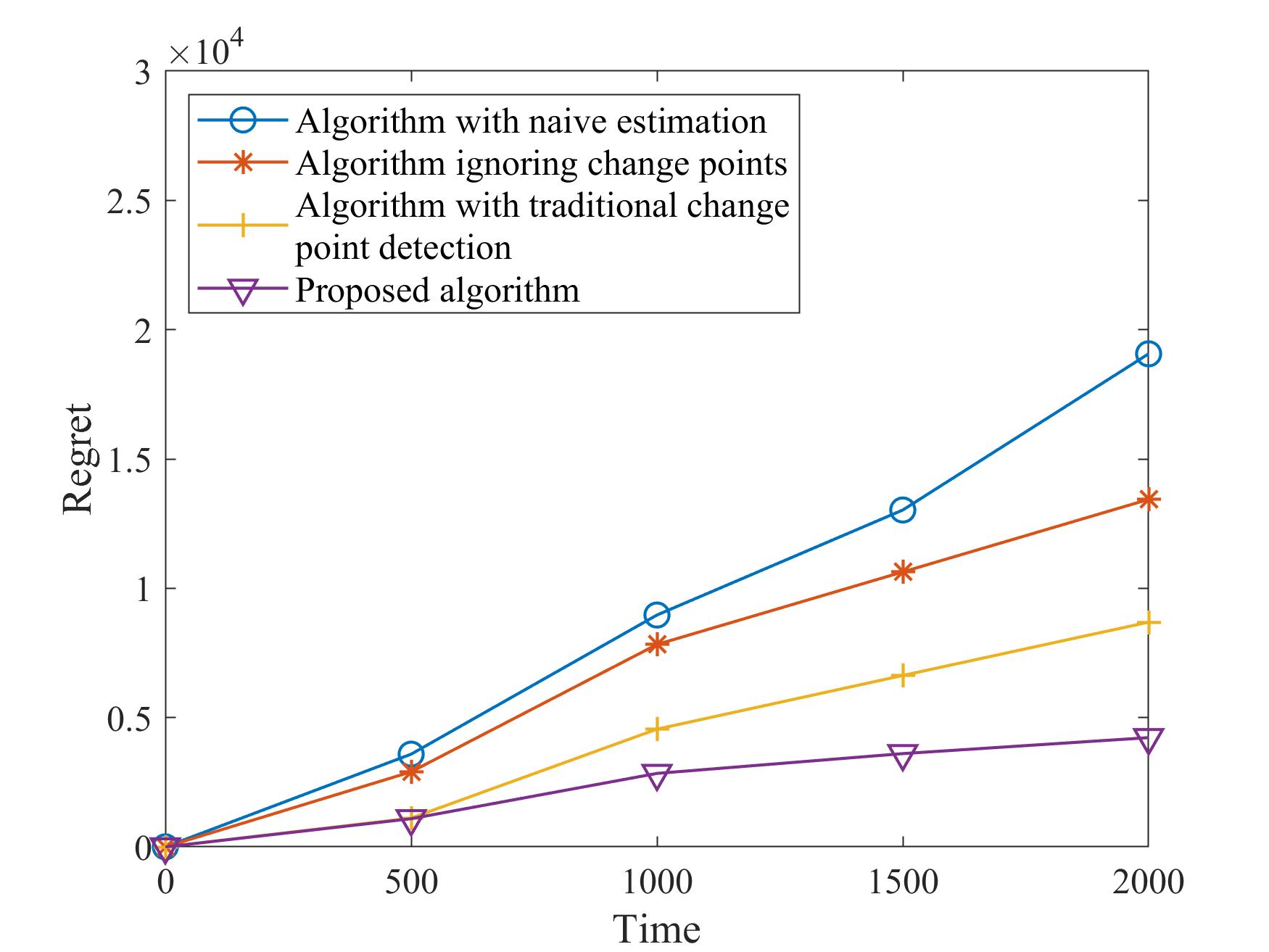}
		\label{fig simulation regret 11}
	\end{minipage}%
	}%
	\subfigure[One origin of fire; $\Lambda^{+}=\Lambda^{-}=100$.]{
		\begin{minipage}[t]{0.5\linewidth}
		\centering
		\includegraphics[width=\textwidth]{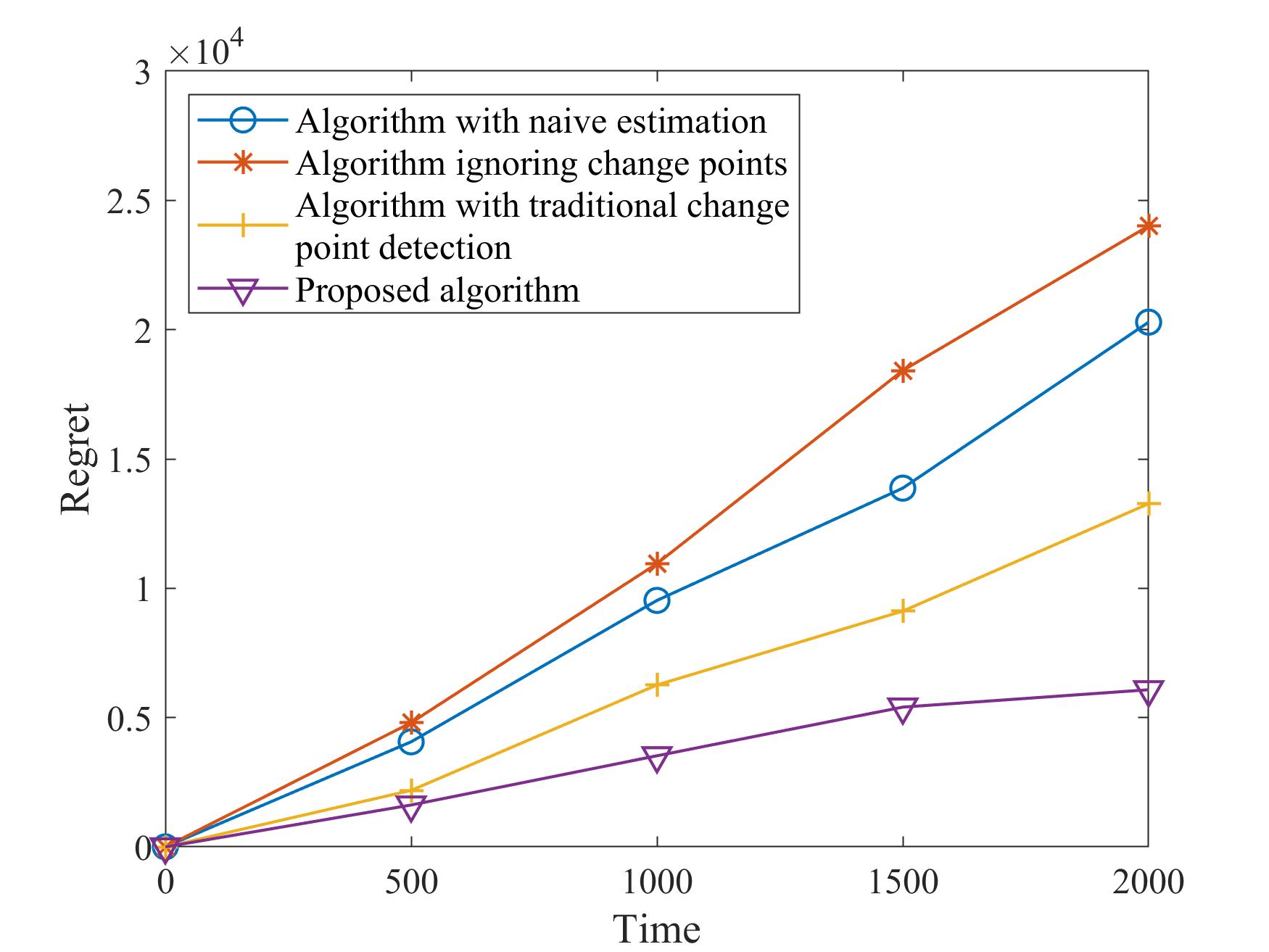}
		\label{fig simulation regret 12}
		\end{minipage}%
	}%
	
	\subfigure[Two origins of fire; $\Lambda^{+}=\Lambda^{-}=50$.]{
	\begin{minipage}[t]{0.5\linewidth}
		\centering
		\includegraphics[width=\textwidth]{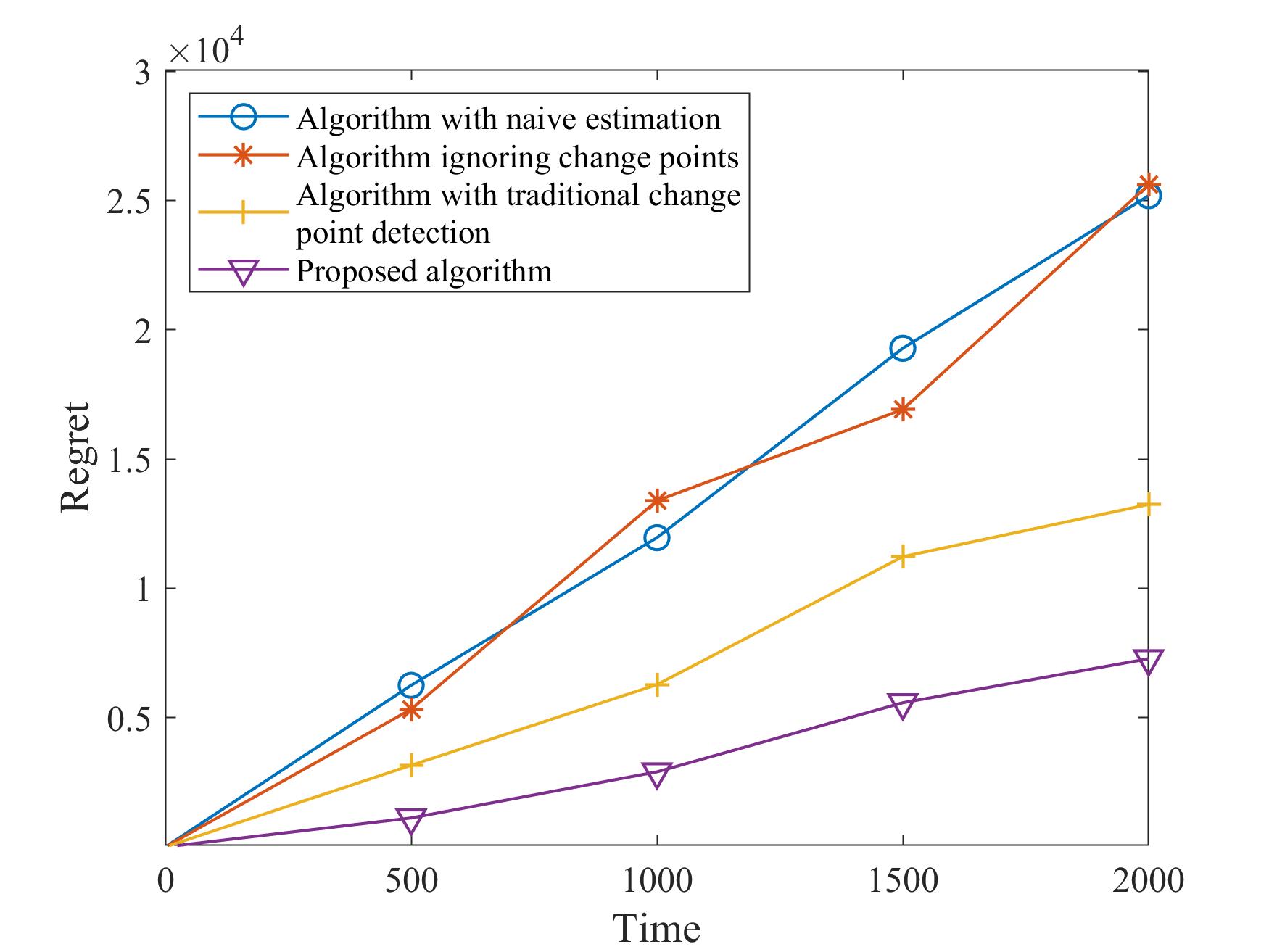}
		\label{fig simulation regret 21}
	\end{minipage}%
	}%
	\subfigure[Two origins of fire; $\Lambda^{+}=\Lambda^{-}=100$.]{
	\begin{minipage}[t]{0.5\linewidth}
		\centering
		\includegraphics[width=\textwidth]{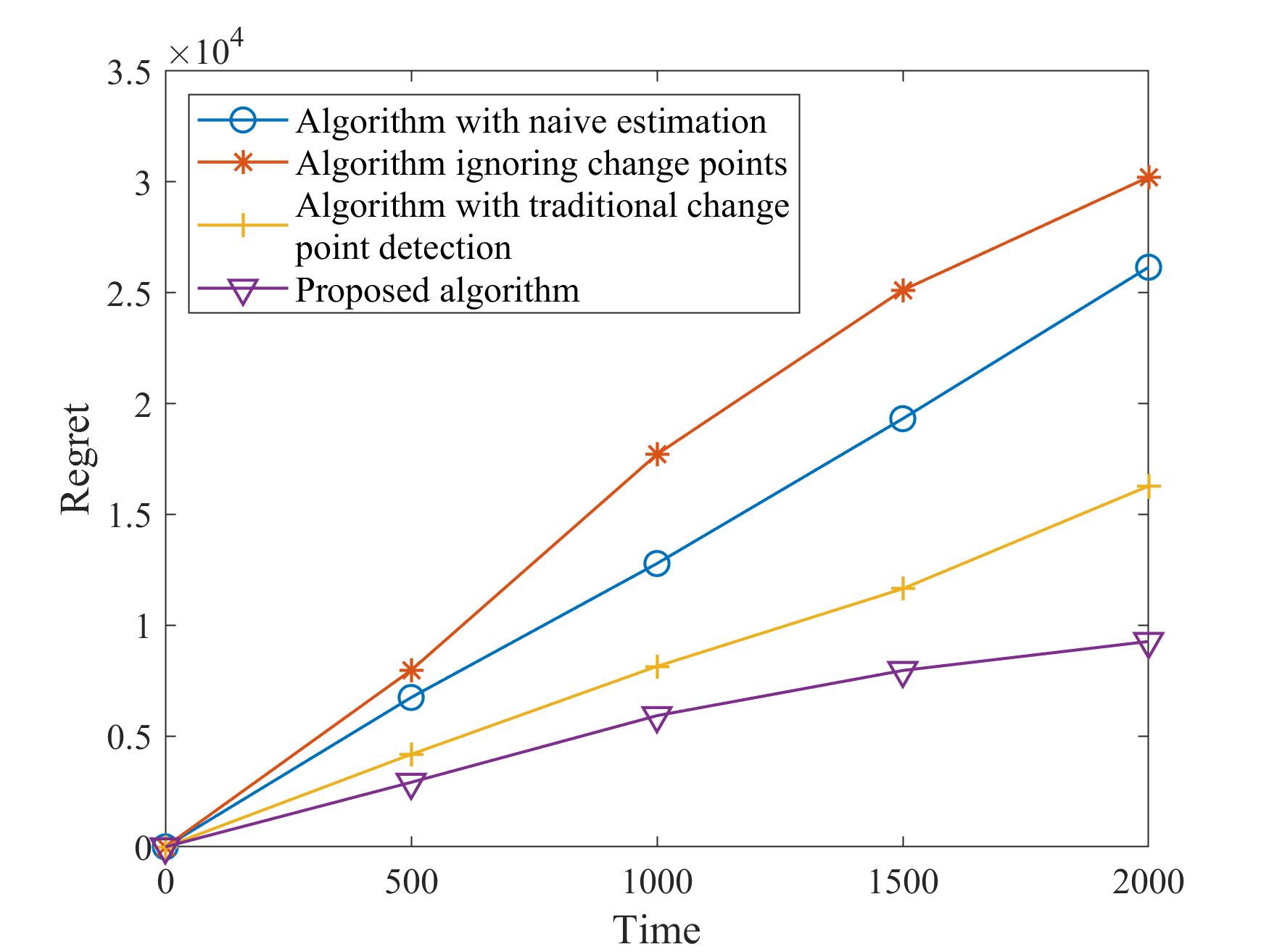}
		\label{fig simulation regret 22}
	\end{minipage}%
	}%
	\caption{The regret functions of the four algorithms for IEEE 11 Bus System.} 
	\label{fig-simulation-regret 11}
\end{figure}

The regret functions in the simulations under all the optimization methods are displayed in Figure~\ref{fig-simulation-regret 11}.
We see that our proposed online algorithm significantly outperforms all three benchmarks, no matter how many origins of bushfires are set in the simulations.
At the end of the simulation horizon, the regret obtained by the benchmarks can be twice to six times of that under our algorithm.
We note that the difference between all the algorithms is not large when $t$ is small.
This may be because, in most simulation replications, it takes some time for the bushfire to spread such that it can affect the electricity network.
Hence, the operations management of the power system is usually not influenced by bushfires at the early stage of simulations.
Moreover, the regrets obtained by the benchmark algorithms do not reveal any sign of convergence over time $t$.
In contrast, the regret by our algorithm shows a sub-linear pattern. 
We can also conclude from Figure \ref{fig-simulation-regret 11} that the performance of our algorithm is very robust when the number of change points increases, in the sense that the regret curve of our algorithm does not change significantly from Figure~\ref{fig simulation regret 11} to Figure~\ref{fig simulation regret 12}. 
This is consistent with Theorem~\ref{theorem regret} that the regret bounds rely on the number of change points by an order of $\sqrt{\Lambda^{+}}$ or $\sqrt{\Lambda^{-}}$. 

We provide some intuitive explanations on why the proposed online change point detection method outperforms the traditional likelihood-ratio-based method for our problem. 
We focus on the sequence 
$\{p^{+}_{t}\}_{t=1}^{\infty}$ for the ease of discussion. 
First, at each time $t$, the traditional method uses $\Gamma^{+}_{t'}$ to  separate the observations from $\tau^{+}$ to $t$ into two sets, i.e., the first $t'-\tau^{+}$ observations and the last $t-t'+1$ observations. 
Then, the traditional method only uses the differences between the averages of these two sets. 
In contrast, the proposed change point detection method checks the differences between the average of all subsets of the observations between $\tau^{+}$ and $t$, which makes our method potentially more sensitive to change points. 
Second, our method uses concentration inequalities to derive the exact probability bounds in (\ref{eq condition p+}). 
These bounds are probably sharper and lead to a more precise detection. 
Last, we show that the probability that any estimation bias exceeds the bound in (\ref{eq condition p+}) can be bounded from above, such that we can prove the convergence of the regret function and obtain the performance guarantee of our algorithm in an online setting.
Such results are generally difficult to derive from the traditional method. 

\subsection{Case Study 2: IEEE 57 Bus System}\label{section simulation study 57}

We also apply our model to the IEEE $57$ Bus System that represents an approximation of a portion of the American Electric Power system (in the U.S.\ Midwest) as it was in the early 1960s. 
The IEEE 57 Bus System is a meshed network with a high resilience to the shutdown of generators due to bushfires.
This system consists of 57 buses, 7 conventional generators, and 42 loads. 
We make some modifications to this 57 Bus System, such as adding several generators using renewable resources, so that the system can represent a modern power system.
We embed this IEEE $57$ Bus System into a grid map $\mathcal{G}$.
According to the actual size of the IEEE $57$ Bus System, the whole grid map $\mathcal{G}$ is $50\times 50$ km$^2$, and the size of each grid cell is $25\times 25$ m$^2$, i.e., $M=N=2000$. 
Appendix~\ref{appen:PS} provides graphical illustrations of the structure of the power system and the coordinates of different components of the smart power system in the grid $\mathcal{G}$.

We separate the whole region into $H=4$ areas. 
Since we have already examined the sensitivity of the algorithm with respect to the number of change points in Section~\ref{section-simulation-study-11}, we fix the number of change points as $100$ in this case study. 
We generate the time for change randomly. 
At each change point, the new values of $p^{+}_{h,\,t}$ and $p^{-}_{h,\,t}$ are chosen randomly from $[0.7,\,0.9]$ and $[0.1,\,0.4]$, respectively, for each area. 
In the IEEE 57 Bus System, there are two types of generators, i.e., fossil-fuel generators and renewable-energy generators. 
We set the capacity and unit cost for different generators as
\begin{align*}
U^\text{P}(i)=\begin{cases}
3, & i \textup{ is a fossil-fuel generator},\\
1, & i \textup{ is a renewable generator},
\end{cases}\quad\quad
C^{\text{P}}(i)=\begin{cases}
5, & i \textup{ is a natural gas plant},\\
2, & i \textup{ is a PV farm or wind farm},
\end{cases}
\end{align*}
and the rest of the parameters are set the same as in Section~\ref{section-simulation-study-11}.
\begin{figure}[h]
	\centering
	\subfigure[Two origins of fire; $\Lambda^{+}=\Lambda^{-}=100$.]{
	\begin{minipage}[t]{0.5\linewidth}
		\centering
		\includegraphics[width=\textwidth]{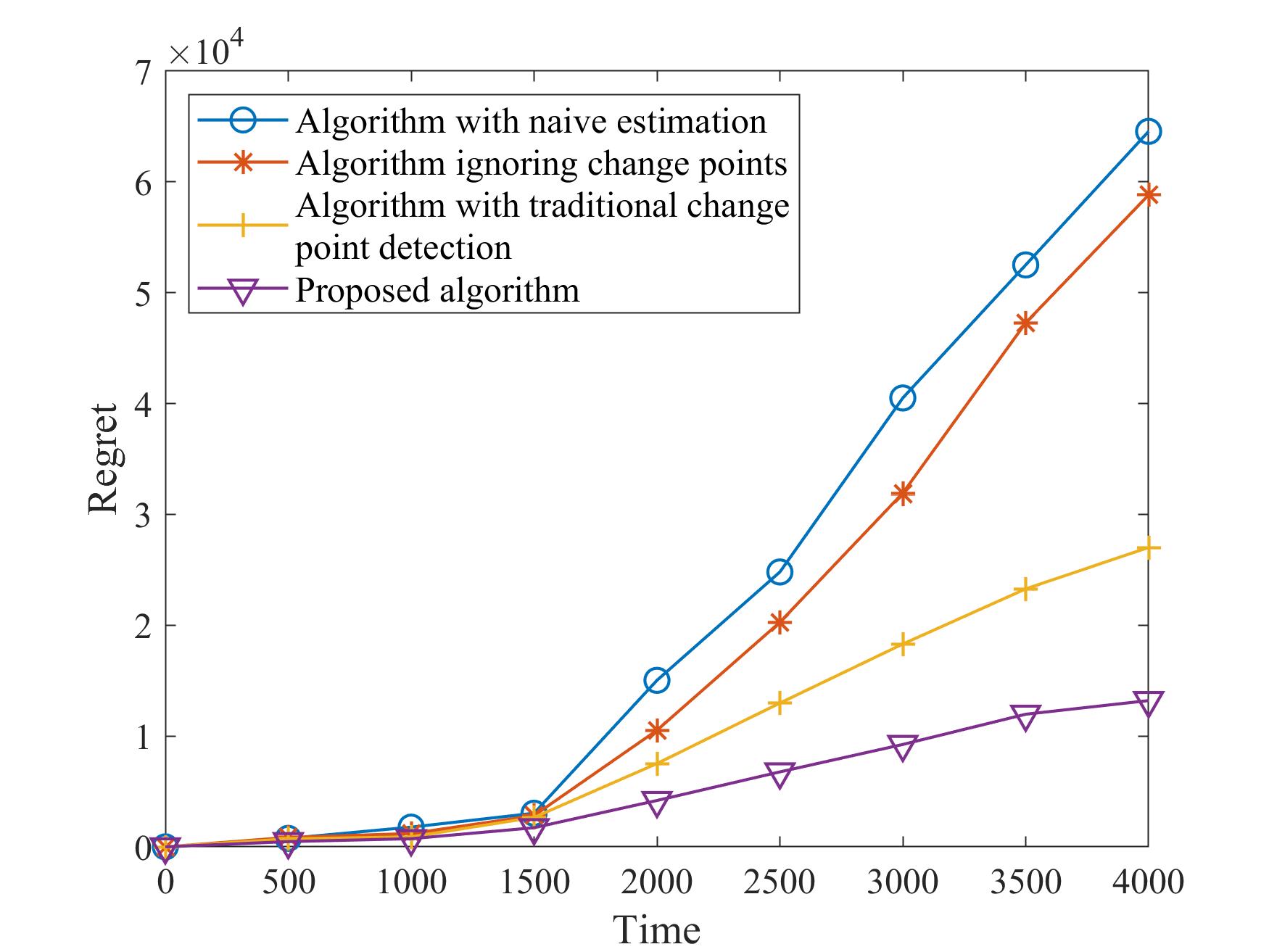}
		\label{fig simulation regret 31}
	\end{minipage}%
	}%
	\subfigure[Four origins of fire; $\Lambda^{+}=\Lambda^{-}=100$.]{
		\begin{minipage}[t]{0.5\linewidth}
		\centering
		\includegraphics[width=\textwidth]{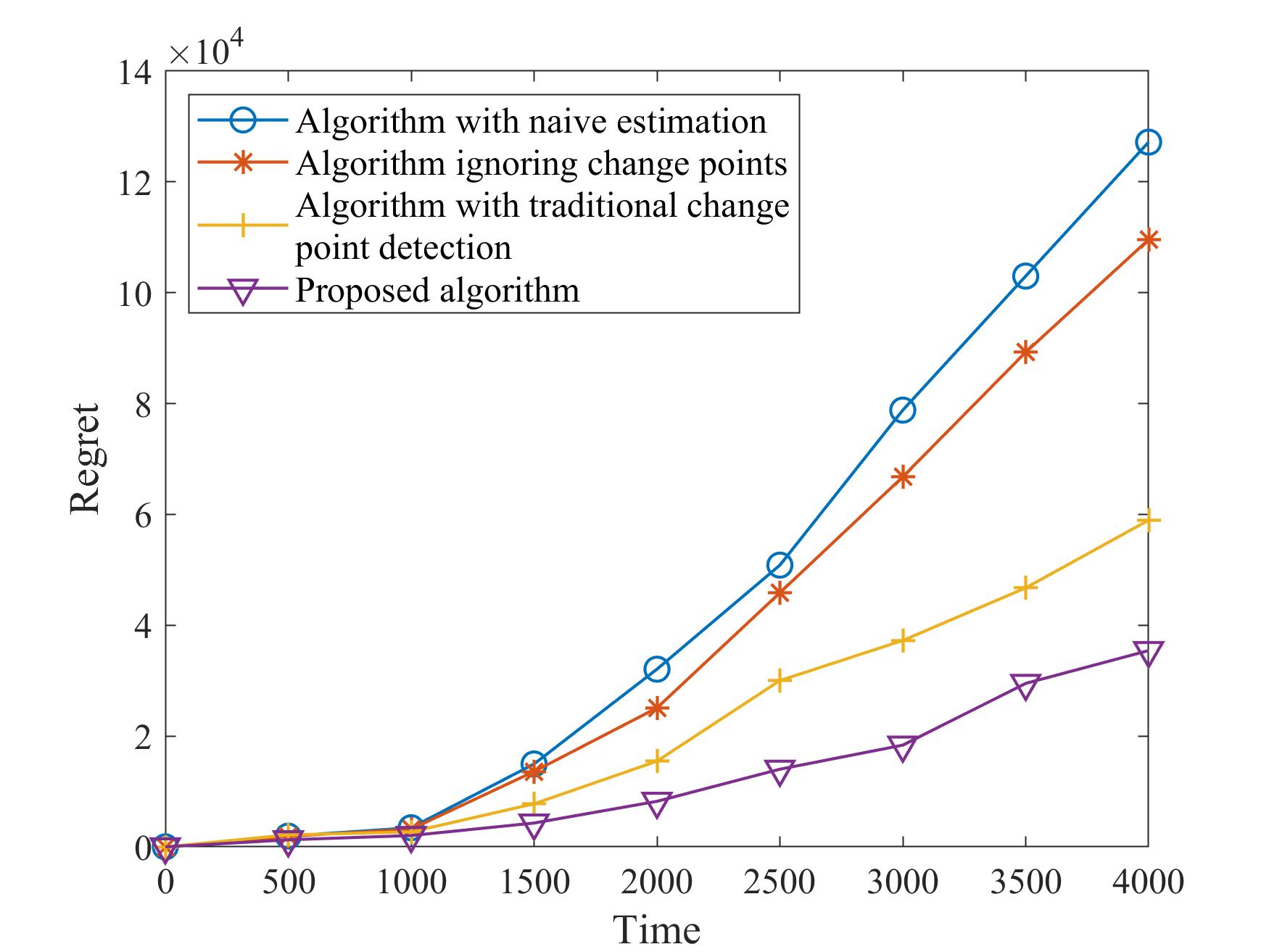}
		\label{fig simulation regret 32}
		\end{minipage}%
	}%
 	\caption{The regret functions of the four algorithms for the IEEE 57 Bus System.} 
	\label{fig-simulation-regret 57}
\end{figure} 

We assess the performance of our model with the benchmarks by setting the numbers of fire origins to two and four and keeping other simulation settings the same as in Section~\ref{section-simulation-study-11}.
When the number of fire origins equals two, we randomly choose two out of the four areas, and a node in each chosen area is randomly selected to be the fire origin.
When the number of fire origins equals four, we randomly select a node in each of the $H=4$ areas as the fire origin.
{Generally, it takes less than a minute to solve Problem~\eqref{eq:power_flow_LP}.}

Figures~\ref{fig simulation regret 31} and
\ref{fig simulation regret 32} show the result when the number of fire origins equals two and four, respectively.
We can see that the IEEE 57 Bus System exhibits higher resilience to bushfires than the IEEE 11-node system, which may be due to the larger scale of the network and its meshed structure. 
Figure~\ref{fig simulation regret 31} shows that, in the early stage when the bushfire does not widely spread, the influence of bushfires on the network is not very significant, and the regrets of all the algorithms are low.
However, the performances of different algorithms can be easily distinguished as time increases. 
The advantage of our proposed algorithm is significant under a long-time horizon.

\section{Conclusions}
\label{section conclusions}

This study successfully bridges the gap between the OPF planning of power systems and the risk of bushfire spread.
The main challenges of the problem are: (i) the model parameters for bushfire spread are \textit{unknown}, and (ii) these parameters are \textit{non-stationary} over time.
We overcame the challenges by developing an online optimization algorithm that plans the power flows of the power system based on the in-situ states of the bushfire locations.
The bushfire states were understood as a ``spatial context'' in our modeling framework, which resembles the contextual information in the RL literature.
To capture the time-varying nature of bushfire spread, we developed our online algorithm based on the state-of-the-art adaptive change point detection approach designed for multi-armed bandit problems.
We provided a theoretical guarantee of our online algorithm by deriving an upper bound of the resulting regret function.
Simulation results show that our online algorithm significantly outperforms existing benchmark algorithms for OPF planning.

In the future, we would relax the sub-Gaussian assumption on the estimation bias to make the model applicable in more scenarios.
In addition, this study uses Moore's neighborhood model to capture the dynamics of bushfire spread, because of its low complexity and high flexibility. 
Nevertheless, more complex bushfire spread models may be used to derive performance guarantees for an online problem.
The model may lead to a relaxation of the piecewise constant assumption of bushfire spread and containment probabilities.
Such a challenging problem is worth investigating in the future. 
It is challenging to replace the DC OPF model with the AC OPF model since the AC OPF optimization problem in non-convex \citep{Kocuk2016str, Byeon2020comm, yang2021robust, aigner2023solving, Gholami2023an}.
Effectively solving such a problem in an online setting is also an interesting future research.


\bibliographystyle{informs2014} 
\bibliography{ref} 

\begin{thebibliography}{}

\bibitem[Abdelmalak and Benidris, 2022]{abdelmalak2022enhancing}
Abdelmalak, M. and Benidris, M. (2022).
\newblock Enhancing power system operational resilience against wildfires.
\newblock {\em IEEE Transactions on Industry Applications}, 58(2):1611--1621.

\bibitem[Abiri-Jahromi et~al., 2020]{abiri2019cyber}
Abiri-Jahromi, A., Kemmeugne, A., Kundur, D., and Haddadi, A. (2020).
\newblock Cyber-physical attacks targeting communication-assisted protection schemes.
\newblock {\em IEEE Transactions on Power Systems}, 35(1):440--450.

\bibitem[AEMO, 2020]{AEMO2020}
AEMO (2020).
\newblock Fact sheet: Explaining load shedding.

\bibitem[Agrawal and Devanur, 2016]{agrawal2016linear}
Agrawal, S. and Devanur, N. (2016).
\newblock Linear contextual bandits with knapsacks.
\newblock In {\em Advances in Neural Information Processing Systems}, pages 1--9.

\bibitem[Aigner et~al., 2023]{aigner2023solving}
Aigner, K.-M., Burlacu, R., Liers, F., and Martin, A. (2023).
\newblock Solving \protect{AC} optimal power flow with discrete decisions to global optimality.
\newblock {\em INFORMS Journal on Computing}, 35(2):458--474.

\bibitem[Alvarez et~al., 2021]{alvarez2021inventory}
Alvarez, A., Cordeau, J.-F., Jans, R., Munari, P., and Morabito, R. (2021).
\newblock Inventory routing under stochastic supply and demand.
\newblock {\em Omega}, 102:102304.

\bibitem[Angelus, 2021]{Angelus2020Dis}
Angelus, A. (2021).
\newblock Distributed renewable power generation and implications for capacity investment and electricity prices.
\newblock {\em Production and Operations Management}, 30(12):4614--4634.

\bibitem[Anokhin et~al., 2021]{Dmitry2021Mo}
Anokhin, D., Dehghanian, P., Lejeune, M.~A., and Su, J. (2021).
\newblock Mobility-as-a-service for resilience delivery in power distribution systems.
\newblock {\em Production and Operations Management}, 30(8):2492--2521.

\bibitem[Aravena et~al., 2023]{Ignacio2023Re}
Aravena, I., Molzhan, D.~K., Zhang, S., Petra, C.~G., Curtis, F.~E., Tu, S., Wachter, Andreas amd~Wei, E., Wong, E., Gholami, A., Sun, K., Sun, X.~A., Elbert, S.~T., Holzer, J.~T., and Veeramany, A. (2023).
\newblock Recent developments in security-constrained ac optimal power flow: Overview of challenge 1 in the arpa-e grid optimization competition.
\newblock {\em Operations Research}, in press.

\bibitem[Auer et~al., 2002]{auer2002nonstochastic}
Auer, P., Cesa-Bianchi, N., Freund, Y., and Schapire, R.~E. (2002).
\newblock The nonstochastic multiarmed bandit problem.
\newblock {\em SIAM journal on computing}, 32(1):48--77.

\bibitem[Auer et~al., 2019]{auer2019adaptively}
Auer, P., Gajane, P., and Ortner, R. (2019).
\newblock Adaptively tracking the best bandit arm with an unknown number of distribution changes.
\newblock In {\em Conference on Learning Theory}, pages 138--158. PMLR.

\bibitem[Awerbuch and Kleinberg, 2008]{awerbuch2008online}
Awerbuch, B. and Kleinberg, R. (2008).
\newblock Online linear optimization and adaptive routing.
\newblock {\em Journal of Computer and System Sciences}, 74(1):97--114.

\bibitem[Badawy and Sozer, 2016]{badawy2016power}
Badawy, M.~O. and Sozer, Y. (2016).
\newblock Power flow management of a grid tied \protect{PV-battery} system for electric vehicles charging.
\newblock {\em IEEE Transactions on Industry Applications}, 53(2):1347--1357.

\bibitem[Barzegar-Kalashani and Mahmud, 2022]{barzegar2022linear}
Barzegar-Kalashani, M. and Mahmud, M.~A. (2022).
\newblock A linear hybrid active disturbance rejection controller design to extenuate powerline bushfires in resonant grounded distribution power systems.
\newblock {\em International Journal of Electrical Power \& Energy Systems}, 142:108192.

\bibitem[Basso et~al., 2021]{basso2021electric}
Basso, R., Kulcs{\'a}r, B., and Sanchez-Diaz, I. (2021).
\newblock Electric vehicle routing problem with machine learning for energy prediction.
\newblock {\em Transportation Research Part B: Methodological}, 145:24--55.

\bibitem[Bertsimas and Mersereau, 2007]{bertsimas2007learning}
Bertsimas, D. and Mersereau, A.~J. (2007).
\newblock A learning approach for interactive marketing to a customer segment.
\newblock {\em Operations Research}, 55(6):1120--1135.

\bibitem[Bertsimas and Ni{\~n}o-Mora, 2000]{bertsimas2000restless}
Bertsimas, D. and Ni{\~n}o-Mora, J. (2000).
\newblock Restless bandits, linear programming relaxations, and a primal-dual index heuristic.
\newblock {\em Operations Research}, 48(1):80--90.

\bibitem[Besbes et~al., 2019]{besbes2019optimal}
Besbes, O., Gur, Y., and Zeevi, A. (2019).
\newblock Optimal exploration--exploitation in a multi-armed bandit problem with non-stationary rewards.
\newblock {\em Stochastic Systems}, 9(4):319--337.

\bibitem[Birchfield et~al., 2016]{birchfield2016statistical}
Birchfield, A.~B., Gegner, K.~M., Xu, T., Shetye, K.~S., and Overbye, T.~J. (2016).
\newblock Statistical considerations in the creation of realistic synthetic power grids for geomagnetic disturbance studies.
\newblock {\em IEEE Transactions on Power Systems}, 32(2):1502--1510.

\bibitem[Birchfield and Overbye, 2018]{birchfield2018techniques}
Birchfield, A.~B. and Overbye, T.~J. (2018).
\newblock Techniques for drawing geographic one-line diagrams: Substation spacing and line routing.
\newblock {\em IEEE Transactions on Power Systems}, 33(6):7269--7276.

\bibitem[BOM, 2020]{BOM2020}
BOM (2020).
\newblock Annual climate statement 2019.

\bibitem[Bubeck et~al., 2012]{bubeck2012regret}
Bubeck, S., Cesa-Bianchi, N., et~al. (2012).
\newblock Regret analysis of stochastic and nonstochastic multi-armed bandit problems.
\newblock {\em Foundations and Trends{\textregistered} in Machine Learning}, 5(1):1--122.

\bibitem[Byeon et~al., 2020]{Byeon2020comm}
Byeon, G., Hentenryck, P.~V., Bent, R., and Harsha, N. (2020).
\newblock Communication-constrained expansion planning for resilient distribution systems.
\newblock {\em INFORMS Journal on Computing}, 32(4):968--985.

\bibitem[Cesa-Bianchi et~al., 2017]{cesa2017algorithmic}
Cesa-Bianchi, N., Gaillard, P., Gentile, C., and Gerchinovitz, S. (2017).
\newblock Algorithmic chaining and the role of partial feedback in online nonparametric learning.
\newblock In {\em Conference on Learning Theory}, pages 465--481. PMLR.

\bibitem[Che et~al., 2018]{che2018cyber}
Che, L., Liu, X., Shuai, Z., Li, Z., and Wen, Y. (2018).
\newblock Cyber cascades screening considering the impacts of false data injection attacks.
\newblock {\em IEEE Transactions on Power Systems}, 33(6):6545--6556.

\bibitem[Chen et~al., 2019]{chen2019distributed}
Chen, W., Ding, D., Dong, H., and Wei, G. (2019).
\newblock Distributed resilient filtering for power systems subject to denial-of-service attacks.
\newblock {\em IEEE Transactions on Systems, Man, and Cybernetics: Systems}, 49(8):1688--1697.

\bibitem[Cheng and Overbye, 2005]{cheng2005ptdf}
Cheng, X. and Overbye, T.~J. (2005).
\newblock \protect{PTDF}-based power system equivalents.
\newblock {\em IEEE Transactions on Power Systems}, 20(4):1868--1876.

\bibitem[Coelho et~al., 2014]{coelho2014thirty}
Coelho, L.~C., Cordeau, J.-F., and Laporte, G. (2014).
\newblock Thirty years of inventory routing.
\newblock {\em Transportation Science}, 48(1):1--19.

\bibitem[Collins et~al., 2018]{collins2018suppression}
Collins, K.~M., Price, O.~F., and Penman, T.~D. (2018).
\newblock Suppression resource decisions are the dominant influence on containment of australian forest and grass fires.
\newblock {\em Journal of environmental management}, 228:373--382.

\bibitem[CSIRO, 2020]{csiro20202019}
CSIRO (2020).
\newblock The 2019--20 bushfires: A \protect{CSIRO} explainer.
\newblock \url{https://www.csiro.au/en/research/natural-disasters/bushfires/2019-20-bushfires-explainer}.

\bibitem[Ding et~al., 2019]{ding2019distributed}
Ding, L., Han, Q.-L., Ning, B., and Yue, D. (2019).
\newblock Distributed resilient finite-time secondary control for heterogeneous battery energy storage systems under denial-of-service attacks.
\newblock {\em IEEE Transactions on Industrial Informatics}, 16(7):4909--4919.

\bibitem[Dong et~al., 2016]{Dong2016Ele}
Dong, C., Ng, C.~T., and Cheng, T. (2016).
\newblock Electricity time‐of‐use tariff with stochastic demand.
\newblock {\em Production and Operations Management}, 26(1):64--79.

\bibitem[Dorrer and Sergey, 2018]{Dorrer2018use}
Dorrer, G. and Sergey, Y. (2018).
\newblock Use of agent-based modeling for wildfire situations simulation.
\newblock {\em 2018 3rd Russian-Pacific Conference on Computer Technology and Applications (RPC)}, pages 1--4.

\bibitem[Drayer and Routtenberg, 2019]{drayer2019detection}
Drayer, E. and Routtenberg, T. (2019).
\newblock Detection of false data injection attacks in smart grids based on graph signal processing.
\newblock {\em IEEE Systems Journal}, 14(2):1886--1896.

\bibitem[Emily, 2023]{Emily2023BBC}
Emily, M. (2023).
\newblock Maui county sues hawaiian electric over wildfire negligence.

\bibitem[Even-Dar et~al., 2009]{even2009online}
Even-Dar, E., Kakade, S.~M., and Mansour, Y. (2009).
\newblock Online markov decision processes.
\newblock {\em Mathematics of Operations Research}, 34(3):726--736.

\bibitem[Fan and Ruszczy{\'n}ski, 2018]{fan2018risk}
Fan, J. and Ruszczy{\'n}ski, A. (2018).
\newblock Risk measurement and risk-averse control of partially observable discrete-time markov systems.
\newblock {\em Mathematical Methods of Operations Research}, pages 1--24.

\bibitem[Foroutan and Salmasi, 2017]{foroutan2017detection}
Foroutan, S.~A. and Salmasi, F.~R. (2017).
\newblock Detection of false data injection attacks against state estimation in smart grids based on a mixture gaussian distribution learning method.
\newblock {\em IET Cyber-Physical Systems: Theory \& Applications}, 2(4):161--171.

\bibitem[Freund and Schapire, 1999]{freund1999adaptive}
Freund, Y. and Schapire, R.~E. (1999).
\newblock Adaptive game playing using multiplicative weights.
\newblock {\em Games and Economic Behavior}, 29(1-2):79--103.

\bibitem[Fryzlewicz, 2014]{fryzlewicz2014wild}
Fryzlewicz, P. (2014).
\newblock Wild binary segmentation for multiple change-point detection.
\newblock {\em The Annals of Statistics}, 42(6):2243--2281.

\bibitem[{Gallo} et~al., 2020]{gallo2020dis}
{Gallo}, A.~J., {Turan}, M.~S., {Boem}, F., {Parisini}, T., and {Ferrari-Trecate}, G. (2020).
\newblock A distributed cyber-attack detection scheme with application to dc microgrids.
\newblock {\em IEEE Transactions on Automatic Control}.

\bibitem[Garivier and Moulines, 2011]{garivier2011upper}
Garivier, A. and Moulines, E. (2011).
\newblock On upper-confidence bound policies for switching bandit problems.
\newblock In {\em International Conference on Algorithmic Learning Theory}, pages 174--188. Springer.

\bibitem[Gas and Company, 2021]{Pacific2021wild}
Gas, P. and Company, E. (2021).
\newblock Wildfire mitigation plan - revised.

\bibitem[Gholami et~al., 2023]{Gholami2023an}
Gholami, A., Sun, K., Zhang, S., and Sun, X.~A. (2023).
\newblock An alternating direction method of multipliers-based distributed optimization method for solving security-constrained alternating current optimal power flow.
\newblock {\em Operations Research}, in press.

\bibitem[Goeke and Schneider, 2015]{goeke2015routing}
Goeke, D. and Schneider, M. (2015).
\newblock Routing a mixed fleet of electric and conventional vehicles.
\newblock {\em European Journal of Operational Research}, 245(1):81--99.

\bibitem[Golari et~al., 2016]{Mehdi2016Mul}
Golari, M., Fan, N., and Jin, T. (2016).
\newblock Multistage stochastic optimization for production‐inventory planning with intermittent renewable energy.
\newblock {\em Production and Operations Management}, 26(3):409--425.

\bibitem[Gungor et~al., 2011]{gungor2011smart}
Gungor, V.~C., Sahin, D., Kocak, T., Ergut, S., Buccella, C., Cecati, C., and Hancke, G.~P. (2011).
\newblock Smart grid technologies: Communication technologies and standards.
\newblock {\em IEEE Transactions on Industrial Informatics}, 7(4):529--539.

\bibitem[Guo et~al., 2018]{guo2018determination}
Guo, Y., Chen, R., Shi, J., Wan, J., Yi, H., and Zhong, J. (2018).
\newblock Determination of the power transmission line ageing failure probability due to the impact of forest fire.
\newblock {\em IET Generation, Transmission \& Distribution}, 12(16):3812--3819.

\bibitem[Gupta et~al., 2016]{Sushil2016Dis}
Gupta, S., Starr, M.~K., Farahani, R.~Z., and Matinrad, N. (2016).
\newblock Disaster management from a pom perspective: Mapping a new domain.
\newblock {\em Production and Operations Management}, 25(10):1611--1637.

\bibitem[Gyorgy and Szepesv{\'a}ri, 2016]{gyorgy2016shifting}
Gyorgy, A. and Szepesv{\'a}ri, C. (2016).
\newblock Shifting regret, mirror descent, and matrices.
\newblock In {\em International Conference on Machine Learning}, pages 2943--2951. PMLR.

\bibitem[Hahn et~al., 2013]{hahn2013cyber}
Hahn, A., Ashok, A., Sridhar, S., and Govindarasu, M. (2013).
\newblock Cyber-physical security testbeds: Architecture, application, and evaluation for smart grid.
\newblock {\em IEEE Transactions on Smart Grid}, 4(2):847--855.

\bibitem[Han et~al., 2022]{han2022multi}
Han, S., Mo, Y., Chen, L., Luo, Z., Foropon, C.~R., and Belal, H. (2022).
\newblock A multi-period closed-loop supply chain network design with circular route planning.
\newblock {\em Annals of Operations Research}, pages 1--39.

\bibitem[Hazan, 2019]{hazan2019introduction}
Hazan, E. (2019).
\newblock Introduction to online convex optimization.
\newblock {\em arXiv preprint arXiv:1909.05207}.

\bibitem[Hazan et~al., 2016]{hazan2016introduction}
Hazan, E. et~al. (2016).
\newblock Introduction to online convex optimization.
\newblock {\em Foundations and Trends{\textregistered} in Optimization}, 2(3-4):157--325.

\bibitem[He et~al., pear]{he2021secure}
He, X., Ren, X., Sandberg, H., and Johansson, K.~H. (to appear).
\newblock How to secure distributed filters under sensor attacks.
\newblock {\em IEEE Transactions on Automatic Control}.

\bibitem[Heyman and Sobel, 2004]{heyman2004superposition}
Heyman, D. and Sobel, M. (2004).
\newblock Superposition of renewal processes.
\newblock {\em Stochastic Models in Operations Research: Stochastic Processes and Operating Characteristics}, page 158.

\bibitem[Higgins et~al., 2020]{higgins2020stealthy}
Higgins, M., Teng, F., and Parisini, T. (2020).
\newblock Stealthy \protect{MTD} against unsupervised learning-based blind \protect{FDI} attacks in power systems.
\newblock {\em IEEE Transactions on Information Forensics and Security}, 16:1275--1287.

\bibitem[Hobbs and Pang, 2007]{hobbs2007nash}
Hobbs, B.~F. and Pang, J.-S. (2007).
\newblock \protect{Nash-Cournot} equilibria in electric power markets with piecewise linear demand functions and joint constraints.
\newblock {\em Operations Research}, 55(1):113--127.

\bibitem[Hong et~al., 2022]{hong2022data}
Hong, W., Wang, B., Yao, M., Callaway, D., Dale, L., and Huang, C. (2022).
\newblock Data-driven power system optimal decision making strategy underwildfire events.
\newblock Technical report, Lawrence Livermore National Lab (LLNL), Livermore, CA (United States).

\bibitem[Hu et~al., 2020]{hu2020detecting}
Hu, Y., Li, H., Luan, T.~H., Yang, A., Sun, L., Wang, Z., and Wang, R. (2020).
\newblock Detecting stealthy attacks on industrial control systems using a permutation entropy-based method.
\newblock {\em Future Generation Computer Systems}, 108:1230--1240.

\bibitem[Huadong and Giovanni, 2019]{mo2019impact}
Huadong, M. and Giovanni, S. (2019).
\newblock Impact of aging and performance degradation on the operational costs of distributed generation systems.
\newblock {\em Renewable energy}, 143:426--439.

\bibitem[Huh and Rusmevichientong, 2014]{huh2014online}
Huh, W.~T. and Rusmevichientong, P. (2014).
\newblock Online sequential optimization with biased gradients: theory and applications to censored demand.
\newblock {\em INFORMS Journal on Computing}, 26(1):150--159.

\bibitem[Hussain et~al., 2019]{hussain2019effort}
Hussain, A., Bui, V.-H., and Kim, H.-M. (2019).
\newblock An effort-based reward approach for allocating load shedding amount in networked microgrids using multiagent system.
\newblock {\em IEEE Transactions on Industrial Informatics}, 16(4):2268--2279.

\bibitem[Ito et~al., 2020]{ito2020evaluation}
Ito, A., Perron, M.~M., Proemse, B.~C., Strzelec, M., Gault-Ringold, M., Boyd, P.~W., and Bowie, A.~R. (2020).
\newblock Evaluation of aerosol iron solubility over \protect{Australian} coastal regions based on inverse modeling: implications of bushfires on bioaccessible iron concentrations in the \protect{Southern Hemisphere}.
\newblock {\em Progress in Earth and Planetary Science}, 7(1):1--17.

\bibitem[Jaillet and Wagner, 2008]{jaillet2008generalized}
Jaillet, P. and Wagner, M.~R. (2008).
\newblock Generalized online routing: New competitive ratios, resource augmentation, and asymptotic analyses.
\newblock {\em Operations Research}, 56(3):745--757.

\bibitem[James et~al., 2018]{yu2018online}
James, J., Hou, Y., and Li, V.~O. (2018).
\newblock Online false data injection attack detection with wavelet transform and deep neural networks.
\newblock {\em IEEE Transactions on Industrial Informatics}, 14(7):3271--3280.

\bibitem[Jiang et~al., 2021]{jiang2021modelling}
Jiang, W., Wang, F., Fang, L., Zheng, X., Qiao, X., Li, Z., and Meng, Q. (2021).
\newblock Modelling of wildland-urban interface fire spread with the heterogeneous cellular automata model.
\newblock {\em Environmental Modelling \& Software}, 135:104895.

\bibitem[Jin et~al., 2018]{jin2018power}
Jin, M., Lavaei, J., and Johansson, K.~H. (2018).
\newblock Power grid \protect{AC}-based state estimation: Vulnerability analysis against cyber attacks.
\newblock {\em IEEE Transactions on Automatic Control}, 64(5):1784--1799.

\bibitem[Johnson and Dey, 2022]{johnson2022scalable}
Johnson, E.~S. and Dey, S.~S. (2022).
\newblock A scalable lower bound for the worst-case relay attack problem on the transmission grid.
\newblock {\em INFORMS Journal on Computing}, 34(4):2296--2312.

\bibitem[Kadir et~al., 2023]{Kadir2023re}
Kadir, S.~U., Majumder, S., Srivastava, A., Chhokra, A., Abhishek~Dubey, H.~N., and Laszka, A. (2023).
\newblock Reinforcement learning based proactive control for enabling power grid resilience to wildfire.
\newblock {\em IEEE Transactions on Industrial Informatics}, in press.

\bibitem[Karafyllidis and Thanailakis, 1997]{karafyllidis1997model}
Karafyllidis, I. and Thanailakis, A. (1997).
\newblock A model for predicting forest fire spreading using cellular automata.
\newblock {\em Ecological Modelling}, 99(1):87--97.

\bibitem[{Khalili} et~al., 2020]{khalili2020dis}
{Khalili}, M., {Zhang}, X., {Cao}, Y., {Polycarpou}, M.~M., and {Parisini}, T. (2020).
\newblock Distributed fault-tolerant control of multiagent systems: An adaptive learning approach.
\newblock {\em IEEE Transactions on Neural Networks and Learning Systems}, 31(2):420--432.

\bibitem[Kim and Lim, 2015]{kim2015robust}
Kim, M.~J. and Lim, A.~E. (2015).
\newblock Robust multiarmed bandit problems.
\newblock {\em Management Science}, 62(1):264--285.

\bibitem[Kim et~al., 2008]{kim2008real}
Kim, Y., Gu, D.-W., and Postlethwaite, I. (2008).
\newblock Real-time path planning with limited information for autonomous unmanned air vehicles.
\newblock {\em Automatica}, 44(3):696--712.

\bibitem[Klima et~al., 2016]{klima2016markov}
Klima, R., Tuyls, K., and Oliehoek, F. (2016).
\newblock Markov security games: Learning in spatial security problems.
\newblock In {\em NIPS Workshop on Learning, Inference and Control of Multi-Agent Systems}, pages 1--8.

\bibitem[Kocuk et~al., 2016a]{Kocuk2016str}
Kocuk, B., Dey, S.~S., and Sun, X.~A. (2016a).
\newblock Strong \protect{SOCP} relaxations for the optimal power flow problem.
\newblock {\em Operations Research}, 64(6):1177--1196.

\bibitem[Kocuk et~al., 2016b]{Burak2016A}
Kocuk, B., Jeon, H., Dey, S.~S., Linderoth, J., Luedtke, J., and Sun, A.~X. (2016b).
\newblock A cycle-based formulation and valid inequalities for dc power transmission problems with switching.
\newblock {\em Operations Research}, 64(4):922--938.

\bibitem[Konstantelos et~al., 2016]{konstantelos2016strategic}
Konstantelos, I., Giannelos, S., and Strbac, G. (2016).
\newblock Strategic valuation of smart grid technology options in distribution networks.
\newblock {\em IEEE Transactions on Power Systems}, 32(2):1293--1303.

\bibitem[Li et~al., 2020]{li2020quantum}
Li, J.-A., Dong, D., Wei, Z., Liu, Y., Pan, Y., Nori, F., and Zhang, X. (2020).
\newblock Quantum reinforcement learning during human decision-making.
\newblock {\em Nature Human Behaviour}, 4(3):294--307.

\bibitem[Li et~al., 2018]{li2018false}
Li, Y., Shi, D., and Chen, T. (2018).
\newblock False data injection attacks on networked control systems: A stackelberg game analysis.
\newblock {\em IEEE Transactions on Automatic Control}, 63(10):3503--3509.

\bibitem[Li and Jia, 2021]{li2021overview}
Li, Y.-F. and Jia, C. (2021).
\newblock An overview of the reliability metrics for power grids and telecommunication networks.
\newblock {\em Frontiers of Engineering Management}, pages 1--14.

\bibitem[Liang et~al., 2016]{liang2016review}
Liang, G., Zhao, J., Luo, F., Weller, S.~R., and Dong, Z.~Y. (2016).
\newblock A review of false data injection attacks against modern power systems.
\newblock {\em IEEE Transactions on Smart Grid}, 8(4):1630--1638.

\bibitem[Littman, 2015]{littman2015reinforcement}
Littman, M.~L. (2015).
\newblock Reinforcement learning improves behaviour from evaluative feedback.
\newblock {\em Nature}, 521(7553):445--451.

\bibitem[Liu, 2019]{liu2019sinr}
Liu, H. (2019).
\newblock \protect{SINR-based} multi-channel power schedule under dos attacks: A stackelberg game approach with incomplete information.
\newblock {\em Automatica}, 100:274--280.

\bibitem[Liu et~al., 2010]{liu2010learning}
Liu, H., Liu, K., and Zhao, Q. (2010).
\newblock Learning in a changing world: Non-bayesian restless multi-armed bandit.
\newblock Technical report, CALIFORNIA UNIV DAVIS DEPT OF ELECTRICAL AND COMPUTER ENGINEERING.

\bibitem[Liu et~al., 2017]{liu2017distributed}
Liu, Z., Wang, L., Wang, J., Dong, D., and Hu, X. (2017).
\newblock Distributed sampled-data control of nonholonomic multi-robot systems with proximity networks.
\newblock {\em Automatica}, 77:170--179.

\bibitem[Loch and Kavadias, 2002]{loch2002dynamic}
Loch, C.~H. and Kavadias, S. (2002).
\newblock Dynamic portfolio selection of npd programs using marginal returns.
\newblock {\em Management Science}, 48(10):1227--1241.

\bibitem[Lu and Wang, 2021]{lu2021online}
Lu, K. and Wang, L. (2021).
\newblock Online distributed optimization with nonconvex objective functions: Sublinearity of first-order optimality condition-based regret.
\newblock {\em IEEE Transactions on Automatic Control}, 67(6):3029--3035.

\bibitem[Lu et~al., 2020]{lu2020real}
Lu, X., Yin, H., Xia, S., Zhang, D., Shahidehpour, M., Zhang, X., and Ding, T. (2020).
\newblock A real-time alternating direction method of multipliers algorithm for nonconvex optimal power flow problem.
\newblock {\em IEEE Transactions on Industry Applications}, 57(1):70--82.

\bibitem[McCarthy et~al., 2012]{mccarthy2012analysis}
McCarthy, G., Plucinski, M., and Gould, J. (2012).
\newblock Analysis of the resourcing and containment of multiple remote fires: the great divide complex of fires, \protect{Victoria, December 2006}.
\newblock {\em Australian Forestry}, 75(1):54--63.

\bibitem[Mena et~al., 2014]{mena2014risk}
Mena, R., Hennebel, M., Li, Y.-F., Ruiz, C., and Zio, E. (2014).
\newblock A risk-based simulation and multi-objective optimization framework for the integration of distributed renewable generation and storage.
\newblock {\em Renewable and Sustainable Energy Reviews}, 37:778--793.

\bibitem[Mersereau et~al., 2009]{mersereau2009structured}
Mersereau, A.~J., Rusmevichientong, P., and Tsitsiklis, J.~N. (2009).
\newblock A structured multiarmed bandit problem and the greedy policy.
\newblock {\em IEEE Transactions on Automatic Control}, 54(12):2787--2802.

\bibitem[Metz and Saraiva, 2018]{metz2018use}
Metz, D. and Saraiva, J.~T. (2018).
\newblock Use of battery storage systems for price arbitrage operations in the 15- and 60-min \protect{German} intraday markets.
\newblock {\em Electric Power Systems Research}, 160:27--36.

\bibitem[Mhanna and Mancarella, 2021]{mhanna2021exact}
Mhanna, S. and Mancarella, P. (2021).
\newblock An exact sequential linear programming algorithm for the optimal power flow problem.
\newblock {\em IEEE Transactions on Power Systems}, 37(1):666--679.

\bibitem[Misra et~al., 2022]{misra2022learning}
Misra, S., Roald, L., and Ng, Y. (2022).
\newblock Learning for constrained optimization: Identifying optimal active constraint sets.
\newblock {\em INFORMS Journal on Computing}, 34(1):463--480.

\bibitem[Mo and Sansavini, 2017]{mo2017dynamic}
Mo, H. and Sansavini, G. (2017).
\newblock Dynamic defense resource allocation for minimizing unsupplied demand in cyber-physical systems against uncertain attacks.
\newblock {\em IEEE Transactions on Reliability}, 66(4):1253--1265.

\bibitem[Moreno et~al., 2022]{moreno2022microgrids}
Moreno, R., Trakas, D.~N., Jamieson, M., Panteli, M., Mancarella, P., Strbac, G., Marnay, C., and Hatziargyriou, N. (2022).
\newblock Microgrids against wildfires: distributed energy resources enhance system resilience.
\newblock {\em IEEE Power and Energy Magazine}, 20(1):78--89.

\bibitem[Moutis and Sriram, 2022]{Moutis2022pmu}
Moutis, P. and Sriram, U. (2022).
\newblock \protect{PMU}-driven non-preemptive disconnection of overhead lines at the approach or break-out of forest fires.
\newblock {\em IEEE Transactions on Power Systems}, 38(1):168--176.

\bibitem[Munawar et~al., 2021]{munawar2021uav}
Munawar, H.~S., Ullah, F., Khan, S.~I., Qadir, Z., and Qayyum, S. (2021).
\newblock \protect{UAV} assisted spatiotemporal analysis and management of bushfires: A case study of the 2020 \protect{Victorian} bushfires.
\newblock {\em Fire}, 4(3):40.

\bibitem[Musleh et~al., 2019]{musleh2019survey}
Musleh, A.~S., Chen, G., and Dong, Z.~Y. (2019).
\newblock A survey on the detection algorithms for false data injection attacks in smart grids.
\newblock {\em IEEE Transactions on Smart Grid}, 11(3):2218--2234.

\bibitem[Neu, 2015]{neu2015explore}
Neu, G. (2015).
\newblock Explore no more: Improved high-probability regret bounds for non-stochastic bandits.
\newblock In {\em Advances on Neural Information Processing Systems 28 (NIPS 2015)}, pages 3150--3158.

\bibitem[Niu et~al., 2016]{niu2016multiple}
Niu, Y.~S., Hao, N., and Zhang, H. (2016).
\newblock Multiple change-point detection: a selective overview.
\newblock {\em Statistical Science}, pages 611--623.

\bibitem[NSW, 2023]{NSWcurrent}
NSW (2023).
\newblock Native vegetation.

\bibitem[Olivares et~al., 2014]{olivares2014centralized}
Olivares, D.~E., Ca{\~n}izares, C.~A., and Kazerani, M. (2014).
\newblock A centralized energy management system for isolated microgrids.
\newblock {\em IEEE Transactions on Smart Grid}, 5(4):1864--1875.

\bibitem[Osband et~al., 2013]{osband2013more}
Osband, I., Russo, D., and Van~Roy, B. (2013).
\newblock (more) efficient reinforcement learning via posterior sampling.
\newblock In {\em Advances in Neural Information Processing Systems}, pages 3003--3011.

\bibitem[Pais et~al., 2021]{pais2021cell2fire}
Pais, C., Carrasco, J., Martell, D., Weintraub, A., and Woodruff, D. (2021).
\newblock Cell2fire: A cell-based forest fire growth model to support strategic landscape management planning. front. for. glob.
\newblock {\em Change}, 4:692706.

\bibitem[Pan et~al., 2021]{Xiang2021Deep}
Pan, X., Zhao, T., Chen, M., and Chen, M. (2021).
\newblock Deepopf: A deep neural network approach for security-constrained dc optimal power flow.
\newblock {\em IEEE Transactions on Power Systems}, 36(3):1725--1735.

\bibitem[Pandey and Gupta, 2017]{pandey2017intelligent}
Pandey, R.~K. and Gupta, D.~K. (2017).
\newblock Intelligent multi-area power control: dynamic knowledge domain inference concept.
\newblock {\em IEEE Transactions on Power Systems}, 32(6):4310--4318.

\bibitem[Papadimitriou and Tsitsiklis, 1999]{papadimitriou1999complexity}
Papadimitriou, C.~H. and Tsitsiklis, J.~N. (1999).
\newblock The complexity of optimal queuing network control.
\newblock {\em Mathematics of Operations Research}, 24(2):293--305.

\bibitem[Papier, 2016]{Papier2016Man}
Papier, F. (2016).
\newblock Managing electricity peak loads in make‐to‐stock manufacturing lines.
\newblock {\em Production and Operations Management}, 25(10):1709--1726.

\bibitem[Parker et~al., 2019]{Parker2019Ele}
Parker, G.~G., Tan, B., and Kazan, O. (2019).
\newblock Electric power industry: Operational and public policy challenges and opportunities.
\newblock {\em Production and Operations Management}, 28(11):2728--2777.

\bibitem[Patsakis et~al., 2018]{patsakis2018optimal}
Patsakis, G., Rajan, D., Aravena, I., Rios, J., and Oren, S. (2018).
\newblock Optimal black start allocation for power system restoration.
\newblock {\em IEEE Transactions on Power Systems}, 33(6):6766--6776.

\bibitem[Peace et~al., 2015]{peace2015fire}
Peace, M., Mattner, T., Mills, G., Kepert, J., and McCaw, L. (2015).
\newblock Fire-modified meteorology in a coupled fire--atmosphere model.
\newblock {\em Journal of Applied Meteorology and Climatology}, 54(3):704--720.

\bibitem[Pena-Ordieres et~al., 2020]{pena2020dc}
Pena-Ordieres, A., Molzahn, D.~K., Roald, L.~A., and W{\"a}chter, A. (2020).
\newblock \protect{DC} optimal power flow with joint chance constraints.
\newblock {\em IEEE Transactions on Power Systems}, 36(1):147--158.

\bibitem[Peng and Sun, 2020]{peng2020switching}
Peng, C. and Sun, H. (2020).
\newblock Switching-like event-triggered control for networked control systems under malicious denial of service attacks.
\newblock {\em IEEE Transactions on Automatic Control}, 65(9):3943--3949.

\bibitem[Peng et~al., 2019]{peng2019survey}
Peng, C., Sun, H., Yang, M., and Wang, Y.-L. (2019).
\newblock A survey on security communication and control for smart grids under malicious cyber attacks.
\newblock {\em IEEE Transactions on Systems, Man, and Cybernetics: Systems}, 49(8):1554--1569.

\bibitem[Perera et~al., 2015]{perera2015emerging}
Perera, C., Liu, C.~H., and Jayawardena, S. (2015).
\newblock The emerging internet of things marketplace from an industrial perspective: A survey.
\newblock {\em IEEE Transactions on Emerging Topics in Computing}, 3(4):585--598.

\bibitem[Plucinski et~al., 2011]{plucinski2011effect}
Plucinski, M., McCarthy, G., Hollis, J., and Gould, J. (2011).
\newblock The effect of aerial suppression on the containment time of \protect{Australian} wildfires estimated by fire management personnel.
\newblock {\em International Journal of Wildland Fire}, 21(3):219--229.

\bibitem[Pogaku et~al., 2007]{pogaku2007modeling}
Pogaku, N., Prodanovic, M., and Green, T.~C. (2007).
\newblock Modeling, analysis and testing of autonomous operation of an inverter-based microgrid.
\newblock {\em IEEE Transactions on power electronics}, 22(2):613--625.

\bibitem[Rana et~al., 2017]{rana2017cyber}
Rana, M.~M., Li, L., and Su, S.~W. (2017).
\newblock Cyber attack protection and control of microgrids.
\newblock {\em IEEE/CAA Journal of Automatica Sinica}, 5(2):602--609.

\bibitem[Rana et~al., 2018]{rana2018smart}
Rana, M.~M., Xiang, W., and Wang, E. (2018).
\newblock Smart grid state estimation and stabilisation.
\newblock {\em International Journal of Electrical Power \& Energy Systems}, 102:152--159.

\bibitem[Rhodes et~al., 2020]{rhodes2020balancing}
Rhodes, N., Ntaimo, L., and Roald, L. (2020).
\newblock Balancing wildfire risk and power outages through optimized power shut-offs.
\newblock {\em IEEE Transactions on Power Systems}, 36(4):3118--3128.

\bibitem[Russo and Van~Roy, 2014]{russo2014learning}
Russo, D. and Van~Roy, B. (2014).
\newblock Learning to optimize via posterior sampling.
\newblock {\em Mathematics of Operations Research}, 39(4):1221--1243.

\bibitem[Ruszczy{\'n}ski, 2010]{ruszczynski2010risk}
Ruszczy{\'n}ski, A. (2010).
\newblock Risk-averse dynamic programming for markov decision processes.
\newblock {\em Mathematical programming}, 125(2):235--261.

\bibitem[Sahraei-Ardakani and Hedman, 2016]{sahraei2016computationally}
Sahraei-Ardakani, M. and Hedman, K.~W. (2016).
\newblock Computationally efficient adjustment of facts set points in dc optimal power flow with shift factor structure.
\newblock {\em IEEE Transactions on Power Systems}, 32(3):1733--1740.

\bibitem[Salehi and Rashidi, 2018]{salehi2018survey}
Salehi, M. and Rashidi, L. (2018).
\newblock A survey on anomaly detection in evolving data: with application to forest fire risk prediction.
\newblock {\em ACM SIGKDD Explorations Newsletter}, 20(1):13--23.

\bibitem[Salehi et~al., 2016]{salehi2016dynamic}
Salehi, M., Rusu, L.~I., Lynar, T., and Phan, A. (2016).
\newblock Dynamic and robust wildfire risk prediction system: an unsupervised approach.
\newblock In {\em Proceedings of the 22nd ACM SIGKDD International Conference on Knowledge Discovery and Data Mining}, pages 245--254.

\bibitem[Shamir, 2011]{shamir2011variant}
Shamir, O. (2011).
\newblock A variant of \protect{Azuma's} inequality for martingales with subgaussian tails.
\newblock {\em arXiv preprint arXiv:1110.2392}.

\bibitem[Shao, 2003]{shao2003mathematical}
Shao, J. (2003).
\newblock {\em Mathematical Statistics}.
\newblock Springer Science \& Business Media.

\bibitem[Shapiro et~al., 2021]{shapiro2021lectures}
Shapiro, A., Dentcheva, D., and Ruszczynski, A. (2021).
\newblock {\em Lectures on Stochastic Programming: Modeling and Theory}.
\newblock SIAM.

\bibitem[Shapiro and Pichler, 2016]{shapiro2016time}
Shapiro, A. and Pichler, A. (2016).
\newblock Time and dynamic consistency of risk averse stochastic programs.

\bibitem[Shen et~al., 2015]{shen2015portfolio}
Shen, W., Wang, J., Jiang, Y.-G., and Zha, H. (2015).
\newblock Portfolio choices with orthogonal bandit learning.
\newblock In {\em IJCAI}, volume~15, pages 974--980.

\bibitem[Shi et~al., 2022]{shi2022enhancing}
Shi, Q., Liu, W., Zeng, B., Hui, H., and Li, F. (2022).
\newblock Enhancing distribution system resilience against extreme weather events: concept review, algorithm summary, and future vision.
\newblock {\em International Journal of Electrical Power \& Energy Systems}, 138:107860.

\bibitem[Simchi-Levi and Xu, 2022]{simchi2022bypassing}
Simchi-Levi, D. and Xu, Y. (2022).
\newblock Bypassing the monster: A faster and simpler optimal algorithm for contextual bandits under realizability.
\newblock {\em Mathematics of Operations Research}, 47(3):1904--1931.

\bibitem[Slivkins et~al., 2019]{slivkins2019introduction}
Slivkins, A. et~al. (2019).
\newblock Introduction to multi-armed bandits.
\newblock {\em Foundations and Trends{\textregistered} in Machine Learning}, 12(1-2):1--286.

\bibitem[Smith and Pat{\'e}-Cornell, 2018]{smith2018cyber}
Smith, M.~D. and Pat{\'e}-Cornell, M.~E. (2018).
\newblock Cyber risk analysis for a smart grid: how smart is smart enough? a multiarmed bandit approach to cyber security investment.
\newblock {\em IEEE Transactions on Engineering Management}, 65(3):434--447.

\bibitem[Sossan et~al., 2018]{sossan2018unsupervised}
Sossan, F., Nespoli, L., Medici, V., and Paolone, M. (2018).
\newblock Unsupervised disaggregation of photovoltaic production from composite power flow measurements of heterogeneous prosumers.
\newblock {\em IEEE Transactions on Industrial Informatics}, 14(9):3904--3913.

\bibitem[Srivas et~al., 2016]{srivas2016wildfire}
Srivas, T., Art{\'e}s, T., De~Callafon, R.~A., and Altintas, I. (2016).
\newblock Wildfire spread prediction and assimilation for farsite using ensemble kalman filtering.
\newblock {\em Procedia Computer Science}, 80:897--908.

\bibitem[Stauffer and Kumar, 2021]{Stauffer2020Im}
Stauffer, J.~M. and Kumar, S. (2021).
\newblock Impact of incorportating returns into pre-disaster deployments for rapid-onset predictable disasters.
\newblock {\em Production and Operations Management}, 30(2):451--474.

\bibitem[Su et~al., 2023]{Su2023qua}
Su, J., Mehrani, S., Dehghanian, P., and Lejeune, M.~A. (2023).
\newblock Quasi second-order stochastic dominance model for balancing wildfire risks and power outages due to proactive public safety de-energizations.
\newblock {\em IEEE Transactions on Power Systems}, in press.

\bibitem[Sutton and Barto, 2018]{sutton2018reinforcement}
Sutton, R.~S. and Barto, A.~G. (2018).
\newblock {\em Reinforcement learning: An introduction}.
\newblock MIT press.

\bibitem[Talebi et~al., 2017]{talebi2017stochastic}
Talebi, M.~S., Zou, Z., Combes, R., Proutiere, A., and Johansson, M. (2017).
\newblock Stochastic online shortest path routing: The value of feedback.
\newblock {\em IEEE Transactions on Automatic Control}, 63(4):915--930.

\bibitem[Tamimi and Vaez-Zadeh, 2008]{tamimi2008optimal}
Tamimi, B. and Vaez-Zadeh, S. (2008).
\newblock An optimal pricing scheme in electricity markets considering voltage security cost.
\newblock {\em IEEE Transactions on Power Systems}, 23(2):451--459.

\bibitem[Todescato et~al., 2020]{todescato2020partition}
Todescato, M., Bof, N., Cavraro, G., Carli, R., and Schenato, L. (2020).
\newblock Partition-based multi-agent optimization in the presence of lossy and asynchronous communication.
\newblock {\em Automatica}, 111:108648.

\bibitem[Trunfio et~al., 2011]{trunfio2011new}
Trunfio, G.~A., D’Ambrosio, D., Rongo, R., Spataro, W., and Di~Gregorio, S. (2011).
\newblock A new algorithm for simulating wildfire spread through cellular automata.
\newblock {\em ACM Transactions on Modeling and Computer Simulation}, 22(1):1--26.

\bibitem[Vanajakumari et~al., 2016]{Manoj2016An}
Vanajakumari, M., Kumar, S., and Gupta, S. (2016).
\newblock An integrated logistic model for predictable disasters.
\newblock {\em Production and Operations Management}, 25(5):791--811.

\bibitem[Wang et~al., 2018]{wang2018risk}
Wang, C., Gao, R., Qiu, F., Wang, J., and Xin, L. (2018).
\newblock Risk-based distributionally robust optimal power flow with dynamic line rating.
\newblock {\em IEEE Transactions on Power Systems}, 33(6):6074--6086.

\bibitem[Wang, 2022]{wang2022bushfire}
Wang, C.-H. (2022).
\newblock Bushfire and climate change risks to electricity transmission networks.
\newblock In {\em Engineering for Extremes}, pages 413--427. Springer.

\bibitem[Wang et~al., 2016a]{wang2016measure}
Wang, J.-F., Zhang, T.-L., and Fu, B.-J. (2016a).
\newblock A measure of spatial stratified heterogeneity.
\newblock {\em Ecological Indicators}, 67:250--256.

\bibitem[Wang et~al., 2020]{wang2020detection}
Wang, X., Luo, X., Zhang, M., Jiang, Z., and Guan, X. (2020).
\newblock Detection and isolation of false data injection attacks in smart grid via unknown input interval observer.
\newblock {\em IEEE Internet of Things Journal}, 7(4):3214--3229.

\bibitem[Wang et~al., 2016b]{wang2016fully}
Wang, Y., Wu, L., and Wang, S. (2016b).
\newblock A fully-decentralized consensus-based admm approach for dc-opf with demand response.
\newblock {\em IEEE Transactions on Smart Grid}, 8(6):2637--2647.

\bibitem[Wei et~al., 2022]{wei2022physics}
Wei, G., Krishnan, V., Xie, Y., Sengupta, M., Zhang, Y., Liao, H., and Liu, X. (2022).
\newblock Physics-informed statistical modeling for wildfire aerosols process using multi-source geostationary satellite remote-sensing data streams.
\newblock {\em arXiv preprint arXiv:2206.11766}.

\bibitem[Whittle, 1988]{whittle1988restless}
Whittle, P. (1988).
\newblock Restless bandits: Activity allocation in a changing world.
\newblock {\em Journal of applied probability}, 25(A):287--298.

\bibitem[Wootton, 2001]{wootton2001local}
Wootton, J.~T. (2001).
\newblock Local interactions predict large-scale pattern in empirically derived cellular automata.
\newblock {\em Nature}, 413(6858):841--844.

\bibitem[Wu et~al., 2022]{wu2022simulation}
Wu, Z., Wang, B., Li, M., Tian, Y., Quan, Y., and Liu, J. (2022).
\newblock Simulation of forest fire spread based on artificial intelligence.
\newblock {\em Ecological Indicators}, 136:108653.

\bibitem[Xavier et~al., 2021]{xavier2021learning}
Xavier, {\'A}.~S., Qiu, F., and Ahmed, S. (2021).
\newblock Learning to solve large-scale security-constrained unit commitment problems.
\newblock {\em INFORMS Journal on Computing}, 33(2):739--756.

\bibitem[Xiang et~al., 2017]{xiang2017coordinated}
Xiang, Y., Wang, L., and Liu, N. (2017).
\newblock Coordinated attacks on electric power systems in a cyber-physical environment.
\newblock {\em Electric Power Systems Research}, 149:156--168.

\bibitem[Xu et~al., 2021]{xu2021bayesian}
Xu, J., Liu, B., Mo, H., and Dong, D. (2021).
\newblock Bayesian adversarial multi-node bandit for optimal smart grid protection against cyber attacks.
\newblock {\em Automatica}, 128:109551.

\bibitem[Xu et~al., 2022]{xu2022online}
Xu, J., Sun, Q., Mo, H., and Dong, D. (2022).
\newblock Online routing for smart electricity network under hybrid uncertainty.
\newblock {\em Automatica}, 145:110538.

\bibitem[Yan et~al., 2016]{yan2016q}
Yan, J., He, H., Zhong, X., and Tang, Y. (2016).
\newblock Q-learning-based vulnerability analysis of smart grid against sequential topology attacks.
\newblock {\em IEEE Transactions on Information Forensics and Security}, 12(1):200--210.

\bibitem[Yang et~al., 2022a]{Yang2022optimizing}
Yang, H., Duque, D., and Morton, D.~P. (2022a).
\newblock Optimizing diesel fuel supply chain operations to mitigate power outages for hurricane relief.
\newblock {\em IISE Transactions}, 54(10):936--949.

\bibitem[Yang et~al., 2021a]{yang2021robust}
Yang, H., Morton, D.~P., Dvijothamand, K., and Martin, A. (2021a).
\newblock Robust optimization for electricity generation.
\newblock {\em INFORMS Journal on Computing}, 33(1):336--351.

\bibitem[Yang et~al., 2022b]{YANG2022118793}
Yang, W., Sparrow, S.~N., Ashtine, M., Wallom, D.~C., and Morstyn, T. (2022b).
\newblock Resilient by design: Preventing wildfires and blackouts with microgrids.
\newblock {\em Applied Energy}, 313:118793.

\bibitem[Yang et~al., 2019]{yang2019distributed}
Yang, W., Zhang, Y., Chen, G., Yang, C., and Shi, L. (2019).
\newblock Distributed filtering under false data injection attacks.
\newblock {\em Automatica}, 102:34--44.

\bibitem[Yang et~al., 2021b]{yang2021criterion}
Yang, Y., Peng, J. C.-H., Ye, C., Ye, Z.-S., and Ding, Y. (2021b).
\newblock A criterion and stochastic unit commitment towards frequency resilience of power systems.
\newblock {\em IEEE Transactions on Power Systems}, 37(1):640--652.

\bibitem[Yassemi et~al., 2008]{yassemi2008design}
Yassemi, S., Dragi{\'c}evi{\'c}, S., and Schmidt, M. (2008).
\newblock Design and implementation of an integrated gis-based cellular automata model to characterize forest fire behaviour.
\newblock {\em ecological modelling}, 210(1-2):71--84.

\bibitem[Yu et~al., 2015]{yu2015analysis}
Yu, K., Ai, Q., Wang, S., Ni, J., and Lv, T. (2015).
\newblock Analysis and optimization of droop controller for microgrid system based on small-signal dynamic model.
\newblock {\em IEEE Transactions on Smart Grid}, 7(2):695--705.

\bibitem[Zeng et~al., 2018]{zeng2018novel}
Zeng, Y., Xiao, X., Li, J., Sun, L., Floudas, C.~A., and Li, H. (2018).
\newblock A novel multi-period mixed-integer linear optimization model for optimal distribution of byproduct gases, steam and power in an iron and steel plant.
\newblock {\em Energy}, 143:881--899.

\bibitem[Zhang et~al., 2018]{zhang2018distributed}
Zhang, H., Meng, W., Qi, J., Wang, X., and Zheng, W.~X. (2018).
\newblock Distributed load sharing under false data injection attack in an inverter-based microgrid.
\newblock {\em IEEE Transactions on Industrial Electronics}, 66(2):1543--1551.

\bibitem[Zhang et~al., 2022]{zhang2021identification}
Zhang, K., Keliris, C., Parisini, T., and Polycarpou, M.~M. (2022).
\newblock Identification of sensor replay attacks and physical faults for cyber-physical systems.
\newblock {\em IEEE Control Systems Letters}, 6:1178--1183.

\bibitem[Zhang and Leung, 2022]{zhang2020joint}
Zhang, S. and Leung, K.-C. (2022).
\newblock Joint optimal power flow routing and vehicle-to-grid scheduling: Theory and algorithms.
\newblock {\em IEEE Transactions on Intelligent Transportation Systems}, (1):1178--1183.

\bibitem[Zhang and Ye, 2020]{zhang2020false}
Zhang, T.-Y. and Ye, D. (2020).
\newblock False data injection attacks with complete stealthiness in cyber--physical systems: A self-generated approach.
\newblock {\em Automatica}, 120:109117.

\bibitem[Zhang and Papachristodoulou, 2015]{zhang2015real}
Zhang, X. and Papachristodoulou, A. (2015).
\newblock A real-time control framework for smart power networks: Design methodology and stability.
\newblock {\em Automatica}, 58:43--50.

\bibitem[Zhu and Mart{\'\i}nez, 2013]{zhu2013distributed}
Zhu, M. and Mart{\'\i}nez, S. (2013).
\newblock On distributed constrained formation control in operator--vehicle adversarial networks.
\newblock {\em Automatica}, 49(12):3571--3582.

\bibitem[Zhu and Martinez, 2013]{zhu2013performance}
Zhu, M. and Martinez, S. (2013).
\newblock On the performance analysis of resilient networked control systems under replay attacks.
\newblock {\em IEEE Transactions on Automatic Control}, 59(3):804--808.

\bibitem[Zolan et~al., 2021]{zolan2021decomposing}
Zolan, A.~J., Scioletti, M.~S., Morton, D.~P., and Newman, A.~M. (2021).
\newblock Decomposing loosely coupled mixed-integer programs for optimal microgrid design.
\newblock {\em INFORMS Journal on Computing}, 33(4):1300--1319.

\bibitem[Zuniga~Vazquez et~al., 2022]{Vazquez2022wild}
Zuniga~Vazquez, D.~A., Qiu, F., Fan, N., and Sharp, K. (2022).
\newblock Wildfire mitigation plans in power systems: A literature review.
\newblock {\em IEEE Transactions on Power Systems}, 37(5):3540--3551.

\end{thebibliography}

\newpage
\begin{APPENDICES}

\noindent
\begin{center}
    {\Large{Electronic Companion to ``Online Planning of Power Flows for Power Systems Against Bushfire Using Spatial Context''}}
\end{center}

\vspace{1em}


\vspace{1em}

\setcounter{figure}{0}  
\renewcommand{\thefigure}{EC.\arabic{figure}}
\setcounter{table}{0}  
\renewcommand{\thetable}{EC.\arabic{table}}
\setcounter{equation}{0}  
\renewcommand{\theequation}{EC.\arabic{equation}}
\setcounter{lemma}{0}  
\renewcommand{\thelemma}{EC.\arabic{lemma}}
\setcounter{theorem}{0}  
\renewcommand{\thetheorem}{EC.\arabic{theorem}}
\setcounter{algorithm}{0}  
\renewcommand{\thealgorithm}{EC.\arabic{algorithm}}
\setcounter{page}{1} 
\renewcommand{\thepage}{EC.\arabic{page}}

\section{Technical Proofs}\label{appen:proof}
\subsection{Proof of Proposition~\ref{proposition lp optimal}}
We show that Problem~\eqref{eq:power_flow_LP} is equivalent to Problem~\eqref{eq:power_flow}. 
By Constraints~\eqref{eq-optimization-online-2} and \eqref{eq-optimization-online-4}, when $\varphi_{t}(i)=1$, we have $\min\{U^\text{P}_{t}(i),\,\alpha(i)\}=\alpha(i)$ and $\min\{U^\text{F}_{t}(j,\,i),\,|\beta(j,\,i)|\}=\beta(j,\,i)$; 
when $\varphi_{t}(i)=0$, we have $\min\{U^\text{P}_{t}(i),\,\alpha(i)\}=0$, $\min\{U^\text{F}(j,\,i),\,|\beta(j,\,i)|\}=0$, and $L_t(i) = 0$.  In such a case, there is no load shedding.
Hence, the shedding load $\texttt{LS}^{\pi}_{t}(i;\, L_{t},\,U^\text{P}_{t},\,U^\text{F}_{t})$ can be rewritten as 
\begin{align*}
\texttt{LS}^{\pi}(i;\,L_{t},\,U^\text{P}_{t},\,U^\text{F}_{t})
= L(i) - \alpha(i) - 
\sum_{j:(j,\,i)\in\mathcal{E}}\varphi_{t}(j,\,i)\cdot\beta(j,\,i).
\end{align*}
Moreover, we have
\begin{align*}
    & \mathbf{E}\left[\texttt{LS}^{\pi}(i;\,L_{t},\,U^\text{P}_{t},\,U^\text{F}_{t})\,|\,\mathcal{B}_{t-1}]\right.\nonumber\\
    =~ & P_{t}(i)\sum_{\mathcal{S}'\in \mathcal{P}({\mathcal{S}(i)})}
    \prod_{j\in\mathcal{S'}}P_{t}(j,\,i)
    \prod_{j\in\mathcal{S}'}\left(1-P_{t}(j,\,i)\right)
    \max \left\{L(i) - \alpha(i) - 
    \sum_{j\in\mathcal{S}(i)\setminus\mathcal{S}'}\beta(j,\,i), 0\right\}\\
    =~ &
    P_{t}(i)\sum_{\mathcal{S}'\in \mathcal{P}({\mathcal{S}(i)})}\rho_{t}(i,\,\mathcal{S}')
    \max \left\{L(i) - \alpha(i) - 
    \sum_{j\in\mathcal{S}(i)\setminus\mathcal{S}'}\beta(j,\,i), 0\right\}.
\end{align*}
Thus,
\begin{align*}
& \mathbf{E}\left[C(\pi;\: L_{t},\,U^\text{P}_{t},\,U^\text{F}_{t})\,|\,\mathcal{B}_{t-1}\right]
\nonumber\\
=~ &
\sum_{i\in\mathcal{S}} C^{\text{P}}(i)\alpha(i)
+ C^{\text{S}}\sum_{i\in\mathcal{S}}\mathbf{E}
\left[\texttt{LS}^{\pi}(i;\,L_{t},\,U^\text{P}_{t},\,U^\text{F}_{t})\,|\,\mathcal{B}_{t-1}\right]\nonumber\\
=~ &
\sum_{i\in\mathcal{S}} C^{\text{P}}(i)\alpha(i)
+  C^{\text{S}} \sum_{i\in\mathcal{S}}\sum_{\mathcal{S}'\in \mathcal{P}({\mathcal{S}(i)})}P_{t}(i) \rho_{t}(i,\,\mathcal{S}')
\max \left\{L(i) - \alpha(i) - \sum_{j\in\mathcal{S}(i)\setminus\mathcal{S}'}\beta(j,\,i), 0\right\}.
\end{align*}
Denote $\alpha^*(i)$, 
$\beta^*(i,j)$, and $H^{*}(i,\,\mathcal{S}')$ as be the corresponding optimal solutions to Problem~\eqref{eq:power_flow_LP}.
Since $C^{\text{S}}P^{i}_{t}\rho_{t}(i,\,\mathcal{S}')\geq0$, the optimal solution $H^{*}(i,\,\mathcal{S}')$ is either $L(i) - \alpha^*(i)- \sum_{j\in\mathcal{S}'}\beta^*(j,i)$ or $0$ for all $i\in\mathcal{S}$ and $\mathcal{S}'\in \mathcal{P}({\mathcal{S}(i)})$, whichever is larger.
This shows the equivalence between the two problems.\Halmos

\subsection{Proof of Theorem \ref{theorem approximate optimal strategy}}
For all $t\geq2$ and $i\in(\mathcal{S},\,\mathcal{E})$, we denote by $\hat{P}_{t}(i)$, $\hat{P}_{t}(i,\,j)$, and $\hat{\rho}_{t}(i,\,\mathcal{S}')$ the estimates of $P_{t}(i)$, $P_{t}(i,\,j)$, and $\rho_{t}(i,\,\mathcal{S}')$ correspondingly, calculated using $\hat{p}^{+}_{t-1}$ and $\hat{p}^{-}_{t}$. By repeatedly using the relation of $\left|(1-x)^{n}-(1-y)^{n}\right|\leq n|x-y|$, $\forall n\in\mathbb{N}^{+},\: 0<x,\,y<1$, 
we have
\begin{align}\label{eq proof of theorem approximate optimal strategy 1}
& \left|\hat{P}_{t}(i)-P_{t}(i)\right|\nonumber\\
=~ &
\Bigg|\prod_{h=1}^{H}\Big(\prod_{d'=1}^{\bar d}
\prod_{j\in(\mathcal{N}_{d'}(i)\cap\mathcal{G}_{h})
\setminus \mathcal{B}_{h,\,t-1}}
(1-p_{h,\,t-1}^+)^{|\mathcal{F}_{t-1}(\mathcal{N}(j))\cap\mathcal{G}_{h}|}
\cdot(p^{-}_{h,\,t-1})^{\sum_{d'=1}^{\bar d}
|\mathcal{F}_{t-1}(\mathcal{N}_{d'}(i))\cap\mathcal{G}_{h}|}
\Big)
\nonumber\\
& - \prod_{h=1}^{H}\Big(\prod_{d'=1}^{\bar d}
\prod_{j\in(\mathcal{N}_{d'}(i)\cap\mathcal{G}_{h})
\setminus \mathcal{B}_{h,\,t-1}}
(1-\hat{p}_{h,\,t-1}^{+})^{|\mathcal{F}_{t-1}(\mathcal{N}(j))\cap\mathcal{G}_{h}|}
\cdot(\hat{p}^{-}_{h,\,t-1})^{\sum_{d'=1}^{\bar d}
|\mathcal{F}_{t-1}(\mathcal{N}_{d'}(i))\cap\mathcal{G}_{h}|}\Big)\Bigg|
\nonumber\\
\leq~ &
\Bigg|\prod_{h=1}^{H}\Big(\prod_{d'=1}^{\bar d}
\prod_{j\in(\mathcal{N}_{d'}(i)\cap\mathcal{G}_{h})
\setminus \mathcal{B}_{h,\,t-1}}
(1-p_{h,\,t-1}^+)^{|\mathcal{F}_{t-1}(\mathcal{N}(j))\cap\mathcal{G}_{h}|}
\cdot(p^{-}_{h,\,t-1})^{\sum_{d'=1}^{\bar d}
|\mathcal{F}_{t-1}(\mathcal{N}_{d'}(i))\cap\mathcal{G}_{h}|}\Big)
\nonumber\\
& - \prod_{h=1}^{H}\Big(\prod_{d'=1}^{\bar d}
\prod_{j\in(\mathcal{N}_{d'}(i)\cap\mathcal{G}_{h})
\setminus \mathcal{B}_{h,\,t-1}}
(1-\hat{p}_{h,\,t-1}^+)^{|\mathcal{F}_{t-1}(\mathcal{N}(j))\cap\mathcal{G}_{h}|}
\cdot(p^{-}_{h,\,t-1})^{\sum_{d'=1}^{\bar d}
|\mathcal{F}_{t-1}(\mathcal{N}_{d'}(i))\cap\mathcal{G}_{h}|}\Big)\Bigg|
\nonumber\\
& +
\Bigg|\prod_{h=1}^{H}\Big(\prod_{d'=1}^{\bar d}
\prod_{j\in(\mathcal{N}_{d'}(i)\cap\mathcal{G}_{h})
\setminus \mathcal{B}_{h,\,t-1}}
(1-\hat{p}_{h,\,t-1}^+)^{|\mathcal{F}_{t-1}(\mathcal{N}(j))\cap\mathcal{G}_{h}|}
\cdot(p^{-}_{h,\,t-1})^{\sum_{d'=1}^{\bar d}
|\mathcal{F}_{t-1}(\mathcal{N}_{d'}(i))\cap\mathcal{G}_{h}|}\Big)
\nonumber\\
& - \prod_{h=1}^{H}\Big(\prod_{d'=1}^{\bar d}
\prod_{j\in(\mathcal{N}_{d'}(i)\cap\mathcal{G}_{h})
\setminus \mathcal{B}_{h,\,t-1}}
(1-\hat{p}_{h,\,t-1}^+)^{|\mathcal{F}_{t-1}(\mathcal{N}(j))\cap\mathcal{G}_{h}|}
\cdot(\hat{p}^{-}_{h,\,t-1})^{\sum_{d'=1}^{\bar d}
|\mathcal{F}_{t-1}(\mathcal{N}_{d'}(i))\cap\mathcal{G}_{h}|}\Big)\Bigg|
\nonumber\\
\leq~ &
\left| \prod_{h=1}^{H}\prod_{d'=1}^{\bar d}
\prod_{j\in(\mathcal{N}_{d'}(i)\cap\mathcal{G}_{h})
\setminus \mathcal{B}_{h,\,t-1}}
(1-p^{+}_{h,\,t-1})^{|\mathcal{F}_{t-1}(\mathcal{N}(j))\cap\mathcal{G}_{h}|}
- \prod_{h=1}^{H}\prod_{d'=1}^{\bar d}
\prod_{j\in(\mathcal{N}_{d'}(i)\cap\mathcal{G}_{h})
\setminus \mathcal{B}_{h,\,t-1}}
(1-\hat{p}^{+}_{h,\,t-1})^{|\mathcal{F}_{t-1}(\mathcal{N}(j))\cap\mathcal{G}_{h}|}
\right|\nonumber\\
& + 
\left|\prod_{h=1}^{H}\left(p^{-}_{h,\,t-1}\right)^{\sum_{d'=1}^{\bar{d}}
\left|\mathcal{F}_{t-1}(\mathcal{N}_{d'}(i))\cap\mathcal{G}_{h}\right|}
-\prod_{h=1}^{H}\left(\hat{p}^{-}_{h,\,t-1}\right)^{\sum_{d'=1}^{\bar{d}}
\left|\mathcal{F}_{t-1}(\mathcal{N}_{d'}(i))\cap\mathcal{G}_{h}\right|}\right|\nonumber\\
\leq~ &
\sum_{h=1}^{H}\sum_{d'=1}^{\bar d}
\sum_{j\in(\mathcal{N}_{d'}(i)\cap\mathcal{G}_{h})
\setminus \mathcal{B}_{h,\,t-1}}
\left| (1-p^{+}_{h,\,t-1})^{|\mathcal{F}_{t-1}(\mathcal{N}(j))\cap\mathcal{G}_{h}|}
-(1-\hat{p}^{+}_{h,\,t-1})^{|\mathcal{F}_{t-1}(\mathcal{N}(j))\cap\mathcal{G}_{h}|}
\right|\nonumber\\
& + 
\left|\sum_{h=1}^{H}\left(p^{-}_{h,\,t-1}\right)^{\sum_{d'=1}^{\bar{d}}
\left|\mathcal{F}_{t-1}(\mathcal{N}_{d'}(i))\cap\mathcal{G}_{h}\right|}
-\sum_{h=1}^{H}\left(\hat{p}^{-}_{h,\,t-1}\right)^{\sum_{d'=1}^{\bar{d}}
\left|\mathcal{F}_{t-1}(\mathcal{N}_{d'}(i))\cap\mathcal{G}_{h}\right|}
\right|\nonumber\\
\leq~ &
\sum_{h=1}^{H}\sum_{d'=1}^{\bar d}
\sum_{j\in\mathcal{N}_{d'}(i)\setminus\mathcal{B}_{t-1}}
\left|\mathcal{F}_{t-1}(\mathcal{N}(j))\cap\mathcal{G}_{h}\right|
\left|p^{+}_{h,\,t-1}-\hat{p}^{+}_{h,\,t-1}\right|
+ \sum_{h=1}^{H}\sum_{d'=1}^{\bar{d}}
\left|\mathcal{F}_{t-1}(\mathcal{N}_{d'}(i))\cap\mathcal{G}_{h}\right|
\left|p^{-}_{h,\,t-1}-\hat{p}^{-}_{h,\,t-1}\right|\nonumber\\
\leq~ &
\sum_{d'=1}^{\bar d}\sum_{j\in\mathcal{N}_{d'}(i)\setminus \mathcal{B}_{t-1}}
\left|\mathcal{F}_{t-1}(\mathcal{N}(j))\right|
\max_{h=1,\ldots,H}\left|p^{+}_{h,\,t-1}-\hat{p}^{+}_{h,\,t-1}\right|
+ \sum_{d'=1}^{\bar{d}}\left|\mathcal{F}_{t-1}(\mathcal{N}_{d'}(i))\right|
\max_{h=1,\ldots,H}\left|p^{-}_{h,\,t-1}-\hat{p}^{-}_{h,\,t-1}\right|\nonumber\\
\leq~ &
16(2\bar{d}+1)^{2}\max_{h=1,\ldots,H}\left|p^{+}_{h,\,t-1}-\hat{p}^{+}_{h,\,t-1}\right|
+ (2\bar{d}+1)^{2}\max_{h=1,\ldots,H}\left|p^{-}_{h,\,t-1}-\hat{p}^{-}_{h,\,t-1}\right|,
\end{align}
where the last inequality holds following the fact of $\left|\mathcal{F}_{t-1}(\mathcal{N}_{d'}(i))\right|
\leq \left|\mathcal{N}_{d'}(i)\right|\leq 8d^{'}$ for all $d^{'}\geq1$. Meanwhile,
\begin{align}\label{eq proof of theorem approximate optimal strategy 2}
& \left|P_{t}(i,\,j) - \hat{P}_{t}(i,\,j)\right| \leq
\left|\prod_{i'\in(i,\,j)}P_{t}(i') - \prod_{i'\in(i,\,j)}\hat{P}_{t}(i')\right|\nonumber\\
\leq &
\left|P_{t}(i^{''})\prod_{i'\in(i,\,j)\setminus\{i^{''}\}}P_{t}(i') - \hat{P}_{t}(i^{''})\prod_{i'\in(i,\,j)\setminus\{i^{''}\}}P_{t}(i')\right| \nonumber \\
& +\left|\hat{P}_{t}(i^{''})\prod_{i'\in(i,\,j)\setminus\{i^{''}\}}P_{t}(i') - \hat{P}_{t}(i^{''})\prod_{i'\in(i,\,j)\setminus\{i^{''}\}}\hat{P}_{t}(i')\right|\nonumber\\
\leq &
\left|P_{t}(i^{''})- \hat{P}_{t}(i^{''})\right|
\prod_{i'\in(i,\,j)\setminus\{i^{''}\}}P_{t}(i')
+ \left|\prod_{i'\in(i,\,j)\setminus\{i^{''}\}}P_{t}(i') - \prod_{i'\in(i,\,j)\setminus\{i^{''}\}}\hat{P}_{t}(i')\right|\hat{P}_{t}(i^{''})\nonumber\\
\leq &
\left|P_{t}(i^{''})- \hat{P}_{t}(i^{''})\right|
+ \left|\prod_{i'\in(i,\,j)\setminus\{i^{''}\}}P_{t}(i') - \prod_{i'\in(i,\,j)\setminus\{i^{''}\}}\hat{P}_{t}(i')\right|.
\end{align}
By using the same process as that in (\ref{eq proof of theorem approximate optimal strategy 2}) recursively and letting $\bar{d}=1$ in (\ref{eq proof of theorem approximate optimal strategy 1}), we have
\begin{align*}
& \left|P_{t}(i,\,j) - \hat{P}_{t}(i,\,j)\right| \leq
\sum_{i'\in(i,\,j)}\left|P_{t}(i^{'})- \hat{P}_{t}(i^{'})\right| \\
\leq &
\sum_{i'\in(i,\,j)}
\left(16\max_{h=1,\ldots,H}\left|p^{+}_{h,\,t-1}-\hat{p}^{+}_{h,\,t-1}\right|
+ \max_{h=1,\ldots,H}\left|p^{-}_{h,\,t-1}-\hat{p}^{-}_{h,\,t-1}\right|\right).
\end{align*}
Similarly, we have
\begin{align}\label{eq proof of theorem approximate optimal strategy 3}
& \left|\rho_{t}(i,\,\mathcal{S}') - \hat{\rho}_{t}(i,\,\mathcal{S}')\right|
\leq \sum_{j\in\mathcal{S}(i)}\left|P_{t}(i,\,j) - \hat{P}_{t}(i,\,j)\right| 
\nonumber\\
\leq &
\sum_{j\in\mathcal{S}(i)}\sum_{i'\in(i,\,j)}
\left(16\max_{h=1,\ldots,H}\left|p^{+}_{h,\,t-1}-\hat{p}^{+}_{h,\,t-1}\right|
+ \max_{h=1,\ldots,H}\left|p^{-}_{h,\,t-1}-\hat{p}^{-}_{h,\,t-1}\right|\right),
\end{align}
for all $i\in\mathcal{S}$ and $\mathcal{S}'\in\mathcal{P}{\mathcal{S}(i)})$. 
Combining (\ref{eq proof of theorem approximate optimal strategy 1}) and 
(\ref{eq proof of theorem approximate optimal strategy 3}) yields
\begin{align}\label{eq proof of theorem approximate optimal strategy 4}
& \left|P_{t}(i)\rho_{t}(i,\,\mathcal{S}') - \hat{P}_{t}(i)\hat{\rho}_{t}(i,\,\mathcal{S}')\right|\nonumber\\
\leq &
(2\bar{d}+1)^{2}\left(16\max_{h=1,\ldots,H}\left|p^{+}_{h,\,t-1}-\hat{p}^{+}_{h,\,t-1}
\right|+ \max_{h=1,\ldots,H}\left|p^{-}_{h,\,t-1}-\hat{p}^{-}_{h,\,t-1}\right|\right)
\nonumber\\
& + 
\sum_{j\in\mathcal{S}(i)}\sum_{i'\in(i,\,j)}
\left(16\max_{h=1,\ldots,H}\left|p^{+}_{h,\,t-1}-\hat{p}^{+}_{h,\,t-1}\right|
+ \max_{h=1,\ldots,H}\left|p^{-}_{h,\,t-1}-\hat{p}^{-}_{h,\,t-1}\right|\right).
\end{align}

Next, we bound the difference between the expected costs for $\pi^{*}_{t}$ and $\hat{\pi}^{*}_{t}$. We let $(\alpha^{*}_{t},\,\Tilde{\beta}^{*}_{t},\,H^{*}_{t})$ and $(\hat{\alpha}^{*}_{t},\,\hat{\beta}^{*}_{t},\,\hat{H}^{*}_{t})$ be the optimal solutions of \eqref{eq:power_flow_LP} corresponding to $\pi^{*}_{t}$ and $\hat{\pi}^{*}_{t}$, respectively. For notational brevity, we let 
$p^{\pm}_{t}=(p^{\pm}_{1,\,t},\cdots,p^{\pm}_{H,\,t})$ and $\hat{p}^{\pm}_{t}$ be its estimator.
To distinguish, we denote by $\mathbf{E}_{p^{+}_{t},\,p^{-}_{t}}\left[C(\pi;\:L_{t},\,U^\text{P}_{t},\,U^\text{F}_{t})\,|\,\mathcal{B}_{t-1}\right]$ and $\mathbf{E}_{\hat{p}^{+}_{t},\,\hat{p}^{-}_{t}}\left[C(\pi;\:L_{t},\,U^\text{P}_{t},\,U^\text{F}_{t})\,|\,\mathcal{B}_{t-1}\right]$ the objective function of \eqref{eq:power_flow_LP} when using $(p^{+}_{t},\,p^{-}_{t})$ and $(\hat{p}^{+}_{t},\,\hat{p}^{-}_{t})$ accordingly for any strategy $\pi$. 
Then, we have
\begin{align}\label{eq proof of theorem approximate optimal strategy 6}
& \mathbf{E}\left[C(\hat{\pi}^{*}_{t};\: L_{t},\,U^\text{P}_{t},\,U^\text{F}_{t})\,|\,\mathcal{B}_{t-1}\right]
-\mathbf{E}\left[C(\pi^{*}_{t};\:L_{t},\,U^\text{P}_{t},\,U^\text{F}_{t})\,|\,\mathcal{B}_{t-1}\right]\nonumber\\
\leq~ &
\left(\mathbf{E}_{p^{+}_{t},\,p^{-}_{t}}\left[C(\hat{\pi}^{*}_{t};
\:L_{t},\,U^\text{P}_{t},\,U^\text{F}_{t})\,|\,\mathcal{B}_{t-1}\right]
-\mathbf{E}_{\hat{p}^{+}_{t},\,\hat{p}^{-}_{t}}\left[C(\hat{\pi}^{*}_{t};
\:L_{t},\,U^\text{P}_{t},\,U^\text{F}_{t})\,|\,\mathcal{B}_{t-1}\right]\right)\nonumber\\
& + 
\left(\mathbf{E}_{\hat{p}^{+}_{t},\,\hat{p}^{-}_{t}}\left[C(\hat{\pi}^{*}_{t};
\:L_{t},\,U^\text{P}_{t},\,U^\text{F}_{t})\,|\,\mathcal{B}_{t-1}\right]
-\mathbf{E}_{\hat{p}^{+}_{t},\,\hat{p}^{-}_{t}}\left[C(\pi^{*}_{t};
\:L_{t},\,U^\text{P}_{t},\,U^\text{F}_{t})\,|\,\mathcal{B}_{t-1}\right]\right)\nonumber\\
& +
\left(\mathbf{E}_{\hat{p}^{+}_{t},\,\hat{p}^{-}_{t}}\left[C(\pi^{*}_{t};
\:L_{t},\,U^\text{P}_{t},\,U^\text{F}_{t})\,|\,\mathcal{B}_{t-1}\right]
-\mathbf{E}_{p^{+}_{t},\,p^{-}_{t}}\left[C(\pi^{*}_{t};
\:L_{t},\,U^\text{P}_{t},\,U^\text{F}_{t})\,|\,\mathcal{B}_{t-1}\right]\right)\nonumber\\
\leq~ &
\left|\mathbf{E}_{p^{+}_{t},\,p^{-}_{t}}\left[C(\hat{\pi}^{*}_{t};
\:L_{t},\,U^\text{P}_{t},\,U^\text{F}_{t})\,|\,\mathcal{B}_{t-1}\right]
-\mathbf{E}_{\hat{p}^{+}_{t},\,\hat{p}^{-}_{t}}\left[C(\hat{\pi}^{*}_{t};
\:L_{t},\,U^\text{P}_{t},\,U^\text{F}_{t})\,|\,\mathcal{B}_{t-1}\right]\right|\nonumber\\
& +
\left|\mathbf{E}_{\hat{p}^{+}_{t},\,\hat{p}^{-}_{t}}\left[C(\pi^{*}_{t};
\:L_{t},\,U^\text{P}_{t},\,U^\text{F}_{t})\,|\,\mathcal{B}_{t-1}\right]
-\mathbf{E}_{p^{+}_{t},\,p^{-}_{t}}\left[C(\pi^{*}_{t};
\:L_{t},\,U^\text{P}_{t},\,U^\text{F}_{t})\,|\,\mathcal{B}_{t-1}\right]\right|\nonumber\\
\leq~ &
C^{\text{S}}\sum_{i\in\mathcal{S}}\sum_{\mathcal{S}'\in\mathcal{P}({\mathcal{S}(i)})}
\left|P_{t}(i)\rho_{t}(i,\,\mathcal{S}') 
-\hat{P}_{t}(i)\hat{\rho}_{t}(i,\,\mathcal{S}')\right|
\left(\hat{H}^{*}_{t}(i,\,\mathcal{S}')+H^{*}_{t}(i,\,\mathcal{S}')\right)\nonumber\\
& +  
C^{\text{S}}\sum_{i\in\mathcal{S}}\sum_{\mathcal{S}'\in \mathcal{P}({\mathcal{S}(i)})}
\left|\left(1-P_{t}(i)\right)\rho_{t}(i,\,\mathcal{S}')
-\left(1-\hat{P}_{t}(i)\right)\hat{\rho}_{t}(i,\,\mathcal{S}')\right|
\left(\hat{M}^{*}_{t}(i,\,\mathcal{S}')+M^{*}_{t}(i,\,\mathcal{S}')\right),
\end{align}
where the second inequality holds because $\hat{\pi}^{*}_{t}$ minimizes $\mathbf{E}_{\hat{p}^{+}_{t},\,\hat{p}^{-}_{t}}\left[C(\pi;\:L_{t},\,U^\text{P}_{t},\,U^\text{F}_{t})\,|\,\mathcal{B}_{t-1}\right]$. Combining (\ref{eq proof of theorem approximate optimal strategy 4}) further yields
\begin{align}
& \mathbf{E}\left[C(\hat{\pi}^{*}_{t};\: L_{t},\,U^\text{P}_{t},\,U^\text{F}_{t})\,|\,\mathcal{B}_{t-1}\right]
-\mathbf{E}\left[C(\pi^{*}_{t};\:L_{t},\,U^\text{P}_{t},\,U^\text{F}_{t}
)\,|\,\mathcal{B}_{t-1}\right]\nonumber\\
\leq~ &
C^{\text{S}}\sum_{i\in\mathcal{S}} |\mathcal{P}({\mathcal{S}(i)})|\left[16(2\bar{d}+1)^{2} + 
8\sum_{j\in\mathcal{S}(i)}\sum_{i'\in(i,\,j)} 1 \right]
\times \left(\hat{H}^{*}_{t}(i,\,\mathcal{S}')+H^{*}_{t}(i,\,\mathcal{S}')\right)
\max_{h=1,\ldots,H}\left|p^{+}_{h,\,t-1}-\hat{p}^{+}_{h,\,t-1}\right|\nonumber\\
& +
C^{\text{S}}\sum_{i\in\mathcal{S}} |\mathcal{P}({\mathcal{S}(i)})|\left[(2d+1)^{2} + 
\sum_{j\in\mathcal{S}(i)}\sum_{i'\in(i,\,j)}1 \right]
\times \left(\hat{H}^{*}_{t}(i,\,\mathcal{S}')+H^{*}_{t}(i,\,\mathcal{S}')\right)
\max_{h=1,\ldots,H}\left|p^{-}_{h,\,t-1}-\hat{p}^{-}_{h,\,t-1}\right|\nonumber \\
\leq~ &
2C^{\text{S}}\sum_{i\in\mathcal{S}}({|\mathcal{P}({\mathcal{S}(i)})|+3})
\left[\Delta(i)+2(2\bar{d}+1)^{2}\right]\left(U^\text{P}(i) + L(i) 
+ \sum_{j:\,(j,\,i)\in\mathcal{E}}U^\text{F}(j,\,i)\right)
\max_{h=1,\ldots,H}\left|p^{+}_{h,\,t-1}-\hat{p}^{+}_{h,\,t-1}\right|\nonumber\\
& +
2C^{\text{S}}\sum_{i\in\mathcal{S}} |\mathcal{P}({\mathcal{S}(i)})|\left[(2d+1)^{2}+\Delta(i)\right]
\left(U^\text{P}(i) + L(i) + \sum_{j:\,(j,\,i)\in\mathcal{E}}U^\text{F}(j,\,i)\right)
\max_{h=1,\ldots,H}\left|p^{-}_{h,\,t-1}-\hat{p}^{-}_{h,\,t-1}\right|,
\end{align}
where the last inequality holds because
\begin{align*}
\max\left\{\hat{H}^{*}_{t}(i,\,\mathcal{S}'),\,H^{*}_{t}(i,\,\mathcal{S}')\right\}
\leq U^\text{P}(i)+L(i)+\sum_{j:\,(j,\,i)\in\mathcal{E}}U^\text{F}(j,\,i).
\end{align*}
Thus, we conclude the proof.
\Halmos

\subsection{Proof of Theorem \ref{theorem regret}}
The proof follows the sketch in Section \ref{section online operational algorithm} and proceeds in five steps.

\noindent\textbf{Step 1. Constructing a good event.}
Recall that $\hat{p}^{+}_{h,\,t}-p^{+}_{h,\,t}$ is a sub-Gaussian random variable.
According to the Hoeffding inequality for sub-Gaussian distributions, given 
$1\leq t_{1}\leq t_{2}\leq T$,
\begin{align}\label{eq proof of theorem regret 1}
\mathrm{Pr}\left\{
\frac{1}{t_{2}-t_{1}+1}\left|\sum_{t=t_{1}}^{t_{2}}\hat{p}^{+}_{h,\,t}
- \sum_{t=t_{1}}^{t_{2}}p^{+}_{h,\,t}\right|\geq\varepsilon\right\}
\leq 2\exp{\left\{-\frac{(t_{2}-t_{1}+1)^{2}\varepsilon^{2}}
{2\sum_{t=t_{1}}^{t_{2}}\nu^{+}_{h,\,t}}\right\}}
\end{align}
holds for all $\varepsilon>0$. 
If we let the right-hand side of (\ref{eq proof of theorem regret 1}) equal $1/(H T^{4})$ and define $p^{+}_{h,\,t_{1},\,t_{2}}\triangleq \sum_{t=t_{1}}^{t_{2}}p^{+}_{h,\,t}$, then
\begin{align*}
\mathrm{Pr}\left\{
\left|\Hat{p}^{+}_{h,\,t_{1},\,t_{2}}- p^{+}_{h,\,t_{1},\,t_{2}}\right|
\geq\frac{4\sqrt{\sum_{t'=t_{1}}^{t_{2}}\nu^{+}_{h,\,t'}\ln{2H\cdot T}}}
{t_{2}-t_{1}+1}\right\}\leq \frac{1}{H T^{4}}.
\end{align*}
Moreover, if we use a union bound on $1\leq t_{2}\leq T$ by fixing $t_{1}$, and then a second union bound on $1\leq t_{1}\leq T$, we have
\begin{align}\label{eq proof of theorem regret 2}
\mathrm{Pr}\left\{\left|\Hat{p}^{+}_{h,\,t_{1},\,t_{2}}- p^{+}_{h,\,t_{1},\,t_{2}}\right|
\geq\frac{4\sqrt{\sum_{t'=t_{1}}^{t_{2}}\nu^{+}_{h,\,t'}\ln{2H T}}}
{t_{2}-t_{1}+1}\right\}\leq \frac{1}{H T^{2}},
\:\: \forall 1\leq t_{1}\leq t_{2}\leq T.
\end{align}
Similarly, if we define $p^{-}_{s_{1},\,s_{2}}\triangleq \sum_{t=s_{1}}^{s_{2}}p^{-}_{t}$, we have
\begin{align}\label{eq proof of theorem regret 3}
\mathrm{Pr}\left\{\left|\Hat{p}^{-}_{h,\,s_{1},\,s_{2}}- p^{-}_{h,\,s_{1},\,s_{2}}\right|
\geq\frac{4\sqrt{\sum_{t'=s_{1}}^{s_{2}}\nu^{-}_{h,\,t'}\ln{2H T}}}{s_{2}-s_{1}+1}\right\}
\leq \frac{1}{H T^{2}},\:\: \forall 1\leq s_{1}\leq s_{2}\leq T.
\end{align}

Combining (\ref{eq proof of theorem regret 2}) and (\ref{eq proof of theorem regret 3}), we define a good event $\Omega$ as follows
\begin{align*}
\Omega\triangleq
& \left\{ 
h=1,\ldots,H,\, 1\leq t_{1}\leq t_{2}\leq T,\, 1\leq s_{1}\leq s_{2}\leq T:\:
\left|\Hat{p}^{+}_{h,\,t_{1},\,t_{2}}- p^{+}_{h,\,t_{1},\,t_{2}}\right|
\leq\frac{4\sqrt{\sum_{t'=t_{1}}^{t_{2}}\nu^{+}_{h,\,t'}\ln{2H T}}}{t_{2}-t_{1}+1}, \right.
\nonumber\\
& \qquad \qquad \left. 
\left|\Hat{p}^{-}_{h,\,s_{1},\,s_{2}}- p^{-}_{h,\,s_{1},\,s_{2}}\right|
\leq\frac{4\sqrt{\sum_{t'=s_{1}}^{s_{2}}\nu^{-}_{h,\,t'}\ln{2H T}}}{s_{2}-s_{1}+1}
\right\}.
\end{align*}

According to (\ref{eq proof of theorem regret 2}) and (\ref{eq proof of theorem regret 3}), $\mathrm{Pr}\left\{\Omega\right\}\geq 1 - 2/T^{2}$. In the rest of the proof, we always restrict our discussions to $\Omega$, unless otherwise specified.

\noindent\textbf{Step 2. Decomposing regret.}
In this step, we decompose the regret function $R(T)$ into the regret in different episodes. Suppose that there are $Q^{+}_{h}$ episodes for $\{p^{+}_{h,\,t}\}_{t=1}^{T}$ and $Q^{-}_{h}$ episodes for $\{p^{-}_{h,\,t}\}_{t=1}^{T}$, $\forall h=1,\ldots,H$. In addition, we let $1=\tau^{+}_{h,\,1}<\tau^{+}_{h,\,2}<\cdots<\tau^{-}_{h,\,Q^{+}}<T$ and $1=\tau^{-}_{h,\,1}<\tau^{-}_{h,\,2}<\cdots<\tau^{-}_{h,\,Q^{-}}<T$ be the exact times that are detected to be change points by Algorithm \ref{algorithm 1} for 
$\{p^{+}_{h,\,t}\}_{t=1}^{T}$ and $\{p^{-}_{h,\,t}\}_{t=1}^{T}$, respectively. According to Theorem \ref{theorem approximate optimal strategy}, for all $t$ such that $\tau^{+}_{h,\,q_{1}}< t\leq \tau^{+}_{h,\,q_{1}+1}$ and $\tau^{-}_{h,\,q_{2}}< t\leq \tau^{-}_{h,\,q_{2}+1}$, where $1\leq q_{1}\leq Q^{+}_{h}$ and 
$1\leq q_{2}\leq Q^{-}_{h}$, we have
\begin{align}\label{eq proof of theorem regret 4}
& \mathbf{E}\left[C(\hat{\pi}^{*}_{t};\: L_{t},\,U^\text{P}_{t},\,U^\text{F}_{t})\,|\,\mathcal{B}_{t-1}\right]
-\mathbf{E}\left[C(\pi^{*}_{t};\:L_{t},\,U^\text{P}_{t},\,U^\text{F}_{t})\,|\,\mathcal{B}_{t-1}\right]\nonumber\\
\leq~  &
K^{+}\max_{h=1,\ldots,H}
\left|p^{+}_{h,\,t-1}-\Hat{p}^{+}_{h,\,\tau^{+}_{q_{1}},\,t-1}\right|
+ K^{-}\max_{h=1,\ldots,H}
\left|p^{-}_{h,\,t-1}-\Hat{p}^{-}_{h,\,\tau^{+}_{q_{2}},\,t-1}\right|.
\end{align}
By combining all episodes together, (\ref{eq proof of theorem regret 4}) implies
\begin{align}\label{eq proof of theorem regret 5}
R(T)\leq
K^{+}\sum_{q=1}^{Q^{+}}\sum_{t=\tau^{+}_{q}+1}^{\tau^{+}_{q+1}}\max_{h=1,\ldots,H}
\left|p^{+}_{h,\,t-1}-\Hat{p}^{+}_{h,\,\tau^{+}_{q},\,t-1}\right|
+ K^{-}\sum_{q=1}^{Q^{-}}\sum_{t=\tau^{-}_{q}+1}^{\tau^{-}_{q+1}}\max_{h=1,\ldots,H}
\left|p^{-}_{h,\,t-1}-\Hat{p}^{-}_{h,\,\tau^{+}_{q},\,t-1}\right|,
\end{align}
where we let $\tau^{+}_{h,\,Q^{+}+1}=\tau^{-}_{h,\,Q^{-}+1}=T$ for notational convenience.

\noindent\textbf{Step 3. Analyzing regret in each episode.}
For any given $h$, we consider a single episode $1\leq q\leq Q^{+}_{h}$ of the sequence 
$\{p^{+}_{h,\,t}\}_{t=1}^{T}$. The same process can be used for $\{p^{-}_{h,\,t}\}_{t=1}^{T}$ to draw a similar result. We suppose that there are $U^{+}_{h,\,q}$ real change points of $\{p^{+}_{h,\,t}\}_{t=1}^{T}$ falling into the time interval $(\tau^{+}_{h,\,q},\,\tau^{+}_{h,\,q+1})$. 
We let these specific change points be $\tau^{+}_{h,\,q}< T_{h,\,q,\,1}<\cdots<T_{h,\,q,\,U^{+}_{q}}<\tau^{+}_{h,\,q+1}$ and introduce $T_{h,\,q,\,U^{+}_{q}+1}=T$ for notational convenience. For all $1\leq u\leq U^{+}_{h,\,q}$, and $T_{h,\,q,\,u}< t\leq T_{h,\,q,\,u+1}$,
\begin{align*}
\left|p^{+}_{h,\,t-1}-\Hat{p}^{+}_{h,\,\tau^{+}_{q},\,t-1}\right|
\leq \left|p^{+}_{h,\,t-1}-\Hat{p}^{+}_{h,\,T_{q,\,u},\,t-1}\right|
+ \left|\Hat{p}^{+}_{h,\,T_{q,\,u},\,t-1}-\Hat{p}^{+}_{h,\,\tau^{+}_{q},\,t-1}\right|.
\end{align*}
Since there is no change point between $T_{h,\,q,\,u}$ and $t$, restricted to $\Omega$, we have
\begin{align*}
\left|p^{+}_{h,\,t-1}-\Hat{p}^{+}_{h,\,T_{q,\,u},\,t-1}\right| 
= \left|p^{+}_{h,\,T_{q,\,u},\,t-1}-\Hat{p}^{+}_{h,\,T_{q,\,u},\,t-1}\right| 
\leq 
\frac{4\sqrt{\sum_{t'=T_{h,\,q,\,u}}^{t-1}\nu^{+}_{h,\,t'}\ln{2HT}}}{t-T_{h,\,q,\,u}}.
\end{align*}
Moreover, since $t$ is not detected to be a change point by Algorithm \ref{algorithm 1}, we have
\begin{align*}
\left|\Hat{p}^{+}_{h,\,T_{q,\,u},\,t-1}-\Hat{p}^{+}_{h,\,\tau^{+}_{q},\,t-1}\right|
\leq 
\frac{4\sqrt{\sum_{t'=T_{h,\,q,\,u}}^{t-1}\nu^{+}_{h,\,t'}\ln{2HT}}}{t-T_{h,\,q,\,u}}
+ \frac{4\sqrt{\sum_{t'=\tau^{+}_{h,\,q}}^{t-1}\nu^{+}_{h,\,t'}\ln{2HT}}}
{t-\tau^{+}_{h,\,q}}.
\end{align*}
Consequently, combining all three inequalities above yields
\begin{align}\label{eq proof of theorem regret 6}
\left|p^{+}_{h,\,t-1}-\Hat{p}^{+}_{h,\,\tau^{+}_{q},\,t-1}\right|
\leq \frac{8\sqrt{\sum_{t'=T_{h,\,q,\,u}}^{t-1}\nu^{+}_{h,\,t'}\ln{2HT}}}
{t-T_{h,\,q,\,u}}
+ \frac{4\sqrt{\sum_{t'=\tau^{+}_{h,\,q}}^{t-1}\nu^{+}_{h,\,t'}\ln{2HT}}}
{t-\tau^{+}_{h,\,q}}.
\end{align}
Meanwhile, noting that there is no change point between $\tau^{+}_{h,\,q}$ and $T_{h,\,q,\,1}$, for all $\tau^{+}_{h,\,q}<t\leq T_{h,\,q,\,1}$, we have 
\begin{align}\label{eq proof of theorem regret 7}
\left|p^{+}_{h,\,t-1}-\Hat{p}^{+}_{h,\,\tau^{+}_{q},\,t-1}\right| \leq
\frac{4\sqrt{\sum_{t'=\tau^{+}_{h,\,q}}^{t-1}\nu^{+}_{h,\,t'}\ln{2HT}}}
{t-\tau^{+}_{h,\,q}}.
\end{align}
Combining (\ref{eq proof of theorem regret 6}) and (\ref{eq proof of theorem regret 7}), we have 
\begin{align*}
& \sum_{t=\tau^{+}_{h,\,q}+1}^{\tau^{+}_{h,\,q+1}}
\left|p^{+}_{h,\,t-1}-\Hat{p}^{+}_{h,\,\tau^{+}_{h,\,q},\,t-1}\right|
\leq 
\sum_{t=\tau^{+}_{h,\,q}+1}^{\tau^{+}_{h,\,q+1}}
\frac{4\sqrt{\sum_{t'=\tau^{+}_{h,\,q}}^{t-1}\nu^{+}_{h,\,t'}\ln{2HT}}}
{t-\tau^{+}_{h,\,q}}
+ \sum_{u=1}^{U^{+}_{h,\,q}}\sum_{t=T_{h,\,q,\,u}+1}^{T_{h,\,q,\,u+1}}
\frac{8\sqrt{\sum_{t'=T_{h,\,q,\,u}}^{t-1}\nu^{+}_{h,\,t'}
\ln{2HT}}}{t-T_{h,\,q,\,u}}\nonumber\\
\leq &
\sum_{t=\tau^{+}_{h,\,q}+1}^{\tau^{+}_{h,\,q+1}}
4\sqrt{\frac{\max_{1\leq t'\leq T}\nu^{+}_{h,\,t'}\ln{2HT}}{t-\tau^{+}_{h,\,q}}}
+ \sum_{u=1}^{U^{+}_{h,\,q}}\sum_{t=T_{h,\,q,\,u}+1}^{T_{h,\,q,\,u+1}}
8\sqrt{\frac{\max_{1\leq t'\leq T}\nu^{+}_{h,\,t'}\ln{2HT}}{t-T_{h,\,q,\,u}}}.
\end{align*}
According to the relation of $\sqrt{1}+\sqrt{1/2}+\cdots+\sqrt{1/n}\leq \sqrt{n+1}$, we have
\begin{align}\label{eq proof of theorem regret 8}
\sum_{t=\tau^{+}_{h,\,q}+1}^{\tau^{+}_{h,\,q+1}}
\left|p^{+}_{h,\,t-1}-\Hat{p}^{+}_{h,\,\tau^{+}_{q},\,t-1}\right|
\leq
4\sqrt{\max_{1\leq t'\leq T}\nu^{+}_{h,\,t'}\ln{2HT}}
\left(\sqrt{\tau^{+}_{h,\,q+1}-\tau^{+}_{h,\,q}} +
2\sum_{u=1}^{U^{+}_{h,\,q}}\sqrt{T_{h,\,q,\,u+1}-T_{h,\,q,\,u}}\right).
\end{align}
%

\noindent\textbf{Step 4. Combining regrets in all episodes.}
Given $h$, we still first consider the sequence $\{p^{+}_{h,\,t}\}_{t=1}^{T}$, and conclude similar results for $\{p^{-}_{h,\,t}\}_{t=1}^{T}$. 
We prove by contradiction that when restricted to $\Omega$, the total number of episodes is smaller than that of the change points, i.e., $Q^{+}_{h}\leq \Lambda^{+}_{h}$.
We show this by proving a sufficient condition: $U^{+}_{h,\,q}\geq 1$. 
We suppose on the contrary that, for some $1\leq q\leq Q^{+}_{h}$, we have $U^{+}_{h,\,q}=0$. By (\ref{eq condition p+}), there exists 
$\tau^{+}_{h,\,q}\leq t_{1}\leq t_{2}< \tau^{+}_{h,\,q+1}$ such that
\begin{align}\label{eq proof of theorem regret 9}
\left|\Hat{p}^{+}_{h,\,t_{1},\,t_{2}}-\Hat{p}^{+}_{h,\,\tau^{+}_{h,\,q},\,\tau^{+}_{h,\,q+1}}\right|
\leq 
\frac{4\sqrt{\sum_{t'=t_{1}}^{t_{2}}\nu^{+}_{h,\,t'}\ln{2HT}}}{t_{2}-t_{1}+1}
+ \frac{4\sqrt{\sum_{t'=\tau^{+}_{h,\,q}}^{\tau^{+}_{h,\,q+1}-1}
\nu^{+}_{h,\,t'}\ln{2HT}}}{\tau^{+}_{h,\,q+1}-\tau^{+}_{h,\,q}}.
\end{align}

Meanwhile, since there is no change point between $\tau^{+}_{h,\,q}$ and $\tau^{+}_{h,\,q+1}$, for all $\tau^{+}_{h,\,q}\leq t< \tau^{+}_{h,\,q+1}$, we have $p^{+}_{h,\,t}=p^{+}_{h,\,\tau^{+}_{h,\,q}}$. 
Then, restricted to $\Omega$, for all $\tau^{+}_{h,\,q}\leq t_{1}
\leq t_{2}< \tau^{+}_{h,\,q+1}$, we have
\begin{align}\label{eq proof of theorem regret 10}
& \left|\Hat{p}^{+}_{h,\,t_{1},\,t_{2}}-\Hat{p}^{+}_{h,\,\tau^{+}_{h,\,q},\,\tau^{+}_{h,\,q+1}}\right|
\leq 
\left|\Hat{p}^{+}_{h,\,t_{1},\,t_{2}} - p^{+}_{h,\,\tau^{+}_{q}}\right| 
+ \left|p^{+}_{h,\,\tau^{+}_{h,\,q}} 
- \Hat{p}^{+}_{h,\,\tau^{+}_{h,\,q},\,\tau^{+}_h,\,{q+1}}\right|
= \left|\Hat{p}^{+}_{h,\,t_{1},\,t_{2}} - p^{+}_{h,\,t_{1},\,t_{2}}\right|
\nonumber\\
& + \left|p^{+}_{h,\,\tau^{+}_{h,\,q},\,\tau^{+}_{h,\,q+1}-1} 
- \Hat{p}^{+}_{h,\,\tau^{+}_{h,\,q},\,\tau^{+}_{h,\,q+1}}\right|
\geq
\frac{4\sqrt{\sum_{t'=t_{1}}^{t_{2}}\nu^{+}_{h,\,t'}\ln{2HT}}}{t_{2}-t_{1}+1}
+ \frac{4\sqrt{\sum_{t'=\tau^{+}_{h,\,q}}^{\tau^{+}_{h,\,q+1}-1}
\nu^{+}_{h,\,t'}\ln{2HT}}}{\tau^{+}_{h,\,q+1}-\tau^{+}_{h,\,q}}.
\end{align}
Based on (\ref{eq proof of theorem regret 9}) and (\ref{eq proof of theorem regret 10}), we conclude by contradiction that $Q^{+}_{h}\leq \Lambda^{+}_{h}$. 
Next, we combine the regrets in all episodes of $\{p^{+}_{h,\,t}\}_{t=1}^{T}$. 
According to \eqref{eq proof of theorem regret 8}, we have
\begin{align*}
\sum_{q=1}^{Q^{+}_{h}}\sum_{t=\tau^{+}_{h,\,q}+1}^{\tau^{+}_{h,\,q+1}}
\left|p^{+}_{h,\,t-1}-\Hat{p}^{+}_{h,\,\tau^{+}_{h,\,q},\,t-1}\right|
\leq 
4\sqrt{\max_{1\leq t'\leq T}\nu^{+}_{h,\,t'}\ln{2HT}}
\left(\sum_{q=1}^{Q^{+}_{h}}\sqrt{\tau^{+}_{h,\,q+1}-\tau^{+}_{h,\,q}} 
+ 2\sum_{q=1}^{Q^{+}_{h}}\sum_{u=1}^{U^{+}_{h,\,q}}
\sqrt{T_{h,\,q,\,u+1}-T_{h,\,q,\,u}}\right).
\end{align*}
Based on the Cauchy-Schwartz inequality 
$\sqrt{x_{1}}+\cdots+\sqrt{x_{n}}\leq \sqrt{n(x_{1}+\cdots+x_{n})}$, and the relationships of that $\sum_{q=1}^{Q^{+}_{h}}(\tau^{+}_{h,\,q+1}-\tau^{+}_{h,\,q})\leq T$,
$\sum_{u=1}^{U^{+}_{h,\,q}}(T_{h,\,q,\,u+1}-T_{h,\,q,\,u})\leq \tau^{+}_{h,\,q+1}
-\tau^{+}_{h,\,q}$ and $\sqrt{\tau^{+}_{h,\,q+1}-T_{h,\,q,\,U^{+}_{h,\,q}}}
\leq \sqrt{T_{h,\,q+1,\,1}-T_{h,\,q,\,U^{+}_{h,\,q}}}$, we have
\begin{align}\label{eq proof of theorem regret 11}
& \sum_{q=1}^{Q^{+}_{h}}\sum_{t=\tau^{+}_{h,\,q}+1}^{\tau^{+}_{h,\,q+1}}
\left|p^{+}_{h,\,t-1}-\Hat{p}^{+}_{h,\,\tau^{+}_{h,\,q},\,t-1}\right|
\leq 
4\sqrt{\max_{1\leq t'\leq T}\nu^{+}_{h,\,t'}\ln{2HT}}
\left(\sqrt{Q^{+}_{h}T} + 2\sqrt{\Lambda^{+}_{h}T}\right)\nonumber\\
\leq &
12\sqrt{\Lambda^{+}_{h}\max_{1\leq t'\leq T}\nu^{+}_{h,\,t'}}\sqrt{T\ln{2HT}}.
\end{align}

We can use a similar procedure to the same proof for $\{p^{-}_{t}\}_{t=1}^{T}$ to obtain the following result:
\begin{align}\label{eq proof of theorem regret 12}
\sum_{q=1}^{Q^{-}_{h}}\sum_{t=\tau^{-}_{h,\,q}+1}^{\tau^{-}_{h,\,q+1}}
\left|p^{-}_{h,\,t-1}-\Hat{p}^{-}_{h,\,\tau^{-}_{h,\,q},\,t-1}\right|
\leq 12\sqrt{\Lambda^{-}_{h}\max_{1\leq t'\leq T}\nu^{-}_{h,\,t'}}\sqrt{T\ln{2HT}}.
\end{align}
%

\noindent\textbf{Step 5. Bounding the total regret.}
According to (\ref{eq proof of theorem regret 11}) and (\ref{eq proof of theorem regret 12}), restricted to $\Omega$, we have
\begin{align*}
R(T) \leq 
12\max_{h=1,\ldots,H}\sqrt{\Lambda^{+}_{h}
\max_{1\leq t'\leq T}\nu^{+}_{h,\,t'}}\sqrt{T\ln{2HT}}
+ 12\max_{h=1,\ldots,H}\sqrt{\Lambda^{-}_{h}
\max_{1\leq t'\leq T}\nu^{-}_{h,\,t'}}\sqrt{T\ln{2HT}}.
\end{align*}
Meanwhile, since 
$\left|p^{+}_{h,\,t-1}-\Hat{p}^{+}_{h,\,\tau^{+}_{h,\,q},\,t-1}\right|\leq1$
and $\left|p^{-}_{h,\,t-1}-\Hat{p}^{-}_{h,\,\tau^{-}_{h,\,q},\,t-1}\right|\leq1$ hold for all $1< t\leq T$ and $h=1,\ldots,H$, we have a natural bound on $R(T)$ as
\begin{align*}
R(T)\leq \left(K^{+} + K^{-}\right)T.
\end{align*}
Therefore, we have
\begin{align*}
& R(T) \leq
\mathrm{Pr}\left(\Omega\right)\cdot
12\left[K^{+}\max_{h=1,\ldots,H}
\sqrt{\Lambda^{+}_{h}\max_{1\leq t'\leq T}\nu^{+}_{h,\,t'}}
+ K^{-}\max_{h=1,\ldots,H}\sqrt{\Lambda^{-}_{h}
\max_{1\leq t'\leq T}\nu^{-}_{h,\,t'}}\right]
\sqrt{T\ln{2T}}\nonumber\\
& + 
\mathrm{Pr}\left(\Omega^{c}\right)\left(K^{+} + K^{-}\right)T
\leq
12\left[K^{+}\max_{h=1,\ldots,H}
\sqrt{\Lambda^{+}_{h}\max_{1\leq t'\leq T}\nu^{+}_{h,\,t'}}
+ K^{-}\max_{h=1,\ldots,H}\sqrt{\Lambda^{-}_{h}
\max_{1\leq t'\leq T}\nu^{-}_{h,\,t'}}\right]\sqrt{T\ln{2T}}\nonumber\\
& + \frac{2\left(K^{+} + K^{-}\right)}{T}.
\end{align*}
Thus, we conclude the proof.
\Halmos

\section{Details on the Power Systems}\label{appen:PS}
Figure \ref{fig 11 node original} shows the structure of the 11-node power system used in our simulation studies in Section~\ref{section simulation study}.
Figure \ref{fig 11 node grid} shows the coordinates of different components of the smart power system in the 11-node system.
Figure \ref{fig 57 node grid} shows the coordinates of different components of the 57-node system.

\begin{figure}[h]
    \centering
    \includegraphics[width=0.525\textwidth]{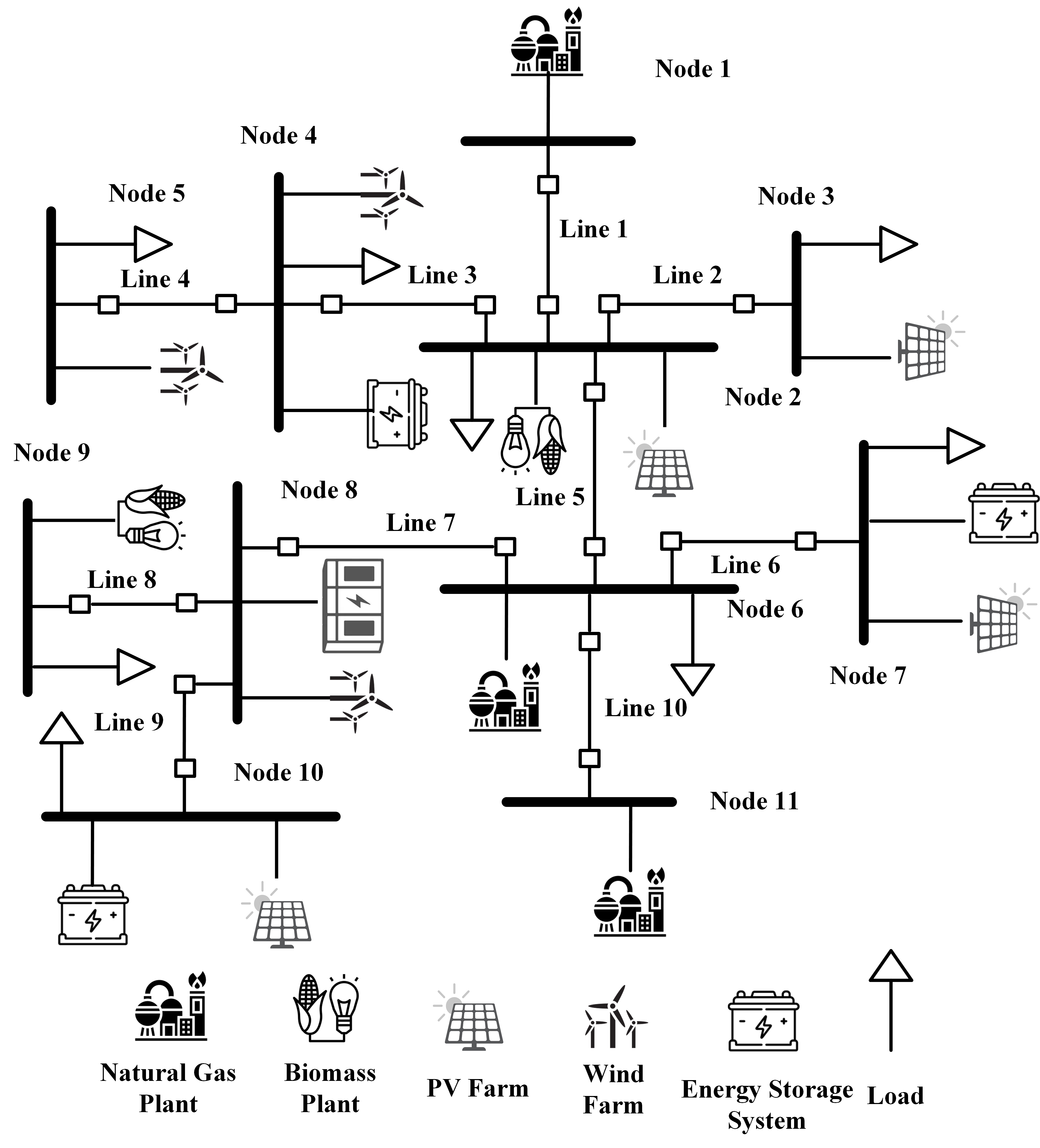}
    \caption{The original $11$-node smart power system model.}
    \label{fig 11 node original}
\end{figure}

\begin{figure}[h]
    \centering
    \includegraphics[width=0.7\textwidth]{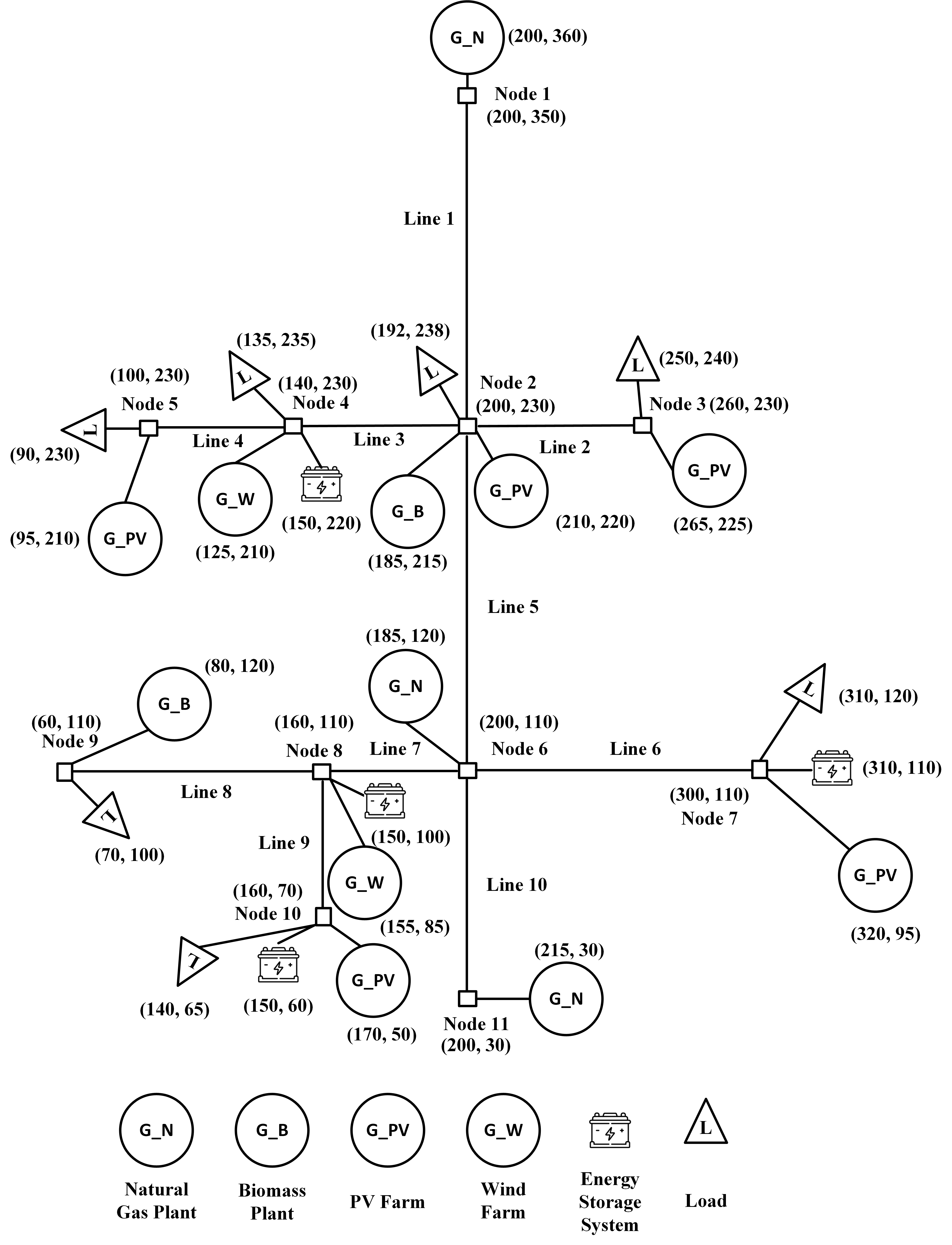}
    \caption{Coordinates of all the nodes of the smart power system in the grid $\mathcal{G}$.}
    \label{fig 11 node grid}
\end{figure}
\begin{figure}[h]
    \centering
    \includegraphics[width=1.0\textwidth]{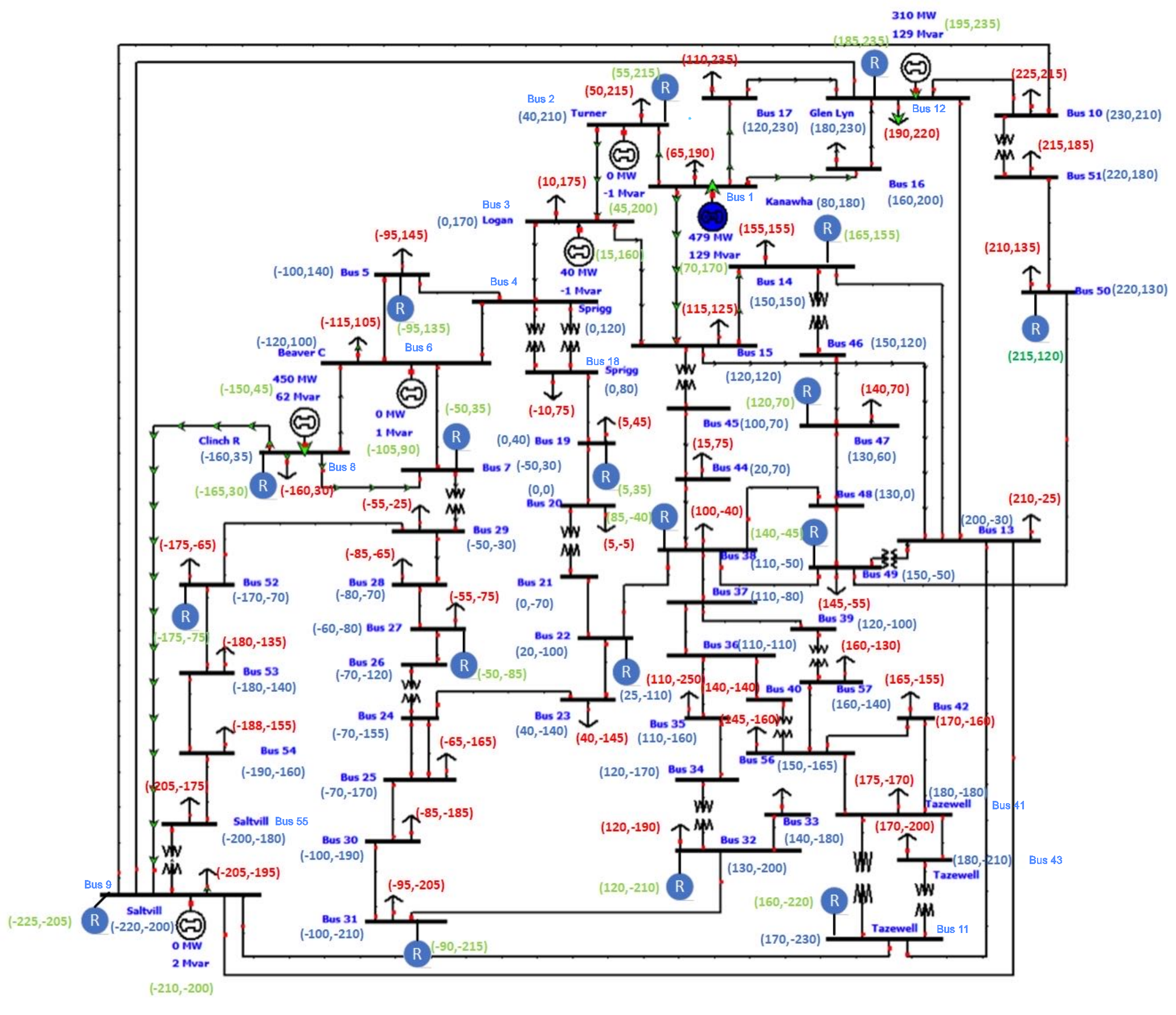}
    \caption{Coordinates of all the nodes of the IEEE 57-bus power system in the grid $\mathcal{G}$.}
    \label{fig 57 node grid}
\end{figure}

\bibliographystyleappendix{informs2014} 

\clearpage

\end{APPENDICES}

\end{document}